\documentclass[10pt,preprint]{aastex}

\usepackage{amsmath}
\usepackage{amssymb}
\usepackage{xspace}
\usepackage{lscape}
\usepackage{graphicx}

\newcommand{\g}{{\ensuremath{\mathrm{g}}}\xspace}
\newcommand{\K}{{\ensuremath{\mathrm{K}}}\xspace}

\newcommand{\km}{{\ensuremath{\mathrm{km}}}\xspace}
\newcommand{\Msun}{{\ensuremath{\mathrm{M}_{\odot}}}\xspace}
\newcommand{\Sec}{{\ensuremath{\mathrm{s}}}\xspace}
\newcommand{\erg}{{\ensuremath{\mathrm{erg}}}\xspace}

\newcommand{\erggs}{{\ensuremath{\erg\,\g^{-1}\,\Sec^{-1}}}\xspace}

\newcommand{\foe}{{\ensuremath{\Ep{51}\,\erg}}\xspace}

\newcommand{\MeV}{{\ensuremath{\mathrm{MeV}}}\xspace}

\newcommand{\ud}{$^\dag$}

\newcommand{\lFig}[1]{{\label{fig:#1}}}

\newcommand{\lTab}[1]{{\label{tab:#1}}}

\newcommand{\pan}[1]{{\textit{#1}}}

\newcommand{\FIGFF}[2]{{\ref{fig:#2}\pan{#1}}}
\newcommand{\Figff}[1]{{\FIGFF{}{#1}}}
\newcommand{\FIG}[2]{{Fig.~\FIGFF{#1}{#2}}}
\newcommand{\Fig}[1]{{\FIG{}{#1}}}

\newcommand{\Figures}[1]{{Figures~\FIGFF{}{#1}}}
\newcommand{\Sectff}[1]{{\ref{sec:#1}}}
\newcommand{\Sect}[1]{{\S~\Sectff{#1}}}

\newcommand{\I}[2]{{\isotope{}{#1}{#2}}}

\newcommand{\Ep}[1]{{\ensuremath{10^{#1}}}}
\newcommand{\E}[1]{{\ensuremath{\powersep\Ep{#1}}}}

\newcommand{\Ye}{{\ensuremath{Y_e}}\xspace}

\newcommand{\Eexp}{{\ensuremath{E_{\mathrm{expl}}}}\xspace}
\newcommand{\Mpist}{{\ensuremath{M_{\mathrm{piston}}}}\xspace}
\newcommand{\gPist}{{\ensuremath{g_{\mathrm{piston}}}}\xspace}

\newcommand{\Rag}{{\ensuremath{(\alpha,\gamma)}}\xspace}
\newcommand{\Ran}{{\ensuremath{(\alpha,\mathrm{n})}}\xspace}

\newcommand{\Rng}{{\ensuremath{(\mathrm{n},\gamma)}}\xspace}

\newcommand{\ltaprx}{\lesssim}

\begin{document}

\title{Nucleosynthesis in Massive Stars With Improved 
Nuclear and Stellar Physics}


\author{T.\ Rauscher}
\affil{Departement f\"ur Physik und Astronomie, Universit\"at Basel, CH-4056
Basel, Switzerland\\
Department of Astronomy and Astrophysics, University of California,
Santa Cruz, CA 95064}
\email{Thomas.Rauscher@unibas.ch}

\author{A.\ Heger}
\affil{Department of Astronomy and Astrophysics, University of
California, Santa Cruz, CA 95064\altaffilmark{1}}
\email{alex@ucolick.org}

\author{R.\ D.\ Hoffman}
\affil{Nuclear Theory and Modeling Group, Lawrence  Livermore National
Laboratory, Livermore, CA 94550}
\email{rdhoffman@llnl.gov}

\author{S.\ E.\ Woosley}
\affil{Department of Astronomy and Astrophysics, University of California,
Santa Cruz, CA 95064}
\email{woosley@ucolick.org}

\altaffiltext{1}{Now at University of Chicago, Department of Astronomy
and Astrophysics, 5640 S.\ Ellis Ave, Chicago, IL 60637}


\begin{abstract}

%% NOTE THAT APJ DOES NOT ALLOW THE REFERENCES (i.e., years!) 
%% IN THE ABSTRACT

We present the first calculations to follow the evolution of all
stable nuclei and their radioactive progenitors in stellar models
computed from the onset of central hydrogen burning through explosion
as Type II supernovae. Calculations are performed for Pop I stars of
15, 19, 20, 21, and 25\,\Msun using the most recently available
experimental and theoretical nuclear data, revised opacity tables,
neutrino losses, and weak interaction rates, and taking into account
mass loss due to stellar winds.  A novel ``adaptive'' reaction network
is employed with a variable number of nuclei (adjusted each time step)
ranging from $\sim700$ on the main sequence to $\gtrsim2200$ during
the explosion.  The network includes, at any given time, all relevant
isotopes from hydrogen through polonium ($Z=84$).  Even the limited
grid of stellar masses studied suggests that overall good agreement
can be achieved with the solar abundances of nuclei between $^{16}$O
and $^{90}$Zr.  Interesting discrepancies are seen in the 20\, \Msun
model and, so far, only in that model, that are a consequence of the
merging of the oxygen, neon, and carbon shells about a day prior to
core collapse.  We find that, in some stars, most of the
``$p$-process'' nuclei can be produced in the convective oxygen
burning shell moments prior to collapse; in others, they are made only
in the explosion. Serious deficiencies still exist in all cases for
the $p$-process isotopes of Ru and Mo.

\end{abstract}


\section{Introduction}
\label{sec:intro}

The nucleosynthetic yields of massive stars are important to many
areas of astronomical research. Besides the inherent interest in
understanding our nuclear origins, the abundances made in supernovae
are used to diagnose models for the explosion and as input to still
grander models for the formation and chemical evolution of galaxies
and the intergalactic medium. They are the target of x-ray
observations of supernova remnants and gamma-ray studies of
radioactivities in the interstellar medium. Some can be used
as cosmochronometers, others power the light curves, still others
appear as anomalous abundances found in tiny meteroitic 
grains in our own solar system.

For these reasons, nucleosynthesis calculations have a long history
and a sizable community that carries them out. Most recently,
nucleosynthesis in massive stars has been studied by
\citet[WW95]{WW95}; \citet{TNH96,lim00} and others. With this paper,
we embark on a new survey, similar to WW95, that will ultimately
include stars of many masses and initial metallicities.  The
characteristics of this new study are improvements in the stellar
physics (mass loss rates, opacities, reaction network, etc.,
\Sect{comp_proc}) and revisions to nuclear reaction rates
(\Sect{mods}) that have occurred during the last eight years. 

This first paper particularly addresses recent improvements in nuclear
physics.  For elements heavier than about silicon, the nuclear level
densities are sufficiently high (provided the particle separation
energies are not too small) that the statistical - or
``Hauser-Feshbach'' - model can be used. Here, in their maiden voyage,
we use rates calculated using the NON-SMOKER code \citep{rtk97,rt98}.
The reaction library, from which the network is drawn, includes all
nuclei from the proton-drip line to the neutron-drip line and elements
up to and including the actinides \citep{RATH}. For elements lighter
than silicon, where they have been measured, results are taken from
the laboratory. Several different compilations are explored. The most
critical choices are the rates for $^{12}$C($\alpha,\gamma)^{16}$O,
$^{22}$Ne($\alpha$,n)$^{25}$Mg, and $^{22}$Ne($\alpha,\gamma)^{26}$Mg.
In order to facilitate comparison, we have chosen a constant value
equal to 1.2 times that of Buchmann (1996) for the
$^{12}$C($\alpha,\gamma)^{16}$O rate in {\sl all} our calculations.
For our {\sl standard} models (defined in $\S$3.1) we further adopt
the lower bound of \citet{kaepp94} for $^{22}$Ne($\alpha$,n)$^{25}$Mg
\citep{HWW01}. In future publications we will explore, in greater
depth, the consequences of different choices for these rates (for
$^{12}$C($\alpha,\gamma)^{16}$O, see also \citealt{WW93,BHW02}).

A novel reaction network is employed, unprecedented in size for
stellar evolution calculations. The network used by WW95, large in its
day, had about 200 nuclides and extended only to germanium.  Studies
using reaction networks of over 5000 nuclei have been carried out for
single zones or regions of stars in order to obtain the $r$-process,
e.g., \citet{CCT85,fre99,kra93}, but ``kilo-nuclide'' studies of
nucleosynthesis in complete stellar models (typically of 1000 zones
each for 20,000 time steps) have not been done before. We describe in
\Sect{dynet} a dynamically evolving network that adds and subtracts
nuclides as appropriate during the star's life to ensure that all
significant nuclear flows are contained. Our present survey uses a
network that has the accuracy of a fixed network of 2500 isotopes.

Section 4 discusses aspects of the stellar evolution that are critical
to the nucleosynthesis and \Sect{results} gives the main results of
our survey. We find overall good agreement of our nucleosynthesis
calculations with solar abundances for intermediate mass elements
(oxygen through zinc) as well as the ``weak component'' of the
$s$-process (A $\ltaprx$ 90), and most of the $p$-process
isotopes. However, there is a systematic deficiency of $p$-process
isotopes below A $\approx$ 125 that is particularly acute for Mo and
Ru, and around A $\approx$ 150.  Possible explanations are discussed
in \Sect{gamma}.  We also find that the nucleosynthesis is at least as
sensitive to the stellar model as to the nuclear physics and, in
particular, find unusual results for a 20\,\Msun model (in the sense
that the results differ greatly from both the sun and those at either
19 or 21\,\Msun).  This is because of the merging of convective
oxygen, neon, and carbon shells that occurred well before collapse in
that model and not in the others (\Sect{results}).

\section{Stellar Physics and Computational Procedure}
\label{sec:comp_proc}

All stellar models were calculated using the implicit hydrodynamics
package, KEPLER (\citealt{WZW78}; WW95), with several improvements to
the physical modeling of stellar structure and to the nuclear reaction
network (see also \citealt{heg01b,HWW01,r_nic}). In the following, we
discuss only those improvements. For further details on the basic
approaches we refer the reader to the previous publications
\citep[e.g.\ see][]{WZW78,WW93,WW95}.

\subsection{New And Improved Physics Since WW95}
\label{sec:mods_stell}

The most important change in stellar physics compared to WW95 is the
inclusion of mass loss.  The prescription employed \citep{NJ90} gives
a mass loss rate that is sensitive to surface temperature and
luminosity.  Operationally, the mass lost in each time step is
subtracted from a stellar zone situated 0.01\,\Msun below the surface
of the star and the surface layers are automatically rezoned (dezoned)
whenever necessary, while conserving mass, momentum, energy, and
composition.  The advection term for the enthalpy flux and the
expansion term due to mass loss can be neglected in these outer layers
since the energy input in the mass loss ($u\nabla\phi$) is small
compared to the luminosity.  
%XXX
The total mass loss of our stars is dominated by the RSG phase and
depends on the modeling of semiconvection and overshooting, and on the
\I{12}C\Rag\I{16}O rate \citep{BHW02}.  We obtain final masses that
lie between the ``normal'' and the ``double'' mass loss rates of
\citet{sch92}.

For temperatures below \Ep8\,\K, the OPAL95 opacities are employed
\citep{IR96}. In particular, these result in a better representation
of the hydrogen envelope in the red giant stage where the ``iron
bump'' in the opacities of \citet{IR96} is known to be important
\citep[e.g.,][]{heg97}. Above \Ep8\,\K, the opacity was the same as in 
WW95 (and Weaver, Zimmerman, \& Woosley 1978).

Plasma neutrino losses were updated to use the rates by \citet[coding
by F.\ Timmes, private communication]{ito96}.  Hydrogen burning and
weak losses during this phase were updated as described in
\citet{HLW00}.  Weak rates and weak neutrino losses in the late
evolution stages now use the rates of \citeauthor{LM00}
(\citeyear{LM00}; see \citealp{heg01a,heg01b}).

Convection is basically treated as presented in \citet{WZW78,WW93}
\citep[see also][for a recent summary]{WHW02}.  However, convective
overshooting (on either side) is now suppressed for isolated
convective zone interfaces to avoid a numerical instability present in
the work of WW95.  Changes to the nuclear reaction rates, both strong
and weak, are discussed in \Sect{mods}.

\subsection{Dynamic Nuclear Reaction Network}
\label{sec:dynet}

As in WW95, two reaction networks are used.  A small network directly
coupled to the stellar model calculation provides the approximate
nuclear energy generation rate \citep{WZW78}, i.e, is solved implicitly
with the Newton-Raphson solver for each time-step in each zone.  A
larger one is used to track the nucleosynthesis.  This smaller network
is essentially the same as in WW95 and WZW78, but with updated nuclear
rates as described in the following sections.  For a study of small
vs.\ large networks and their ability to accurately and efficiently
provide input values of the nuclear energy generation rate during
advance stages of stellar evolution see \citet{THW00}.

The nucleosynthesis is coupled to convection in an ``operator split''
mode.  At the end of each time-step for the evolutionary model, the
large reaction network is called, for the existing conditions
(temperature and density), and the chemical species are diffusively
mixed. To save computer time, the composition is not updated in zones
where the temperature is too low for any nuclear activity during the
previous time step, though every zone participates in convective
mixing, where appropriate, every time step.

One of the major improvements over WW95 and other stellar models is
that, for the first time, the synthesis of all nuclides of any
appreciable abundance is followed simultaneously in an \emph{adaptive
network}.  Using a library containing rate information for 4,679
isotopes from hydrogen to astatine, the ``adaptive'' network
automatically adjusts its size to accommodate the current nuclear
flows.  This means the constitution of the network evolves to
accommodate the most extreme thermodynamic conditions present anywhere
in the model.  If the mass fraction of an isotope exceeds a parameter,
here \Ep{-18}, anywhere in the star, the neighboring isotopes, and all
to which that species might decay, are added.  Similarly, if the
abundance of an isotope drops below \Ep{-24}, it is removed (unless it
is along the decay chain of an abundant nucleus).  Because different
zones may become convectively coupled at unpredictable times, the same
network must be used throughout the star. The network includes all
strong reactions involving nucleons, $\alpha$-particles, and photons plus a
few ``special'' reactions for light isotopes (like the triple $\alpha$
process, $^{12}$C+$^{12}$C, etc.; for details see WW95) as well as all
weak interactions - electron capture, positron decay, and beta-decay.
The network is solved implicitly using a sparse matrix inverter
\citep{TWW95}.

For our 15\,\Msun star, for example, the network initially contained
645 isotopes during hydrogen burning, including 283 stable or
long-lived species (like \I{40}K or \I{180}{Ta}). This number grew to
$\gtrsim740$ isotopes at the end of central helium burning (to
accommodate the $s$-process, $\gtrsim850$ during carbon (shell)
burning, $\gtrsim1,050$ during oxygen burning, $\gtrsim1,230$ during
oxygen shell burning, and $\gtrsim1,400$ at the presupernova
stage. When the supernova shock hit the base of the helium shell
causing a weak $r$-process, the network reached its maximum size,
$\gtrsim2,200$.  In total 2,435 different isotopes were included at
one time or another.  A network plot is shown in Fig.\
\ref{fig:S15net}.

The major limitation of this network is that it purposefully does not
include elements beyond astatine, i.e., the heavy $r$-process and
fission-cycling could not be followed. Given the current uncertainties
in the explosion mechanism and our simplified treatment, a description
of the high-entropy zones close to the proto-neutron star is beyond
the scope of this paper. Thus, we did not calculate the $r$-process that
might occur in the neutrino wind \citep{woo94}.  Another limitation in
the current implementation of the network is that it only includes one
state per isotope, which limits its accuracy for a few isotopes like
\I{26}{Al} and \I{180}{Ta}.

\section{Nuclear Physics}
\label{sec:mods}

Since 1993, when the nuclear physics used in WW95 was ``frozen'',
there have been numerous revisions to nuclear reaction rates. In this
work we are presenting our choice of a thoroughly updated standard
rate set, including most recent experimental data and theoretical
results. Numerically, the greatest fraction of nuclear information is
theoretical, coming from a statistical model calculation \citep{RATH},
described in \Sect{hf}. These theoretical rates are supplemented by
experimental values where they are known. Details are discussed in the
following sections and the experimental rates employed are summarized
in Table \ref{tab:rrrm}. The table is truncated at $^{42}$Ca because
above that, with the few exceptions indicated, all rates are either
from \citet{bao00} for \Rng reactions (see Table \ref{tab:macs} for a
listing of these nuclei) or \citet{RATH}.

The weak rates used here are discussed in \Sect{weak}.

\subsection{Experimental Rates Below Silicon}
\label{sec:light}

Nuclear reactions involving elements lighter than silicon are
particularly important both for nucleosynthesis and determining the
stellar structure.  Our standard set of light element reaction rates
begins with \citet{CF88} as updated by \citet{HWW01} and \citet{id01}.
Further minor modifications were made to the rates $^1$H(n,$\gamma$)D
and $^3$He(n,$\gamma$)$^4$He \citep{REACLIB95}. Table \ref{tab:rrrm}
gives the sources of all charged particle reaction rates not taken
from \citet{RATH}. The proper references can be found in Table
\ref{tab:rrrm_expl}.

Some important rates, e.g., those of \citet{id01}, have been given by
these authors in tabular form and not as fitted functions of
temperature.  In Appendix \ref{app:fit} we describe a novel procedure
that we shall follow for all such tabulations in the future and which
we recommend to the community.  The bulk of the temperature
sensitivity is extracted from the rate using a simple fitting function
that does \emph{not} by itself give the necessary accuracy across the
temperature grid. The ratios of the actual rate to the fitting
function are then carried as a table in the computer and can be
interpolated much more accurately than the rate itself.

Besides models that used this standard set of nuclear physics (the
``S''-series of models; 15, 19, 20, 21, and 25\,\Msun), we also
present calculations using two other rate sets. This was done to
facilitate the comparison of different choices of reaction rates in
otherwise identical stellar models.  One other set was the NACRE
compilation of charged particle rates \citet{ang99} (Model series
``N''; for 15, 20, and 25\,\Msun). For one 25\, \Msun star, we used
the network and rates of \citet{HWW01}. Since that work only included
nuclear data up to about mass 110, the recalculation here used a
similar static network of 477 nuclides (Set ``H''; for 25\,\Msun
only).

\subsubsection{$^{12}$C($\alpha,\gamma)^{16}$O}
\label{sec:c12ag}

Of utmost importance for nucleosynthesis is the rate adopted for
$^{12}$C($\alpha$,$\gamma$)$^{16}$O.  The same value was used in all
studies reported in this paper (sets S, N, H) since variation of this
single rate would alter the stellar model and obscure the sensitivity
to the other nuclear physics.  The rate previously used in WW95 and
\citet{HWW01} was that of \citet{CF88} multiplied by 1.7.  Here, the
more recent evaluation of \citet{Buc96,Buc00} was used as a basis for
the temperature dependence, but the overall rate was multiplied by a
factor of 1.2 to bring the recommended value $S(300)=146$ keV barn
into better accord with our standard $S(300)=170$ keV barn
\citep{WW93}.  This value and temperature dependence is also
consistent with recent measurements by \citet{kun01a,kun01b}.

\subsubsection{$^{22}$Ne($\alpha,$n)$^{25}$Mg and
$^{22}$Ne($\alpha,\gamma$)$^{26}$Mg}
\label{sec:ne22an}

The reaction $^{22}$Ne($\alpha,$n)$^{25}$Mg, acting in competition
with $^{22}$Ne($\alpha,\gamma$)$^{26}$Mg, is critical for determining
the strength of the helium-burning $s$-process in massive stars
\citep[e.g.,][]{kaepp94}. Here, while experimenting with several
choices, the standard set employed is the lower limit of
\citeauthor{kaepp94} (\citeyear{kaepp94}; the same rates as used by
\citealt{HWW01}). This choice is in reasonable agreement with more
recent work by \citet{jae01}.  Following an early recommendation by
\citet{wiepriv}, only the resonance at 828 keV was considered in the
rate for $^{22}$Ne($\alpha$,n)$^{25}$Mg and the resonance at 633 keV
was ignored.  Further, the 828 keV resonance itself was given a
strength equal to its 1 $\sigma$ lower limit, 164 $\mu$eV.  The rate
for $^{22}$Ne($\alpha,\gamma$)$^{26}$Mg was that recommended by
K\"appeler et al., but with the strength for the 633 keV resonance
multiplied by 0.5.  The modified \citet{kaepp94} rates were merged
with the rate given by \citet{CF88}, which was used for temperatures
$T_9 \geq 0.6$.  For model series N, we used the same
\I{12}C($\alpha$,$\gamma$) rate ($1.2\times$ Buchmann 2000), but the
recommended values for $^{22}$Ne($\alpha$,$\gamma$)$^{26}$Mg,
$^{22}$Ne($\alpha,$n)$^{25}$Mg, and all other rates provided by NACRE.

\subsection{Experimental Rates Above Silicon}
\label{sec:inter}

Additional experimental rates for nuclei heavier than silicon are
given in Table \ref{tab:rrrm}. These are largely drawn from
\citet{hw92}. The entries in Table \ref{tab:rrrm} are referenced in
Table \ref{tab:rrrm_expl}.

Recent measurements of the reactions
$^{70}$Ge($\alpha$,$\gamma$)$^{74}$Se \citep{fue96} and
$^{144}$Sm($\alpha$,$\gamma$)$^{148}$Gd \citep{som98} are of great
importance for the $\gamma$-process yields.  Especially the
predictions of the latter reaction were found to be very sensitive to
the optical model $\alpha$ potential used \citep{WH90,rto96}.  For our
rate library, the resulting experimental rates of both reactions were
fitted to the format described in \citep{RATH}.  The experimental
information was also used to recalculate other rates involving the
same $\alpha$ potentials. See \Sect{hf} for a further discussion.

\subsection{Hauser-Feshbach Rates}
\label{sec:hf}

For those cases where experimental information was lacking and the
level density was sufficiently high (typically $A > 24$) we employed
the Hauser-Feshbach rates obtained using the NON-SMOKER code
\citep{rtk97,rt98}.  A library of theoretical reaction rates
calculated with this code and fitted to an analytical function ---
ready to be incorporated into stellar model codes --- was published
recently, in the following (and in Table \ref{tab:rrrm}) referred to
as RATH \citep{RATH,rt01}.  It includes binary reaction rates
involving nucleons, alpha-particles, and photons interacting with all
possible targets from neon to bismuth and all isotopes of these
elements from the proton to neutron drip-lines. It is thus the most
extensive published library of theoretical reaction rates to date. For
the network described here, we utilized the rates based on the FRDM
set as these provide the most reliable description around the valley
of stability.

Partition functions were also taken from \citet{RATH}, but were
converted to the format defined in \citet{HWFZ76} to be used in
KEPLER.  This was achieved by fitting them with the appropriate
functions.  Low-lying nuclear levels had to be used for a few cases
and in such a case the same information as for the NON-SMOKER
calculation was utilized \citep{rt01}.

Recent investigations underline the fact that the $\alpha$+nucleus
optical potential for intermediate and heavy targets is not well
understood at astrophysically relevant energies. Although $\alpha$
capture itself will be negligible for highly charged nuclei, the
optical potential still is a necessary ingredient to determine the
reverse ($\gamma$,$\alpha$) reaction which is important in the
$\gamma$-process \citep{WH90,rto96}.  Two $\alpha$ capture reactions
have been studied experimentally close to the relevant energy
range. While the reaction $^{70}$Ge($\alpha$,$\gamma$)$^{74}$Se
\citep{fue96} was essentially well predicted by theory and needed only
a small adjustment of the optical potential,
$^{144}$Sm($\alpha$,$\gamma$)$^{148}$Gd exhibited strong deviations
from previous estimates \citep{som98}. As stated above, the
experimental rates were implemented in RATH format in our rate
library.  Moreover, for reasons of consistency, all reactions
involving the channels $\alpha$+$^{70}$Ge and $\alpha$+$^{144}$Sm were
recalculated with the statistical model code NON-SMOKER, utilizing the
optical potentials derived from the capture data \citep{som98}. The
resulting fit parameters in RATH format are given in Table
\ref{tab:alpha}.

Of special interest are $\alpha$-capture reactions on self-conjugate
($N=Z$) target nuclei. The probabilities for these reactions are
suppressed by isospin effects and require special treatment in any
theoretical model. Capture data is scarce, even for lighter nuclei.
Recently, \citet{rtgw00} published a semi-empirical evaluation of
resonance data (i.e.\ $\alpha$ resonances taken not only from
($\alpha$,$\gamma$) reactions but also from other approaches) and
compared it to predictions made with the code NON-SMOKER which
includes an improved treatment of the isospin suppression effect
\citep{rt98}. Reasonable agreement was found around $T_9=1-2$. At
lower temperatures either the statistical model was not applicable or
had problems with the prediction of the optical $\alpha$ potential
(similar as discussed above), depending on the considered
reaction. Since we need a reliable rate across the whole temperature
range, a mixed approach was chosen: below a certain temperature
$T_{\rm match}$ the contributions of single resonances -- taken from
\citet{rtgw00} -- are added, above $T_{\rm match}$ the statistical
model rate renormalized to the experimental value at $T_{\rm match}$
is used. Table \ref{tab:iso} lists the parameters and temperatures.

Also important are the rates for neutron capture. These directly
affect both the neutron budget (acting as neutron "poisons"), and the
abundance of all $s$-process isotopes, including many species below
the iron group (see Table 3 of \citealt{WHW02}).  Where available, the
theoretical rates of RATH were supplemented by experiment using the
\citet{bao00} compilation of recommended neutron capture cross
sections.  Because only 30 keV MACS (Maxwellian Averaged Cross
Sections) are given in that reference, we renormalized the fits given
in RATH in order to obtain the same MACS values at 30 keV, thus
maintaining the (weak) temperature dependence of the theoretical
rates.  The normalization factors are given in Table
\ref{tab:macs}. Both forward and reverse rates of RATH are multiplied
by the same factor. For targets below Ne, for which statistical model
calculations cannot be applied with any accuracy, a $1/v$ dependence
of the cross section -- leading to a constant rate -- was assumed,
unless other experimental information was available.

\subsection{Weak Interactions}
\label{sec:weak}

The experimental $\beta^-$, $\beta^+$, and $\alpha$ decay rates of
\citeauthor{NWC5} (\citeyear{NWC5}; calculated from the laboratory
ground state half-life), and their respective branching ratios were
implemented.  Where feasible, a temperature-dependent weak rate was
coded that couples the ground state to a shorter lived excited state
(both assumed to be in thermal equilibrium, \citealt{clayton68}).
Further experimental $\beta^-$ decay rates were taken from
\citeauthor{kratzpriv} (\citeyear{kratzpriv}; see also
\citealt{moe97}).  For all other targets, we used the theoretical
$\beta^-$ and $\beta^+$ rates of \citet{moe97}.  As a special case, we
implemented a temperature-dependent $^{180\rm m}$Ta decay rate
\citep{end99}.

Usually, the ground state rates are a lower bound to the actual weak
decay rates. Where fitted functions are available, we also utilize
temperature and density-dependent weak rates
\citep{ffn1,ffn2,ffn3,ffn4} accounting for a continuum of excited
states.  An important change of the weak interaction rates for $45
\leq A \leq 65$ is brought about by the recent work of \citet{LM00}.
Where information is available, we use Langanke \& Martinez rates in
preference to Fuller et al. Their inclusion leads to interesting
changes in the presupernova structure (see \Sect{presup}), but not so
much in the abundances outside the iron core.

Neutrino losses are a critical aspect of stellar evolution in massive
stars beginning with carbon burning. The dominant losses before
silicon burning are due to thermal processes (chiefly
pair-annihilation), which provide a loss term that is very roughly
proportional to T$^9$ in the range of interest for advanced burning
stages \citep{clayton68}. This temperature sensitivity, combined with
the need to burn heavier fuels at higher temperatures to surmount the
increasing charge barriers, is what leads to the rapid decrease in
lifetime to burn a given fuel, with obvious consequences for
nucleosynthesis. We include the latest treatment \citep{ito96}.

The neutrino flux of a core-collapse supernova is high enough to
contribute to the nucleosynthesis of certain rare elements and
isotopes.  In this so-called $\nu$-process, inelastic neutral-current
scattering of a neutrino leads to the formation of an excited daughter
nuclide which then decays by particle emission.  Rare isotopes with
highly abundant ``neighbors'' (or neighbors of their radioactive
progenitors) can be significantly produced by this process.  As
previously used by WW95, we adopt the rates of \citet{woo90}.

\section{Stellar Evolution}
\label{sec:struc}

\subsection{Presupernova Evolution}
\label{sec:presup}

Table \ref{tab:struc_models} summarizes the presupernova properties of
the new models. The helium, carbon-oxygen, and neon-oxygen cores are
defined as the enclosed mass where hydrogen, helium, and carbon mass
fractions, first drop below 1\,\%.  The silicon core is defined by
where silicon becomes more abundant than oxygen and the iron core by
where the sum of the mass fractions of iron group nuclei first exceeds
50\,\% (all criteria applied moving inward).  The deleptonized core
is the region where the number of electrons per baryon, $Y_{\rm e}$,
drops below 0.49.

Revisions in opacity and the introduction of mass loss generally lead
to smaller helium cores which also tend to decrease the mass of the
carbon-oxygen and the silicon cores.  Note, however, that the absolute
values of these core masses depend on many uncertainties, in
particular, in the efficiencies of mixing processes in the stellar
interior - semiconvection, overshooting, and rotationally induced
mixing (not included here; cf.\ also \citealt{imb01}).  For example,
the helium core of the new model S25 is about one solar mass smaller
than in the equivalent 25\,\Msun model of WW95.  A model that was
computed with the new opacity tables, but without mass loss, had about
half a solar mass smaller helium core.  Thus, we attribute the other
half solar mass of decrease in helium core mass to the action of mass
loss. Of course, the two effects are not entirely independent.

As a result of the reduced helium core size, our new models generally 
have lower carbon-oxygen and oxygen-neon cores.  Due to the interaction
of the different phases of shells burning, the sizes of the ``inner
cores'' do not always monotonically change with the size of the helium
core, though a general trend is followed \citep{WHW02}.

The change in the weak rates \citep{LM00}, important after central
oxygen burning, leads to a $2-3\,\%$ increase in the central value of
$Y_{\rm e}$ at the time of core collapse (over what would have
resulted using \citep{ffn1}), and the ``deleptonized core'' tends to
contain less mass.  More importantly, we find $30-50\,\%$ higher
densities in the region $m=1.5-2\,\Msun$ which may affect the core
collapse supernova mechanism due to correspondingly higher
ram-pressure of the infalling matter (cf.\ \citealt{Jan01}).  Further
details concerning the effect of the new weak rates are discussed in
\citet{heg01b}.

\subsection{Supernova Explosions}

The most recent multi-dimensional calculations of core collapse and
supernova explosion still offer little guidance as to the exact
placement of the mass cut, the entropy and $Y_e$ of the innermost
ejecta, or even if a given model will explode (Herant et al. 1994;
Burrows, Hayes, \& Fryxell 1995; Janka \& M\"uller 1996; Mezzacappa et
al. 1998; Fryer \& Heger 2000). Nucleosynthesis studies must still
parameterize the explosion as best they can.
% XXX
In the present paper supernova explosions were simulated, as in WW95,
by a piston that first moved inwards for 0.45 s to a radius of 500 km,
and then rebounded to a radius of 10\,000 km.  For the inward motion,
the initial velocity is the local velocity of the corresponding mass
shell at the time of the presupernova model and the acceleration of
the piston is a constant fraction of the actual local gravitational
acceleration, $G\Mpist/r$.  The arbitrary fraction is chosen such that
the piston arrives at a radius of 500\,\km in 0.45\,\Sec.  The
subsequent outward movement also is a ballistic trajectory in a
gravitational field given by a different constant fraction of the
actual local gravitational acceleration.  Now the factor is chosen
such that an explosion of 1.2\E{51}\,\erg of kinetic energy in the
ejecta (measured at infinity) results in the 15 and 19\,\Msun models.
% XXX
This much kinetic energy is commonly assumed for SN~1987A (e.g., Woosley
1988), but it could have been very different in other supernovae.
% XXX 
For the heavier stars, this relatively modest energy gives large
amounts of ``fallback'', so that much of the interesting
nucleosynthesis falls into the neutron star (see also WW95).
Therefore, an alternate prescription that resulted in larger energies
was used for the 20, 21, and 25\,\Msun stars.  The energy there was
adjusted (increased) until the ejecta contained about 0.1\, \Msun of
\I{56}{Ni}.  
% XXX
This is comparable to the mass of $^{56}$Ni, $\sim 0.07$ \Msun,
commonly adopted for SN~1987A (e.g., Arnett et al. 1989).
% XXX 
For Model S25P, which had the same presupernova evolution
as Model S25, a still more powerful explosion was calculated that
ejected about 0.2\,\Msun of \I{56}{Ni} (see Tables
\ref{tab:struc_models} and \ref{tab:r}).

The final mass cut outside the piston was determined by the mass that
had settled on the piston at $2.5\times 10^4$ s after core collapse.
Note that the amount of fallback resulting from this prescription
depends on both the initial location of the piston used as well as its
energy.  In particular, the yields of $^{44}$Ti and $^{56}$Ni are very
sensitive to the ``final mass cut'' often determined by the
fallback.

The neutrino process ($\nu$-process) during the supernova explosion
was implemented using the same prescription as in WW95 and using the
same cross sections.  We used a neutrino pulse characterized by a
total energy of 3\E{53}\,\erg decaying exponentially on a time-scale
of 3\,\Sec. The neutrinos were assumed to have a mean energy of
4\,\MeV for the electron neutrinos and 6\,\MeV for the $\mu$ and
$\tau$ neutrinos (different from WW95 who used 8\,\MeV for the $\mu$
and $\tau$ neutrinos) The lower value is recommended by Myra \&
Burrows (1990).

\section{Nucleosynthesis Results}
\label{sec:results}

Yields were determined for 15, 19, 20, 21, and 25\,\Msun stars 
(Series S; Models S15, S19, $\ldots$) with
initial solar composition (\citealt{AG89}; see also Table \ref{tab:s}) 
and the standard rate set (Table \ref{tab:rrrm}). 
Identical stellar models having 15, 20, 25\,\Msun were also
calculated using the NACRE rate set (Set N; see also Table
\ref{tab:rrrm}). That is, all reactions given by NACRE were
substituted for their counterparts, except for
$^{12}$C($\alpha,\gamma)^{16}$O; all rates not given by NACRE were
left the same.  A single 25\,\Msun star was calculated that employed
the rate set of \citeauthor{HWW01} (\citeyear{HWW01}; Set H; see
\Sect{light}) which is much smaller than our current network.

\Figures{S15pf} -- \Figff{S25pf} show the production factors after the
explosion and the decay of all unstable species (except $^{40}$K and
$^{180}$Ta) in our ``standard'' S-series.  The abundances edited are
those outside the mass cut given as ``remnant mass'' in 
Table \ref{tab:struc_models}.  The
resulting abundances, including all those lost to winds,
have been divided by their solar (i.e., initial) values.  Isotopes
of each element are drawn in the same color and connected by lines.
The production factor of $^{16}$O -- the dominant ``metal'' yield of
massive stars -- is used as a fiducial point to provide a band of
acceptable agreement of $\pm 0.3$ dex relative to its value
(\textsl{dashed} and \textsl{dotted lines}).

These yields are also given in Table~\ref{tab:y} and are available
electronically from the authors.  Table \ref{tab:r} gives the yields
of all radioactivities still having appreciable abundance at
2.5\E4\,\Sec, the time of the mass cut determination 
(Table \ref{tab:struc_models}). For a
few isotopes, the edits include progenitors that have not decayed at
that time.  For example, \I{57}{Co} is almost all produced initially
as \I{57}{Ni} and results from its decay.  Consequently, the
\I{57}{Co} yield as given in Table~\ref{tab:r} is, (only) slightly,
higher than that of \I{57}{Ni}.

\Fig{Sc25pf}, for Model S25, allows the reader to gauge the importance
of explosive $vs.$ pre-explosive nucleosynthesis for various isotopes
in a 25\, \Msun star. The pre-supernova production factors of model S21
are shown in Fig.\ \ref{fig:S21cpf}.

\Figures{SN15pf}, \Figff{SN20pf}, and \Figff{SN25pf} show the
resulting post-explosive production factors of the NACRE runs relative
to our standard set (e.g., yields of Model N15 divided by yields of
Model S15) and \Fig{SH25pf} gives the same comparison for the rate set
H for the 25 \Msun star. The reaction network of \citet{HWW01} only
reached up to Ru.

Additionally, starting from the presupernova stage of Model S25 and
using the standard rate set, we followed a more powerful explosion
that gave twice the amount of \I{56}{Ni} (Model S25P). The results
are shown along with the others in Tables \ref{tab:y} and
\ref{tab:r}.  \Fig{SP25pf} gives a direct comparison of
nucleosynthesis in the model with high explosion energy (S25P)
relative to the one with lower (standard Model S25) energy.

\subsection{Production From Light Elements to the Fe Peak}

The light isotopes \I2H and \I3{He} as well as the elements Li, Be,
and B were destroyed during pre-main sequence and main sequence
evolution.  Though some \I3{He} is initially made, it is destroyed
again in the inner parts of the star. Different for WW95, fragile
isotopes can additionally be preserved in the stellar wind, especially
in the more massive stars, resulting in a slightly increased yield
of, e.g., \I3{He} in the S25 star compared to model S25A of WW95.
However, substantial amounts of $^7$Li and $^{11}$B were
created, along with $^{19}$F, by the $\nu$-process during the
explosion (\Fig{Sc25pf}).  The significant underproduction of \I{17}O
is a result of the revised reaction rates for
\I{17}O(p,$\alpha$)\I{14}N and \I{17}O(p,$\gamma$)\I{18}F
\citep{HWW01}.

Nucleosynthesis from Ca to Fe shows considerable scatter which only
partly relates to the nuclear rates. Yields in this region are
particularly sensitive to the details of the explosion and fall back
as can be seen in the comparison between S25 and S25P. The higher
explosion energy mostly alters the iron group (\Fig{SP25pf}).  In
particular, the yields of \I{44}{Ca}, \I{48}{Ti}, \I{56}{Fe},
\I{57}{Fe}, \I{59}{Co} and \I{58,60-62}{Ni} are significantly
enhanced.  Lighter nuclei produced further out in the star and heavier
nuclides made by the $s$-process are not greatly affected by the
explosion (hence \Fig{SP25pf} does not extend to high atomic mass).

\subsection{The 19, 20, and 21 \Msun Models}

It is necessary to discuss the 19, 20, and 21 \Msun models separately
because of the peculiar evolution of the 20 \Msun model. Model S20 is
at the transition mass (for our choice of
$^{12}$C($\alpha,\gamma)^{16}$O and convection theory) where stars
change from exoergic convective carbon core burning at their centers
(less than 20 \Msun) to stars where central carbon burning never
generates an excess of energy above neutrino losses (though carbon
shell burning always does).  We show the history of the convective
structure and energy generation for models S15, S20, and S25
in the Kippenhahn plots given in
Figs.\ \ref{fig:S15cnv}--\ref{fig:S25cnv}. Model S20 exhibits a strong
overproduction of several elements between Si and V, especially
isotopes of Cl, K, and V (Fig.\ \ref{fig:S20pf}).  Interestingly, Cr,
Mn, and the light Fe-isotopes are underproduced.  This is atypical and
is due to a stellar structure effect which appears, for the five
stellar masses considered, only in this model.  In specific, Model S20 
experienced 
the merging of the convective oxygen, neon, and carbon shells (cf.\
also \citealt{BA94}) well before (about one day) the end of the star's
life, during the core contraction phase from central silicon burning
till silicon shell ignition (Fig.\ \ref{fig:S20cnv}).  
The merged shells carry neutron sources
such as $^{22}$Ne and especially $^{26}$Mg to depths where they burn
rapidly and provide a strong source of free neutrons.  Capture of
these neutrons is responsible for the largest overproductions.

To illustrate that this feature is confined to models close to 20
\Msun, we also computed 19 and 21 \Msun models (Figs.\
\ref{fig:S19pf} and \ref{fig:S21pf}). In the 19 \Msun model,
$^{23}$Na, $^{38,40\!}$Ar, $^{42,46}$Ca, and $^{84}$Sr are enhanced
whereas the other elements follow the expected trend when compared
to models S15 and S25. Interestingly, S19 does not
show any traces of a $\gamma$-process up to mass 152.  In the 21 \Msun
model, $^{23}$Na is overproduced as in S19, but S, Cl, and the odd K
isotopes are produced less. Otherwise a ``standard'' pattern is
emerging. Also, similar $p$-process features (as in S15 and S25) are
emerging, with the same Mo-Ru deficiency.

Clearly the solar abundances have not originated in stars of any
single mass and calculations of Galactic chemical evolution must use
many more stars (and with a range of initial metallicities) 
than the five presented here.

\subsection{The $s$-Process}
\label{sec:sproc}

Nuclei above the iron group up to about $A=90$ are produced in massive
stars mainly by the $s$-process.  When these yields are combined with
those of metal-poor stars that contribute correspondingly less
$s$-process, it is helpful if they are somewhat large, say at the
factor of two level, compared to those for primary species like
oxygen.  For current choices of rates, our $s$-process yields are,
overall, consistent with this requirement.  There is significant
overproduction of the $s$-process products in the range $70\leq A\leq
90$ in the 25 $M_\odot$, but this is partly offset, for many isotopes,
by a more consistent production (relative to $^{16}$O) in the 15\,
\Msun model.  This is because of the well known tendency of higher
mass stars to be more effective in burning $^{22}$Ne (Prantzos,
Hashimoto, \& Nomoto 1990).

In terms of specific isotopes, $^{64}$Zn is underproduced in all cases
studied.  This nucleus may be a product of the neutrino wind from the
proto-neutron star \citep{hof96} not simulated here.  The
overabundance of the neutron-rich nickel isotopes, $^{61,62,64}$Ni,
and other $s$-process products in the $A=60-90$ mass range has been
observed before \citep{TWW95,HWW01} and is still not well understood.
This overproduction is especially pronounced in the 25\, \Msun model.
% XXX
Chiefly due to a reduced $^{22}$Ne($\alpha$,$\gamma$) rate in our
``standard'' rate set, we obtain a smaller $s$-process overproduction
in S25 compared to H25.

While the overproduction may be related to residual uncertainties in the
stellar model, this is a place where the nuclear physics might also be
suspect. For $^{62}$Ni in particular, the neutron capture rate given
in \citet{bao00} is about a factor of three lower than that given
previously in \citet{bao}. Using the earlier rate, more $^{62}$Ni
would be destroyed by neutron capture bringing the production factor
down into the acceptable range. Both recommended rates are based on
different extrapolations of the same experimental thermal neutron
capture cross section. Both extrapolations assume s-waves, but the
more recent one includes the estimated effect of a sub-threshold
resonance \citep{beer01}.  Such extrapolations have large
uncertainties, especially for a heavy nucleus where resonance
contributions can be expected already at around 30 keV. Therefore, it
is important to measure the cross sections of the Ni isotopes directly
in the relevant energy range.

The sensitivity of the $s$-process to changes in the charged particle
reaction rates can be seen by comparing to the results obtained with
rate set N (NACRE). The ratio of the production factors from sets S
and N is shown in Figs.\ \ref{fig:SN15pf}--\ref{fig:SN25pf}. The
overproduction of the problematic Ni isotopes 
is less pronounced with set N, but the
production of all nuclides between Ni and Pd is reduced.  This is
mainly due to the different \I{22}{Ne}\Ran\I{25}{Mg} and
$^{22}$Ne($\alpha,\gamma)^{26}$Mg rates, particularly the latter.
Consequently, the $p$-isotopes at $A>100$ are also produced less because of
the decreased production of seed nuclei in the $s$-process.
To underline the fact that the ($\alpha$,$\gamma$) and ($\alpha$,n)
reactions on $^{22}$Ne are the main source of the differences, in 
Fig.\ \ref{fig:SM25pf} we
show the result of a test calculation using the NACRE set but replacing
the two rates in question by our standard rates as given in \Sect{ne22an}.

The large uncertainty in a few NACRE rates allows the existence of a
much stronger $s$-process. 
Recent work \citep{crza00} claims that,
with a \I{22}{Ne}\Ran\I{25}{Mg} rate enhanced by a factor of
$100-1000$ over the recommended NACRE value, the well-known problem of
the underproduction of the $p$-isotopes of Mo and Ru might be cured.
However, leaving aside the important question of whether such a large
variation can be tolerated given more recent laboratory data \citep{jae01},
a dramatic alteration in rates of this sort would have
consequences, not only for the $p$-process, but for the production of
numerous nuclei between Fe and Ru (Figs.\
\ref{fig:SN15pf}--\ref{fig:SN25pf}). An intolerably strong $s$-process
may result. A strongly enhanced \I{22}{Ne}\Ran\I{25}{Mg} rate might
also pose problems for the $s$-process in AGB stars. We defer a detailed
numerical study of this and related questions to another paper, but
certainly a more accurate determination of the cross sections for
$^{22}$Ne interacting with $\alpha$-particles should have a very high
priority in the nuclear astrophysics laboratory.

On the other hand, a comparison of Models H25 and S25 (\Fig{SH25pf})
shows considerable variations in the $s$-process, especially for
individual isotopes, despite the fact that both studies used the same
rates for $^{22}$Ne($\alpha,$n)$^{25}$Mg and
$^{22}$Ne($\alpha,\gamma)^{26}$Mg. This is because H25 is the only
case where the neutron-capture cross sections along with all
Hauser-Feshbach rates were different.  All other studies changed only
the mass of the star, explosion energy, or charged-particle rate
set. The size of the variations in \Fig{SH25pf} - up to a factor of 5
in some cases where the network of H25 was still adequate - suggests
that there is still a lot of work to be done in the nuclear
laboratory. For example, the capture rates were about a factor of
two {\sl lower} at $s$-process temperatures for the Sr isotopes in
Model H25 and up to three times {\sl greater} for the $s$-process
isotopes of Mo. Inclusion of $^{16}$O as a neutron poison in Model
S25, and not in H25, as well as a larger cross section for the neutron
poison $^{26}$Mg in Model S25 also contributed to making the
$s$-process in S25 a little weaker.

Above $A=100$ the $s$-process does very little, though there
are redistributions of some of the heavy nuclei.  This has a minor 
impact on the $\gamma$-process, as mentioned above. Most of the
$s$-process above mass $90$ is believed to come from AGB stars.

\subsection{The n-Process}

The base of the helium shell has long been promoted as a possible site
for fast neutron capture processes as the supernova shock front passes
\citep{hil81,TCC78}. In our present models a slight redistribution of
heavy mass nuclei was found at the base of the helium shell, including
significant production of the gamma-ray line candidate
$^{60}$Fe. Integrated over the star however, the production of either
the $r$-process in general, or an appreciable subset of $r$-process
nuclei above mass 100 was negligible compared with other species.
Quite a few $r$-process isotopes above the iron group, but lighter
than mass A = 88 were made chiefly in the carbon and neon shells.

\subsection{The $\gamma$-Process}
\label{sec:gamma}

The production of the ``$p$-process'' nuclei results from
photo-disintegration of heavy nuclei during hydrostatic and explosive
oxygen and neon burning.  This is more properly called the
$\gamma$-process \citep{Ar76,WH78,RAP90,ray95}.  The present
calculations are the first to follow the $\gamma$-processes through the
presupernova stages and the supernova explosion in a self-consistent
fashion. Here the $\gamma$-process operates in stellar regions that
previously experienced the $s$-process, with the "seeds" being provided 
by the initial "solar" distribution of these Pop I stars plus any
additional production ($64 \le A \le 88$) that occured in-situ 
prior to explosion (Figure \ref{fig:S21cpf}).

For the 15, 21, and 25\,\Msun models, in the mass ranges $124\leq A
\leq 150$ and $168\leq A\leq 200$, the proton-rich heavy isotopes are
produced in solar abundance ratios within about a factor of two
relative to \I{16}O (Figs.\ \ref{fig:S15pf}, \ref{fig:S21pf}, and
\ref{fig:S25pf}).  Below $A=124$ and between $150\leq A\leq 165$ the
production of the proton-rich isotopes is down by about a factor of
three to four.  Similar trends can be found in all our models,
although with different magnitudes.  The total production of the
proton-rich isotopes increases for higher entropy in the oxygen shell,
i.e., with increasing mass of the helium core, as one can see in the
25\,\Msun star, but also depends on details of stellar structure and
the composition of the star at the time of core collapse.

It is interesting to note that in some stars, production of the
$p$-process nuclei occurs to varying extents in the oxygen burning
shell \emph{before} the explosion.  For example, in the 25\,\Msun
star, $p$-nuclei with $A < 90$ are made before the explosion (as also
noted by \citet{HWW01}), but essentially none for $A > 90$
(Fig.\ \ref{fig:Sc25pf}).  In the
21\,\Msun star a large production of $p$-nuclei at $A > 90$ occurs
before the explosion (Fig.\ \ref{fig:S21cpf}).  This pre-explosive
production is even more pronounced in the 20\,\Msun model where the
carbon and oxygen shells merged.  Indeed, some of the production
factors of $p$-nuclei in the 20\,\Msun model are so large that they
will remain important even if this is a comparatively rare event.  The
15\,\Msun star shows a significant $\gamma$-process in the
$A=160\ldots200$ region before the explosion, but not around $A=130$.
In the 19\,\Msun star, essentially no $\gamma$-process occurs before
the collapse. The details of the pre-explosive $p$-production depend, of
course, on the adopted convection model \citep{BA94}.

Once again, the diversity of nucleosynthetic outcomes for stars of
comparable mass is highlighted.  A fine grid of masses must be
calculated to correctly weight all these contributions.  Because it
depends on the extent of prior $s$-processes, the depth and possible
merging of convective shells in the last hours of the stars life, and
the strength of the explosion, the $\gamma$-process yields of stars
can vary wildly.  Ultimately this may make the $\gamma$-process an
important diagnostic of stellar evolution.

In terms of nuclear physics, it should be noted that the
($\gamma$,$\alpha$)/($\gamma$,n) branching at $^{148}$Gd, which
determines the production ratio $^{144}$Sm/$^{146}$Sm
\citep{WH90,rto96}, is now known to much better accuracy than in
previous investigations (see \Sect{hf}).  Although the experiment of
\citet{som98} did not quite reach the relevant energy window, it
highly improved on the necessary extrapolation and yielded an S-factor
which was several orders of magnitude lower than previous estimates.
The remaining uncertainty is almost entirely due to the
$^{148}$Gd($\gamma$,n)$^{147}$Gd branch.  Obviously, the total
production of $^{144}$Sm and $^{146}$Sm is still sensitive to a number
of photodisintegration rates only known theoretically.

\subsubsection{The case of $^{180}$Ta}

The production factors of the isotope $^{180}$Ta, the rarest stable
nuclear species in the solar abundance pattern, needs special
consideration. In the 25\,\Msun model $^{180}$Ta shows a slight
overproduction, despite our taking into account its destruction by
de-excitation into the short-lived ground state through thermal
excitation into an intermediate state \citep{end99}.  However, in the
calculation we do not explicitly follow the population of ground and
isomeric state and therefore what is found is rather the sum of the
produced $^{180\rm g}$Ta+$^{180\rm m}$Ta.  The nucleus $^{180}$Ta is
peculiar in the way that its ground state has a half-life of only
8.152 h, much shorter than the half-life of the isomeric state with
$T_{1/2}>1.2\times 10^{15}$ y.  In order to determine the fraction of
the long-lived isomer in the total yield one would have to know the
population of ground and isomeric states. In Appendix \ref{app:ta} we
show how to arrive at an estimate of the state population based on the
experiment of \citet{end99}. It is concluded that about 30--50\% of
the produced $^{180}$Ta are actually in the isomeric state $^{180\rm
m}$Ta. Therefore, our production factors and yields have to be
renormalized by that factor.  This brings the production factor of
this isotope down into the acceptable range for all stellar models.

\section{Summary and Conclusions}

Using a nuclear reaction network of unprecedented size,
nucleosynthesis has been investigated in several stellar models in the
mass range 15 \Msun to 25 \Msun.  The models include the best
currently available nuclear and stellar physics.  For the first time,
it was also possible to self-consistently follow the $\gamma$-process
up to Bi.

Overall good agreement can be achieved with the solar abundances of
nuclei between $^{16}$O and $^{90}$Zr. This good agreement is, to
first order, independent of the reaction rate set employed; our current standard,
\citet{ang99} or \citet{HWW01}, though several key nuclear
uncertainties are identified. In addition to the well-known need for
greater accuracy in the rate for $\alpha$-capture on \I{12}C, the
rates for \I{22}{Ne}($\alpha$,n)\I{25}{Mg} and
$^{22}$Ne($\alpha,\gamma)^{26}$Mg are critical. We also urge a
re-examination of some of the neutron capture cross sections for the
isotopes of nickel.

For the $p$-isotopes, two regions of atomic mass are found where those
isotopes are underproduced, $92 \leq A \leq124$ and $150\leq A\leq 165$. It
remains unclear whether this deficiency is due to nuclear cross
sections, stellar physics, or if alternative (additional) $p$-process
scenarios have to be invoked. However, we find that part of the
$p$-nuclides may be produced in convective oxygen shell burning during
the last hour of the star's life. The remainder is made explosively.

Interesting and unusual nucleosynthetic results are found for
one particular 20 \Msun model due to its special stellar
structure. This effect, a merging of heavy element shells late in the stars
evolution, seems to be confined to a narrow range of masses. In
particular it is not seen in 19 and 21 \Msun models. However, we have
explored a very limited set of masses and those only in one spatial 
dimension (for caveats see Bazan \& Arnett 1994).

\acknowledgments

We are grateful to Tom Weaver for his central role in developing the
Kepler computer code and to Frank Timmes for providing the sparse
matrix inverter we used for the large network.  This research was
supported, in part, by the DOE (W-7405-ENG-48 and SciDAC)), the
National Science Foundation (AST 97-31569, INT 97-26315), the Alexander
von Humboldt Foundation (FLF-1065004), and the Swiss National Science
Foundation (2000-061822.00). T.R. acknowledges support by a PROFIL
professorship from the Swiss National Science foundation (grant
2124-055832.98).

\appendix
\section{A New Approach to Fitting Reaction Rates}
\label{app:fit}

Frequently, experimentalists find it easiest to provide reaction rate
data in tabular form, but there are several issues that make using
such tables difficult for the stellar modeler. First is the issue of
accuracy. Most charged particle reaction rates change by many orders
of magnitude over narrow temperature ranges, making direct
interpolation difficult. To improve accuracy a fine temperature grid
is required. Coupled with the vast number of reaction rates required
in a large reaction network, the memory storage issues alone have
historically forced the designers of astrophysical data bases to adopt
fits to reaction rates, and accept a (marginal) loss of accuracy. This
is especially true for theoretical (Hauser-Feshbach) reaction rates,
which are often smooth enough to be accurately fit. Reaction rates
that use a standard form, some combination of powers of temperature in
a single exponential for example, are also particularly efficient to
calculate on the machine. This is an important consideration when
computing many thousands of rates in every zone of a star at every time
step.

The most important rates (usually those on targets lighter than
silicon, and especially those that play a dominant role in energy
generation) have, until now, been fitted to analytical functions
\citep[][who provide both tables and fits]{CF88,ang99}. Over the years
these formulae have become increasingly diverse and complicated.
These important reactions are small in number, and in principle can be
used in tabular form.

We propose an approach that takes advantage of the best features of
both approaches - analytic fits and tables - while functioning
efficiently on the machine at a modest cost in memory allocation.  All
of the charged particle reaction rates from the recent compilation of
\citet{id01} were fit this way and used in our calculations.

The reaction rate as a function of temperature $\lambda (T_9)$,
provided in tabular form by the experimenter, is first fit to an
analytic function chosen for its accuracy, speed in evaluation, and
approximately correct physical behavior at low temperatures. For the
charged particle reactions treated in \citet{id01}, we used Equation
(8) of \citet{wfhz} (always fit in the exoergic direction):

\begin{eqnarray}
\label{woofit}
\lambda_{jk} &=& T_9^{-2/3}{\rm exp}[A - ({\rm TAU}/T_9^{1/3}) \nonumber \\ 
&\times& (1 + BT_9 + CT_9^2 + DT_9^3)]
\end{eqnarray}
where ${\rm TAU}=4.2487(Z_I Z_j {\hat A}_j)^{1/3}$ reflects the
Coulomb barrier for a charged particle in the entrance channel of
reaction $I(j,k)L$, $Z_I$ and $Z_j$ are the charges of the target and
incident particle, ${\hat A}_j$ is the reduced mass of the compound system. 
This fit is intended to contain the bulk of the temperature
dependence of the rate, but often does {\sl not}, by itself,
constitute an acceptable fit over the tabulated temperature range,
especially if individual resonances are important.  But also available
from the fitting process are the residuals at each tabulated temperature.
The logarithm of the ratios of the actual rate to 
the rate predicted by 
the fitting function are carried as a table in the computer and can be
interpolated much more accurately than the rate itself. Such a
procedure is directly analogous - for rates - to the traditional
representation of cross sections as a value times an ``S-factor''
which contains the zeroth order Coulomb penetration function. Indeed
the low temperature behavior of the analytic fit function is precisely
that of a reaction rate calculated with a constant S-factor.  Typical
accuracy achieved at non-tabulated grid points is better than 10\% at
temperatures where the rate is important.

\section{Population of ground and isomeric state in $^{180}$Ta}
\label{app:ta}

The ground state of $^{180}$Ta has $J^{\pi}=1^+$ whereas the isomeric
state is a $J^{\pi}=9^-$ state.
Because of the spin and parity assignments, the isomeric
state cannot directly decay into the ground state but when the nucleus
is thermally excited it can be depopulated via an intermediate state
which lies above the isomeric state.  The temperature-dependent
half-life derived in \citet{end99} is based on the condition that the
nucleus is in thermal equilibrium with the photon bath at a given
temperature. In an explosive scenario $^{180}$Ta is produced at a
temperature sufficiently high to provide thermal equilibrium. During
freeze-out the populations of the states will remain in equilibrium as
long as the temperature is high enough to sufficiently feed the ground
state. Below a critical temperature $T^{\rm crit}$, the de-excitation
of the isomeric state will not be fast enough to compensate for the
decay of the ground state and the isotope drops out of equilibrium. From
that moment isomer and ground state have to be considered as two
distinct species.  Therefore, we have to take the population ratio at
the lowest temperature (i.e., $T^{\rm crit}$) before equilibration
ceases in order to determine how much $^{180\rm m}$Ta actually
remains.

A time-dependent calculation of the transition probabilities and the
speed of the process is very involved. However, an estimate of $T^{\rm
crit}$ can be found by using the result of \citet{end99}. The
effective half-life curve (figure 4 in that paper) shows three
different regimes: i) at $T_9>0.35$ the states are fully equilibrated
and the effective half-life is essentially the half-life of the ground
state; ii) at $T_9<0.15$ the two states are fully decoupled and the
contribution of the ground state to the effective half-life is
negligible; iii) the intermediate region with $0.15\leq T_9\leq 0.35$
is a transitional region in which the communication between the two
states quickly ceases and equilibrium is not well established.

In thermal equilibrium the population $P_{\rm
iso}$ of the isomer relative to the ground state is given by
\begin{equation}
P_{\rm iso}=\frac{\left(2J_{\rm iso}+1\right)\exp \left(-E_{\rm
iso}/kT\right)}{\left(2J_{\rm
g.s.}+1\right)}=\frac{19}{3}e^{\frac{-0.8738}{T_9}} \quad .
\end{equation}
The effective decay rate is given by
\begin{equation}
\lambda_{\rm eff}=\lambda_{\rm g.s.}+P_{\rm iso} \lambda_{\rm iso}\quad.
\end{equation}
Thus, for $T_9=T_9^{\rm crit}=0.35$ there would be about 0.52 times as much
$^{180\rm m}$Ta than $^{180\rm g}$Ta, i.e., we have to divide the final
total abundance of $^{180}$Ta by three to get the surviving abundance of
the isomer.

In order to get proper amounts of $^{180\rm g}$Ta and $^{180\rm m}$Ta
one would have to know how much time is spent in the intermediate
region. If that region is covered quickly, it should be safe to take
$T_9^{\rm crit}=0.35$. A choice of $T_9^{\rm crit}=0.4$ appears
reasonable to compensate for neglecting the intermediate phase and any
additional production in that phase. This would lead to a relative
abundance of $P_{\rm iso}=0.71 P_{\rm g.s.}$.

The temperature $T_9^{\rm crit}$ at which equilibrium is left does not 
strongly depend on the g.s.\ half-life.
The largest uncertainty comes from the
excitation energy of the intermediate state which is not well
determined experimentally. However, based on the \citet{end99} results
we can assume an upper limit of $^{180\rm m}$Ta to be half of the
$^{180}$Ta produced in our models and an educated guess would be
between 0.3 and 0.5 of the produced $^{180}$Ta. 




%% Generally speaking, only the figure captions, and not the figures
%% themselves, are included in electronic manuscript submissions.
%% Use \figcaption to format your figure captions. They should begin on a
%% new page.

\clearpage

%% No more than seven \figcaption commands are allowed per page,
%% so if you have more than seven captions, insert a \clearpage

%% There must be a \figcaption command for each legend. Key the text of the
%% legend and the optional \label in curly braces. If you wish, you may
%% include the name of the corresponding figure file in square brackets.
%% The label is for identification purposes only. It will not insert the
%% figures themselves into the document.
%% If you want to include your art in the paper, use \plotone.
%% Refer to the on-line documentation for details.

\begin{figure}
\epsscale{0.9} 
\plotone{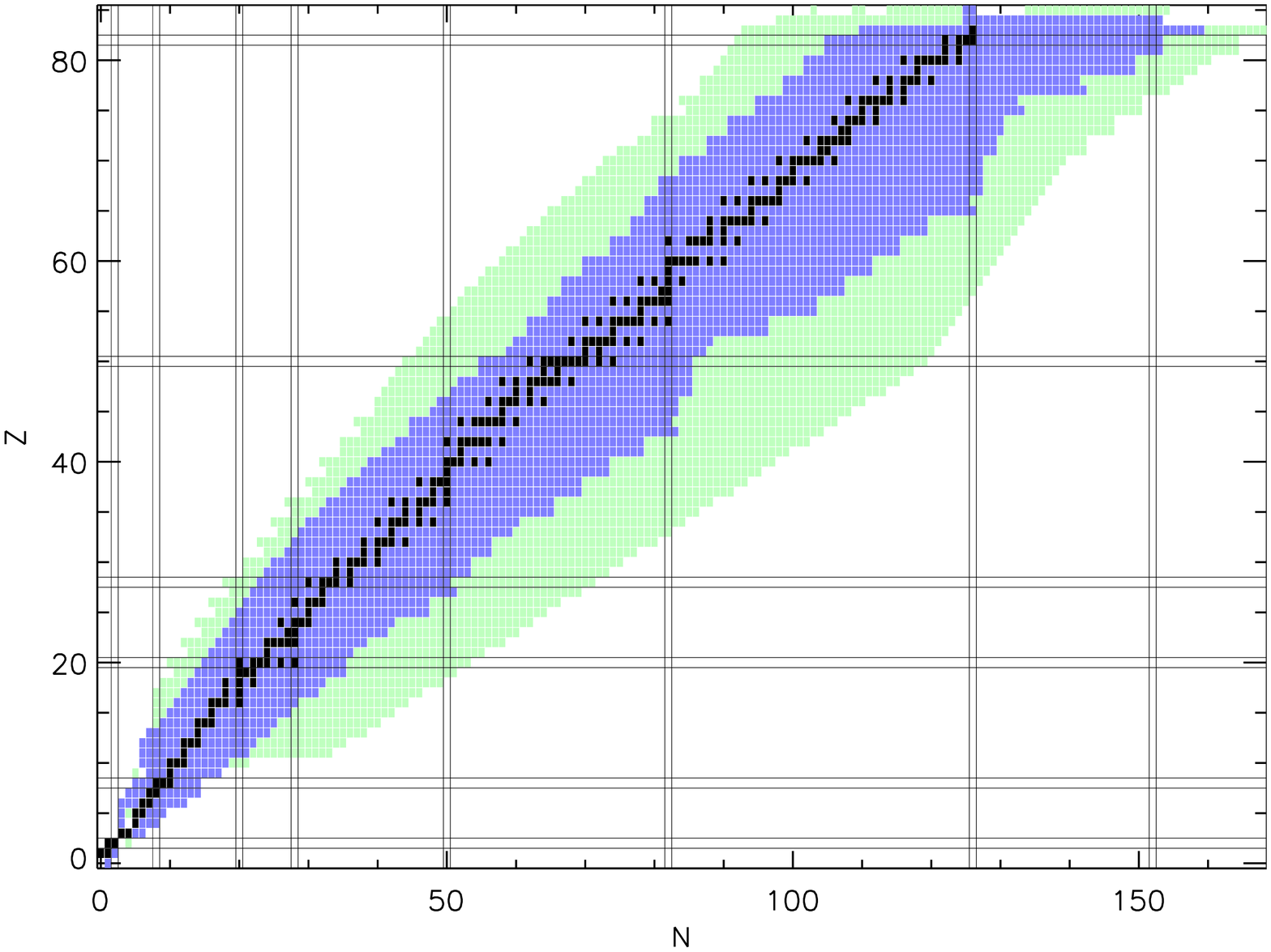}
\figcaption[f1.eps]{\lFig{S15net} Isotopes in our data base
(\textsl{green}), used by model S15 (\textsl{blue}) and
stable/long-lived isotopes (\textsl{black}).}
\end{figure}

\clearpage

\begin{figure}
\epsscale{0.775} 
\plotone{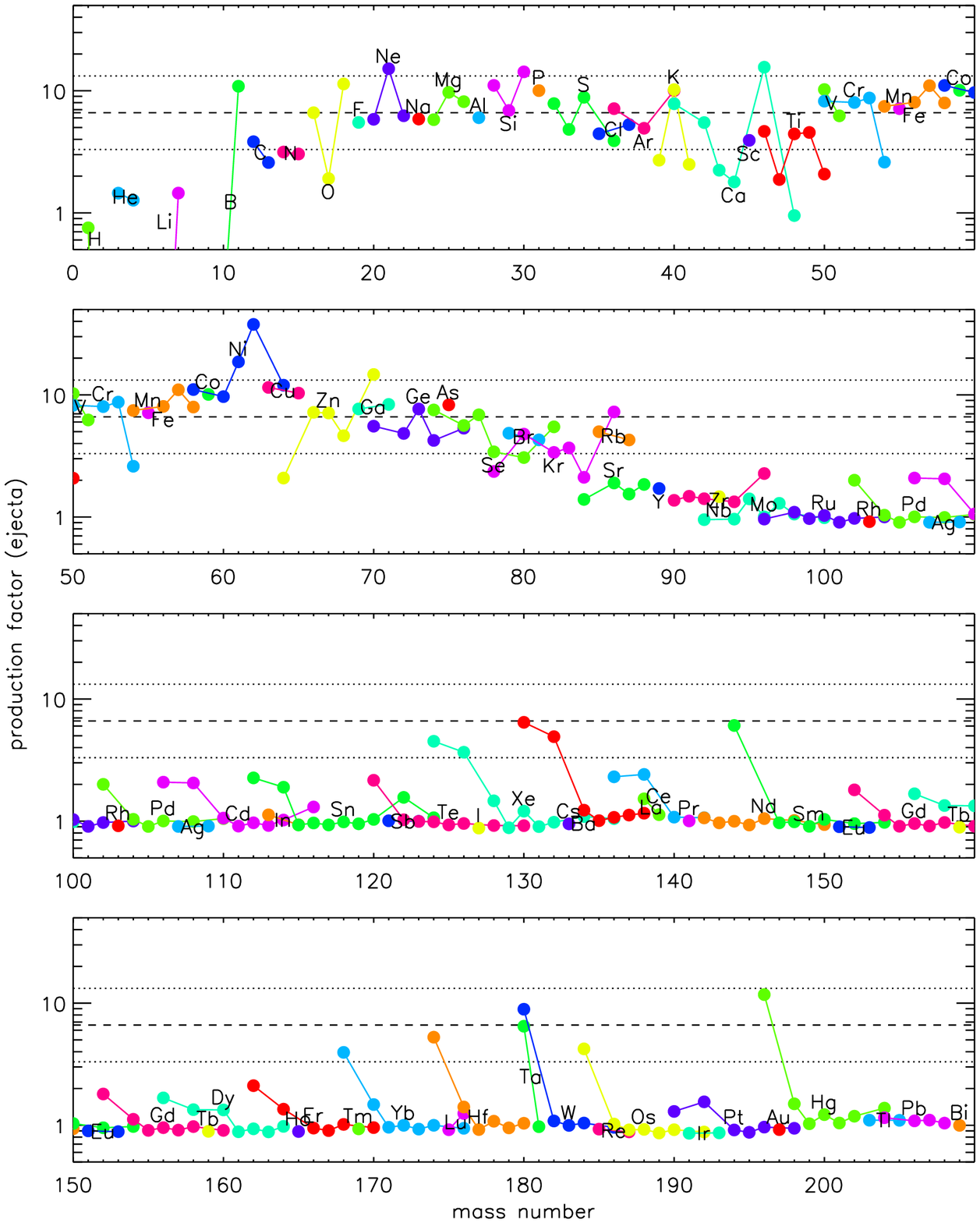} 

\figcaption[f2.eps]{\lFig{S15pf} Post-explosive production factors
following the decay of all radioactivities in Model S15. Comparison is
relative to solar abundances \citep{AG89}.  Note the consistent
production at the factor of ten level of most of the isotopes from A =
16 to 88 as well as a large fraction of the $p$-process isotopes. The
relative overproduction of $^{61,62}$Ni may indicate some lingering
uncertainty in the destruction cross sections for these species by
(n,$\gamma$). Many of the $p$-isotopes, especially those in the mass
range A = 90 to 130 are underproduced.  This may also improve with the
inclusion of other models, but seems to be chronic and may indicate an
incomplete understanding of the $s$-process in massive stars (the
$s$-process is the seed from which the $p$-process is made). Other
deficiencies are discussed in Woosley \& Weaver (1995) and Woosley,
Heger, \& Weaver (2002).}
\end{figure}

\clearpage

\begin{figure}
\epsscale{0.8} 
\plotone{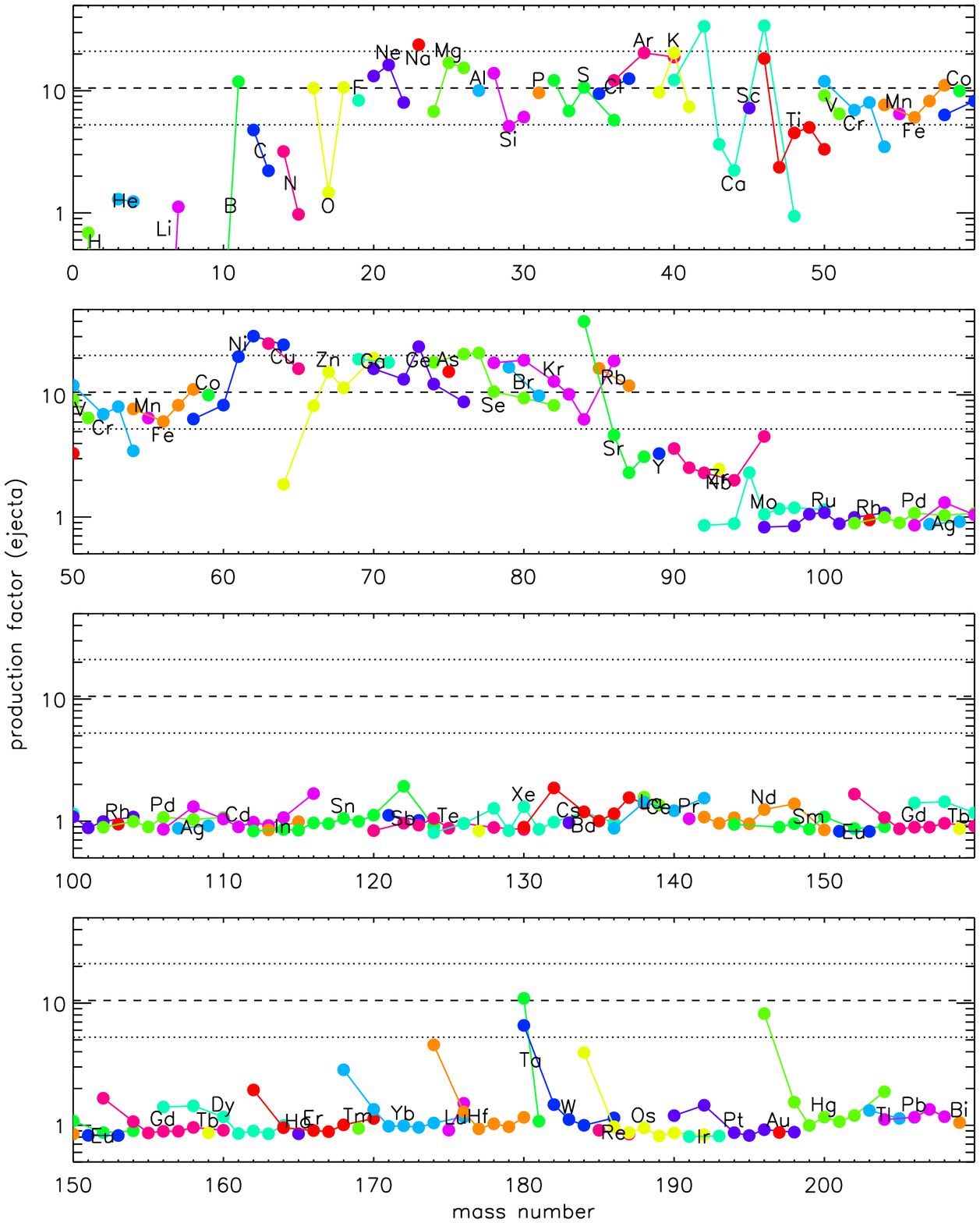} 

\figcaption[f3.eps]{ \lFig{S19pf} Similar to \Fig{S15pf} but for
Model S19, a 19\,\Msun star.  Note that the $p$-process around $A=180$
is weaker than in Model S15, and around $A=130$ is essentially
missing.  The reason is that before the explosion the star had a
convective shell close to the end of oxygen burning in which all
isotopes with $A\gtrsim100$ had already been largely destroyed -- so
the explosion cannot make heavy $\gamma$-process here.  The neon-rich
layer above was located too far out to become hot enough and make the
$p$-process nuclei near $A=130$.  It became hot enough, however, to have some
partial explosive neon burning at its bottom and make a little of the
$p$-process isotopes of the $A=180$ group, though less than we
typically observed in the other models.  }
\end{figure}

\clearpage

\begin{figure}
\epsscale{0.8} \plotone{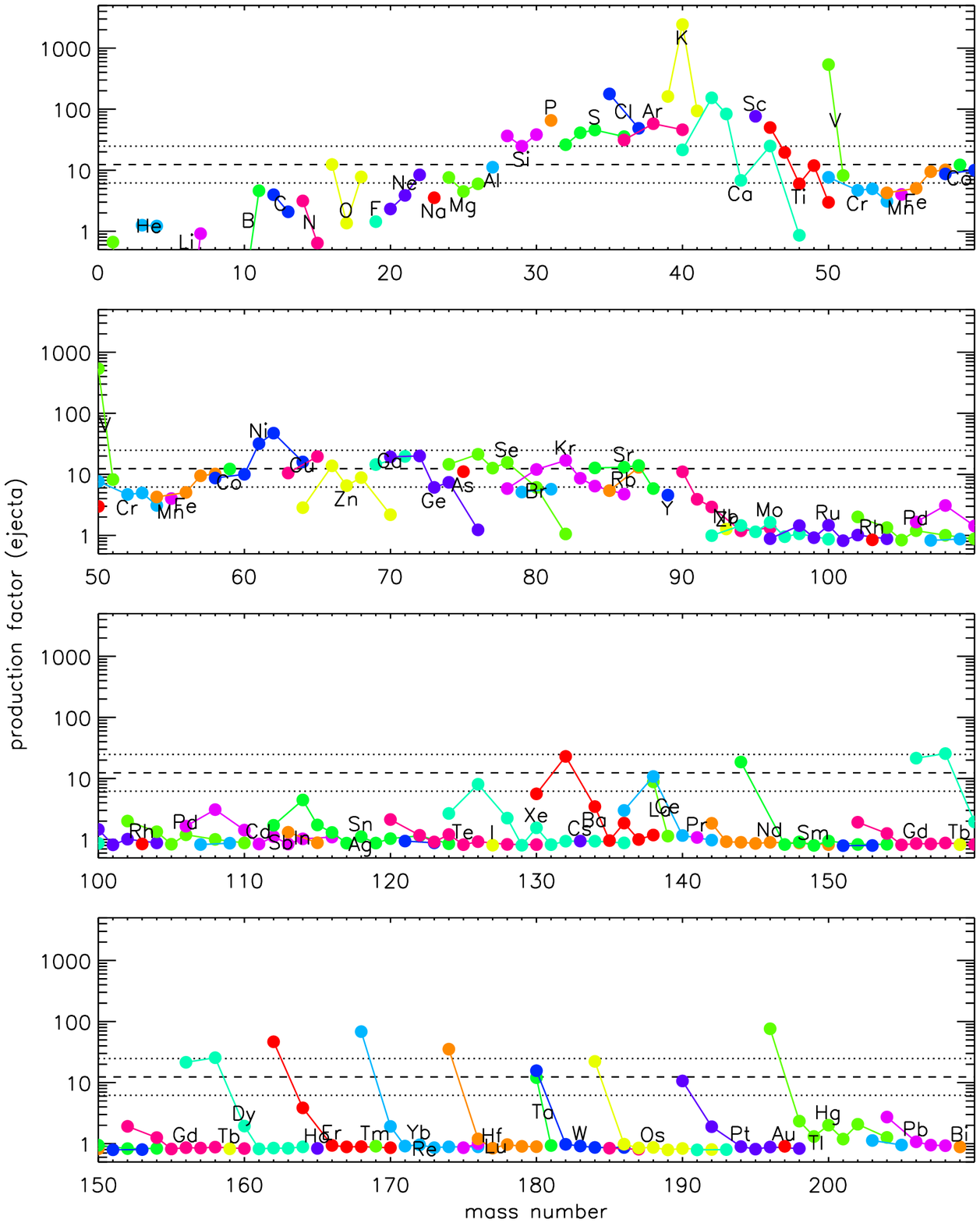} \figcaption[f4.eps]{\lFig{S20pf}
Similar to \Fig{S15pf} but for Model S20, a 20 \Msun star. Note the
very large overproduction of the $s$-process here, especially of
$^{40}$K and $^{50}$V.  These overproductions are unique within the 5
masses studied (15, 19, 20, 21, and 25) and show the importance of
integrating over a large number of masses.  The $s$-process is overly
strong here because of a link-up and mixing of the convective carbon,
neon, and oxygen shells about a day before the supernova.  Note
that despite this vigorous $s$-process, $p$-process isotopes are still
underproduced from A = 92 to 124.}
\end{figure}

\clearpage
\begin{figure}
\epsscale{0.8} 
\plotone{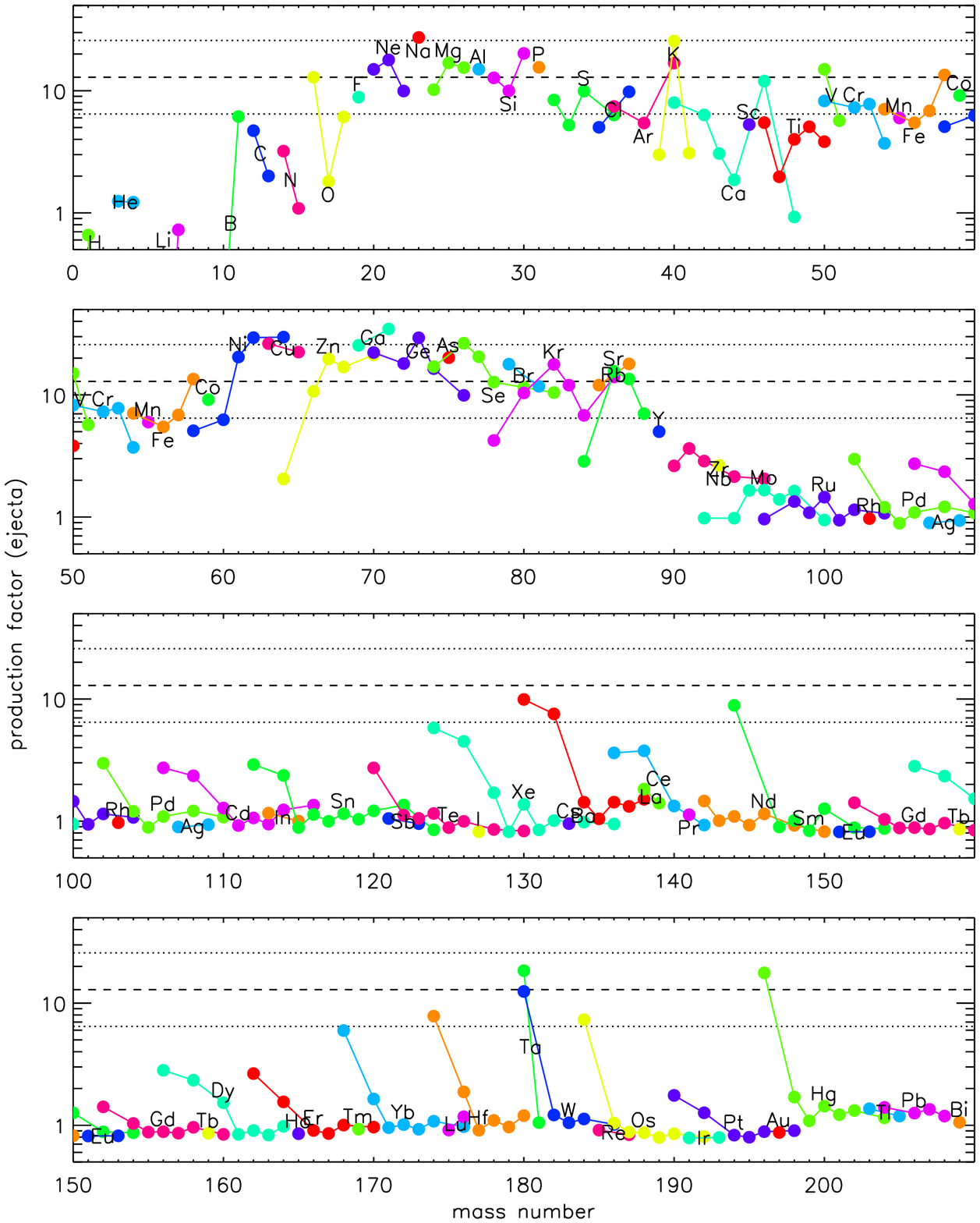}
\figcaption[f5.eps]{\lFig{S21pf} Similar to \Fig{S15pf}, but for
Model S21, a fully evolved and exploded 21 \Msun star. 
These results look much more like those
for a 25 \Msun star than for a 20 \Msun star.}
\end{figure}

\clearpage

\begin{figure}
\epsscale{0.8} 
\plotone{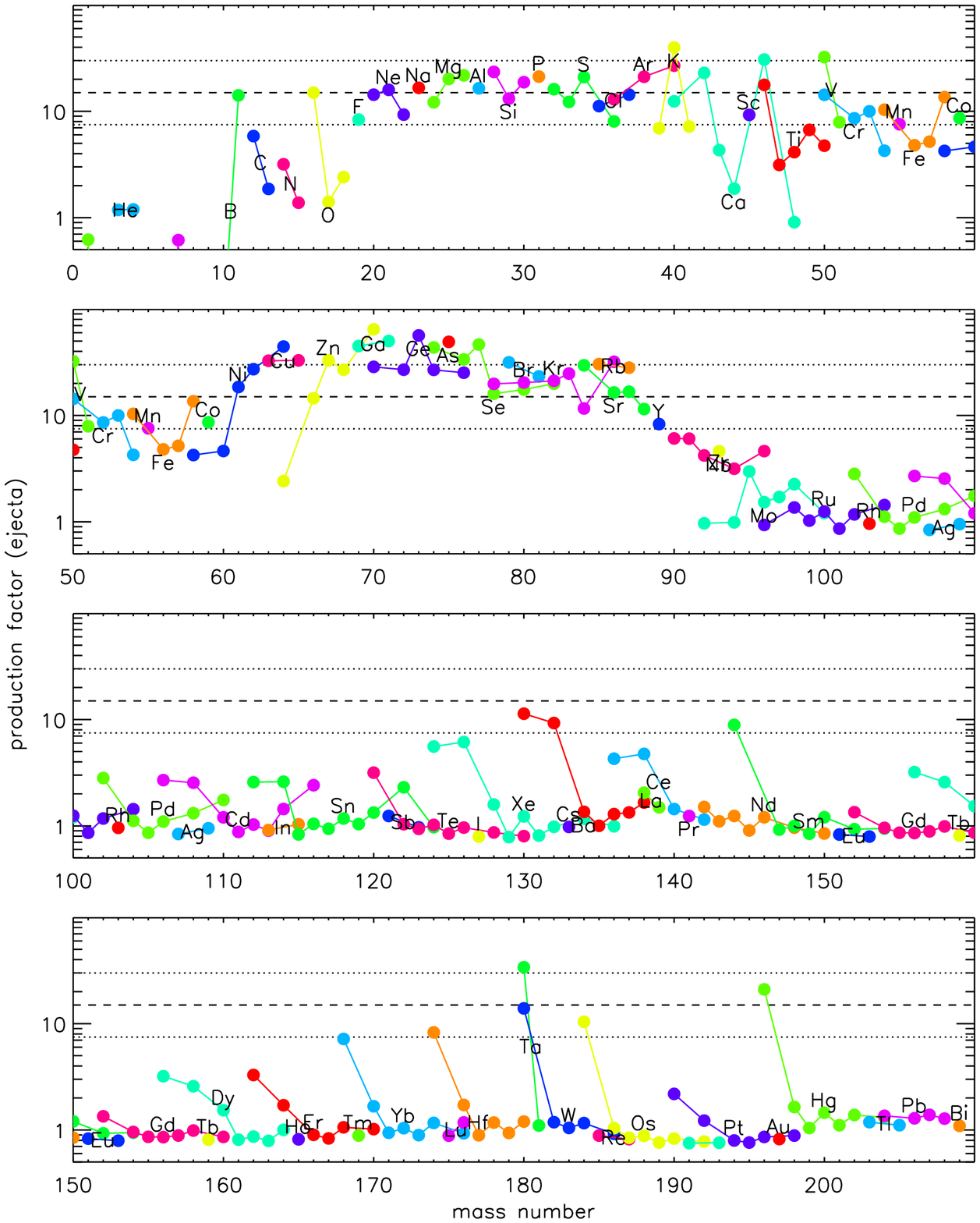} 

\figcaption[f6.eps]{\lFig{S25pf} Similar to \Fig{S15pf}, but for
Model S25, our standard 25 \Msun star. The $s$-process is overproduced
but not badly, especially if one is to average these yields 
with lower mass stars
like S15 and lower metallicity stars. As previously noted by
\citet{HWW01}, the common co-production of many $r$-, $s$-, and
$p$-isotopes from $A = 60$ to 88 is striking. Production of iron-group
elements would be higher in an explosion with greater energy
(\Fig{SP25pf}).}
\end{figure}

\clearpage

\begin{figure}
\epsscale{0.8} \plotone{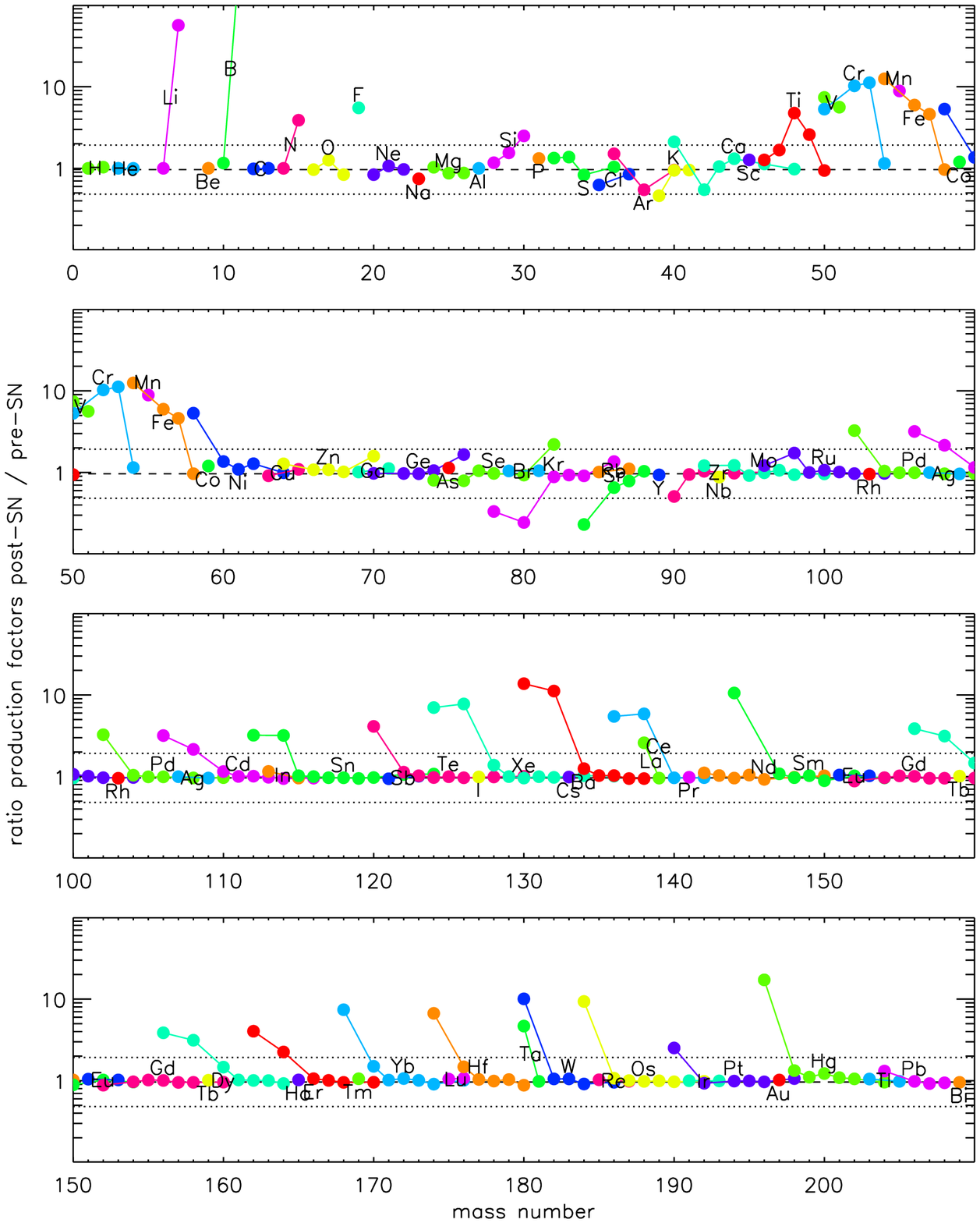}
\figcaption[f7.eps]{\lFig{Sc25pf} Ratio of the decayed production
factors of Model S25 after explosion to those before the explosion.
In both cases we only consider the mass layers ejected in the
supernova explosion and include the abundances ejected in the wind.
Most pronounced is the production of the $\nu$-process nuclei 
($^{7}$Li, $^{11}$B, $^{15}$N, and $^{19}$F), most of the the iron group,
and the $p$-nuclei (with $A\ge 110$) made during the 
operation of the $\gamma$-process. The abundance of the light $p$-nuclei 
$^{78}$Kr and $^{84}$Sr are reduced from their peak levels made prior to
the explosion.}
\end{figure}

\clearpage

\begin{figure}
\epsscale{0.8} \plotone{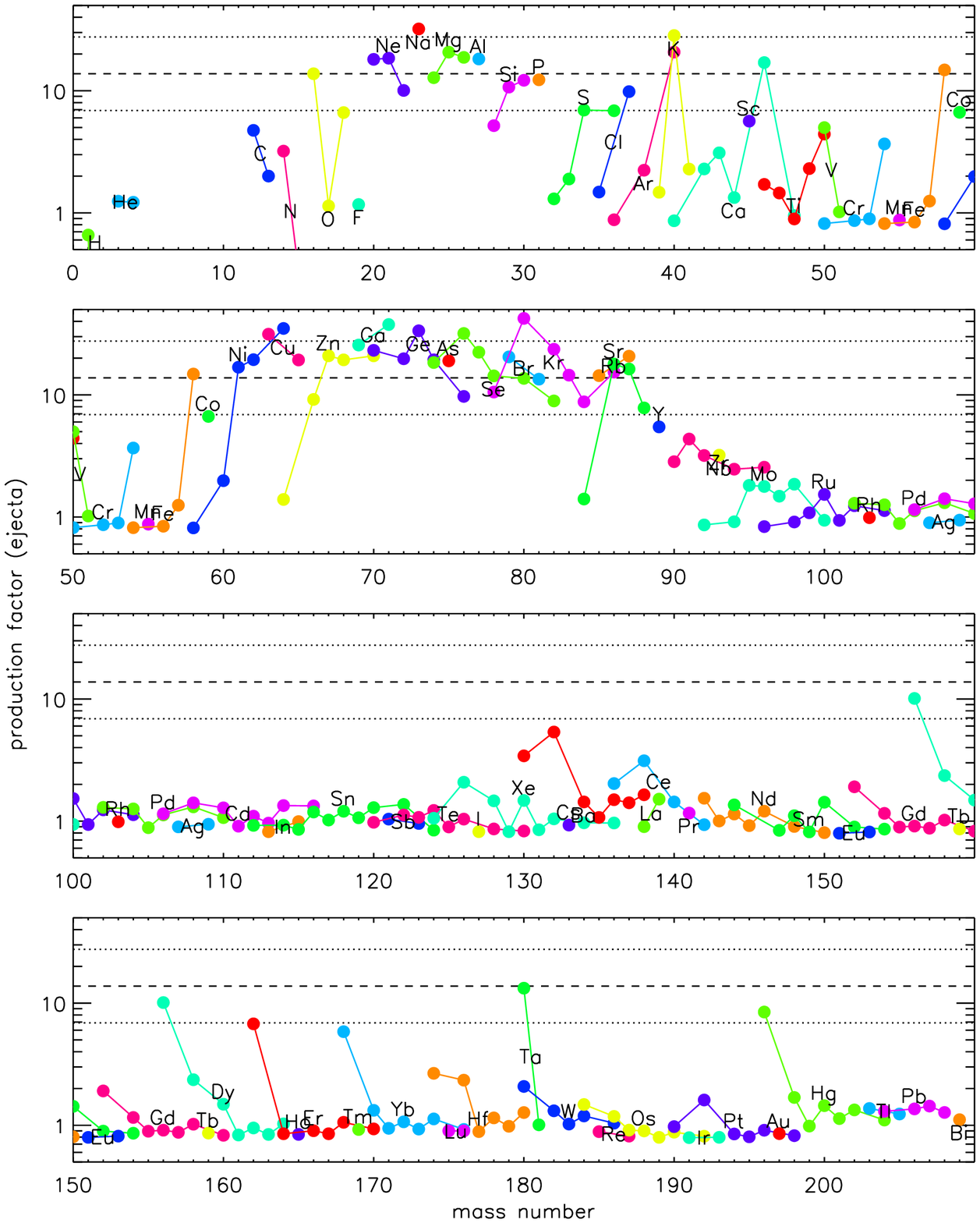}

\figcaption[f8.eps]{\lFig{S21cpf} Similar to \Fig{S15pf}, but
shows the {\sl presupernova} production factors of Model S21, a 21
\Msun star.  Only the mass outside of the baryonic remnant mass (Table
6), including winds, is considered.  One can already see significant
$\gamma$-process at $A\approx130$ and $A=155\ldots200$.}
\end{figure}

\clearpage
\begin{figure}
\epsscale{0.8} 
\plotone{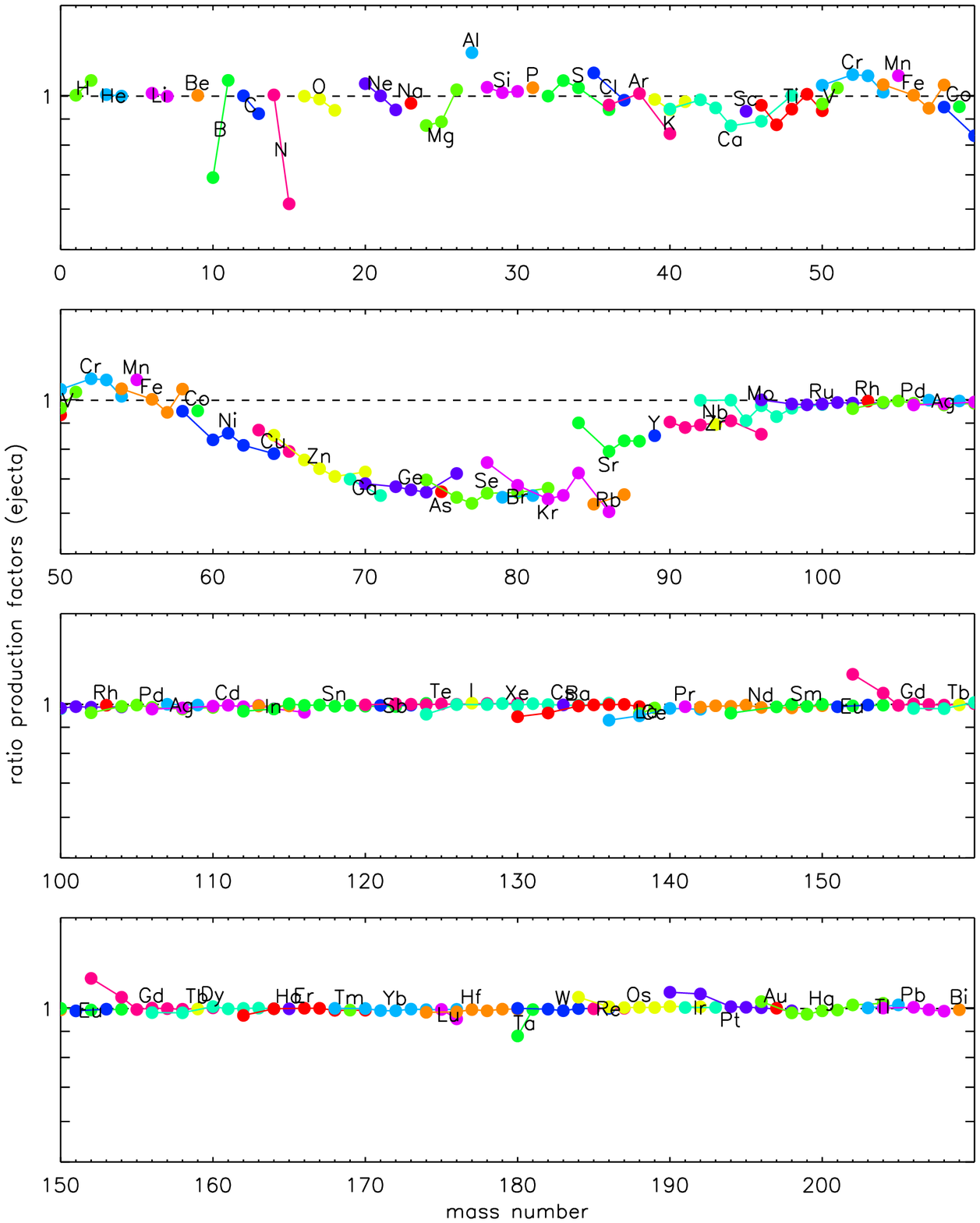}
\figcaption[f9.eps]{\lFig{SN15pf} Production factors
of the 15\,\Msun Model N15 (NACRE) divided by S15 (standard). Note a
weaker $s$-process in the run using the NACRE rates. This is chiefly
because of their choice of a larger rate for the reaction
$^{22}$Ne($\alpha,\gamma)^{26}$Mg.}
\end{figure}

\clearpage

\begin{figure}
\epsscale{0.8} 
\plotone{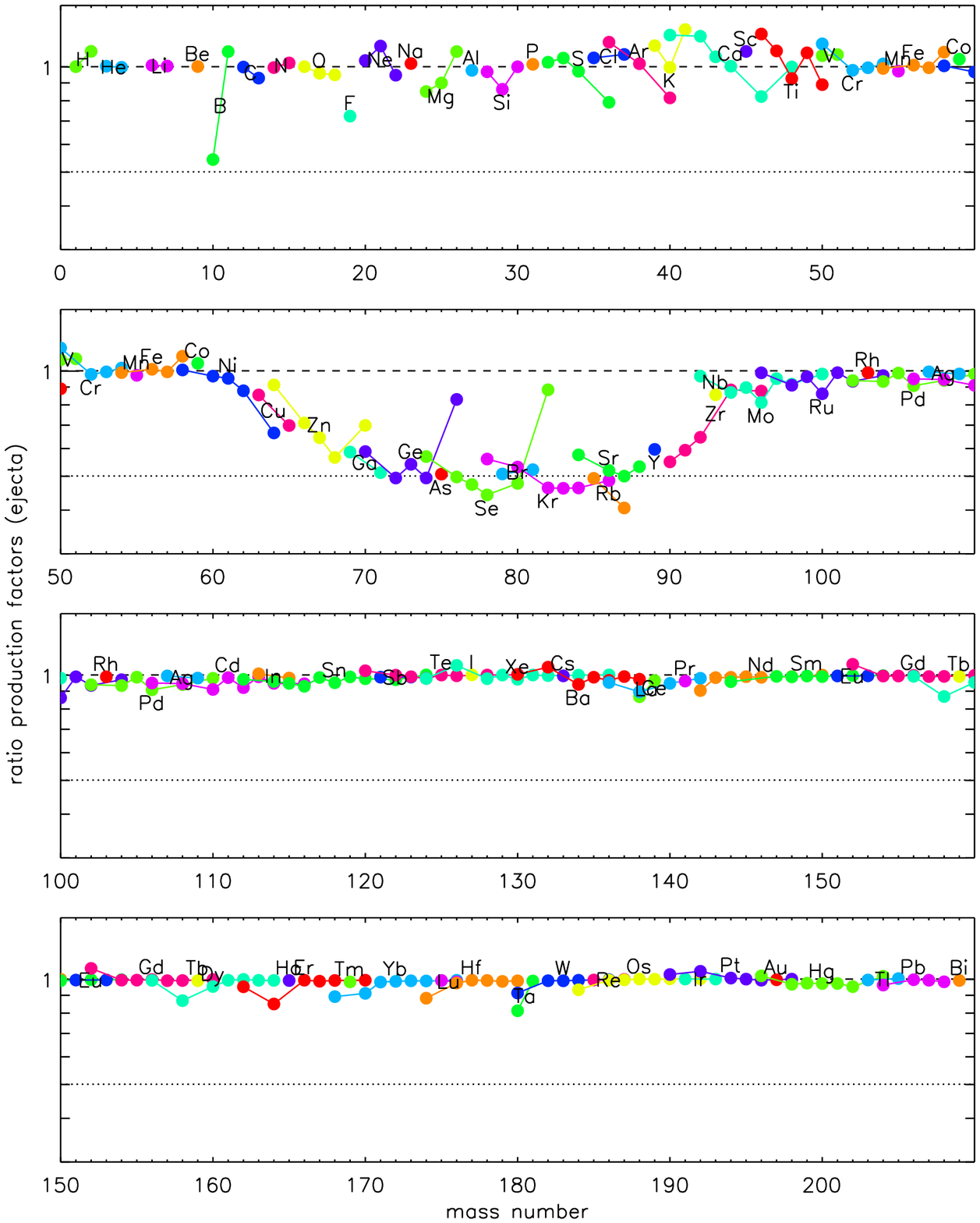}
\figcaption[f10.eps]{\lFig{SN20pf} Similar to \Fig{SN15pf} but shown
are the production factors
of the 20\,\Msun Model N20 (NACRE) divided by S20 (standard).}
\end{figure}

\clearpage

\begin{figure}
\epsscale{0.8} 
\plotone{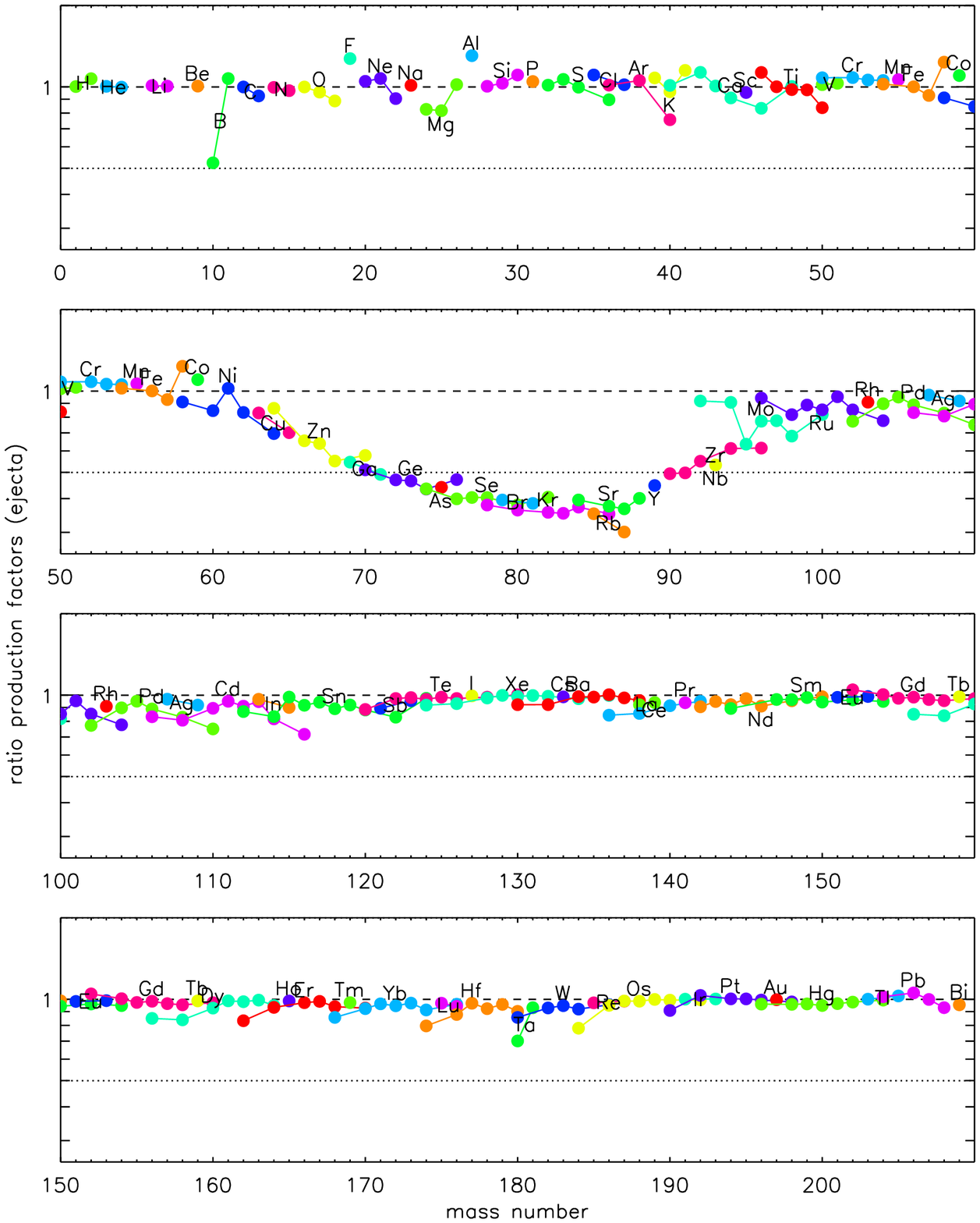}
\figcaption[f11.eps]{\lFig{SN25pf} Similar to \Fig{SN15pf} but shown
are the production factors
of the 25\,\Msun Model N25 (NACRE) divided by S25 (standard).}
\end{figure}

\clearpage

\begin{figure}
\epsscale{0.8} \plotone{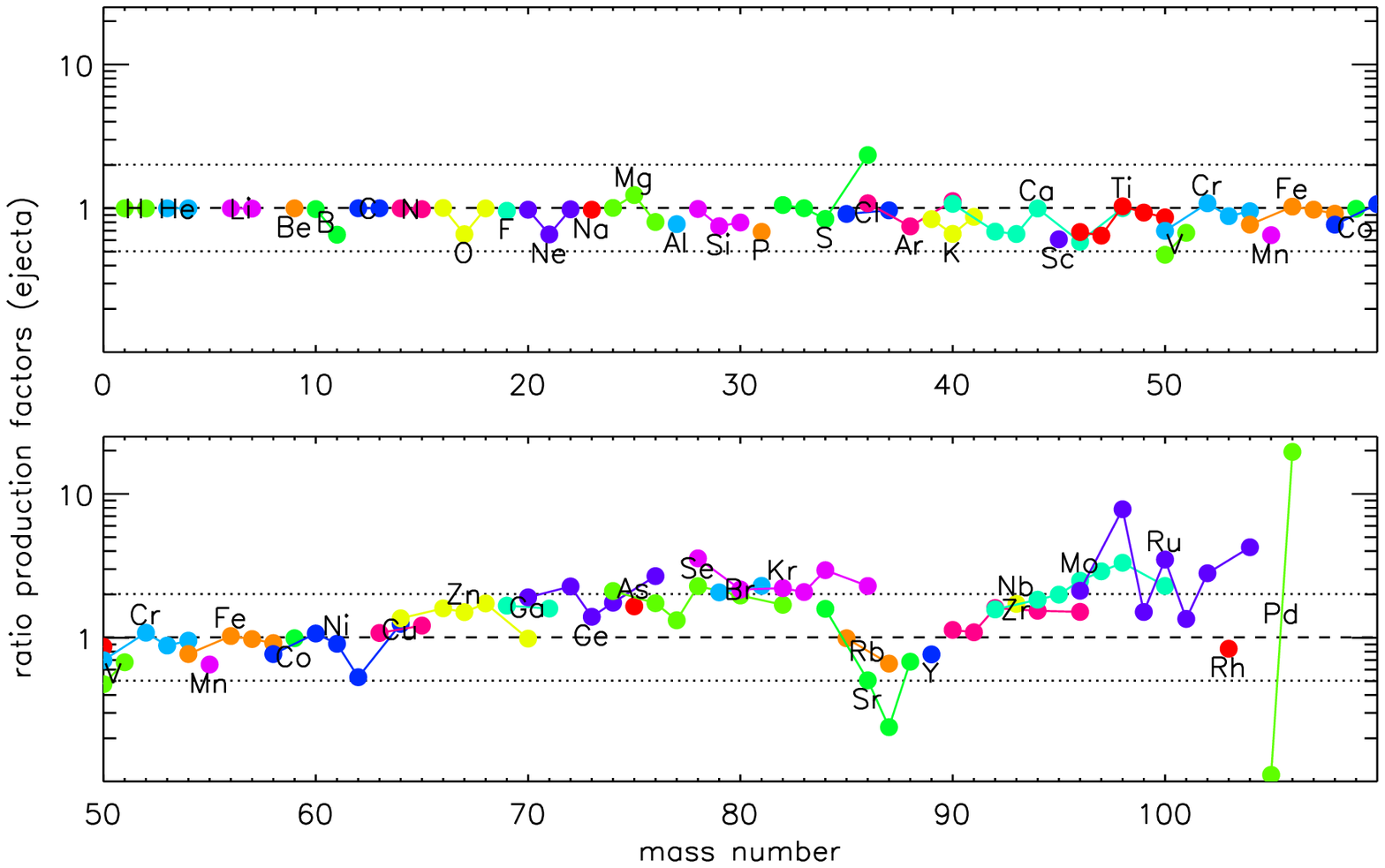}

\figcaption[f12.eps]{\lFig{SH25pf} Ratio of the production factors
for a model that used the rate set of \citet{HWW01}, H25, divided by
those of the standard Model S25.  Both models have an identical
stellar structure and evolution; only the network and reaction rates used
to calculate nucleosynthesis were changed. Most of the differences,
which can be quite significant, are due to differences in the cross
sections for \Rng in the two studies. In particular, H25 had {\sl
larger} destruction cross sections for the $s$-process isotopes of Sr
(in the case of $^{87}$Sr, the \Rng rate was 2.6 times greater) and
{\sl smaller} cross sections for the isotopes of Mo. The increased
production of $^{98}$Ru, a $p$-process isotope, reflects the larger
abundance of seeds heavier than A = 98. The abundance of this isotope
and heavier ones are not accurately calculated in H25 because of the
truncated network which ended at Ru.
%The dip in production beyond \I{84}{Kr} is caused by a three times
%lower \Rng cross section of \I{84}{Kr}. \I{86}{Kr} is still high,
%because the \Rng rate of \I{85}{Kr} is three times higher while the
%\Rng rate on \I{86}{Kr} is not much changed.
}
\end{figure}

\clearpage

\begin{figure}
\epsscale{0.8} 
\plotone{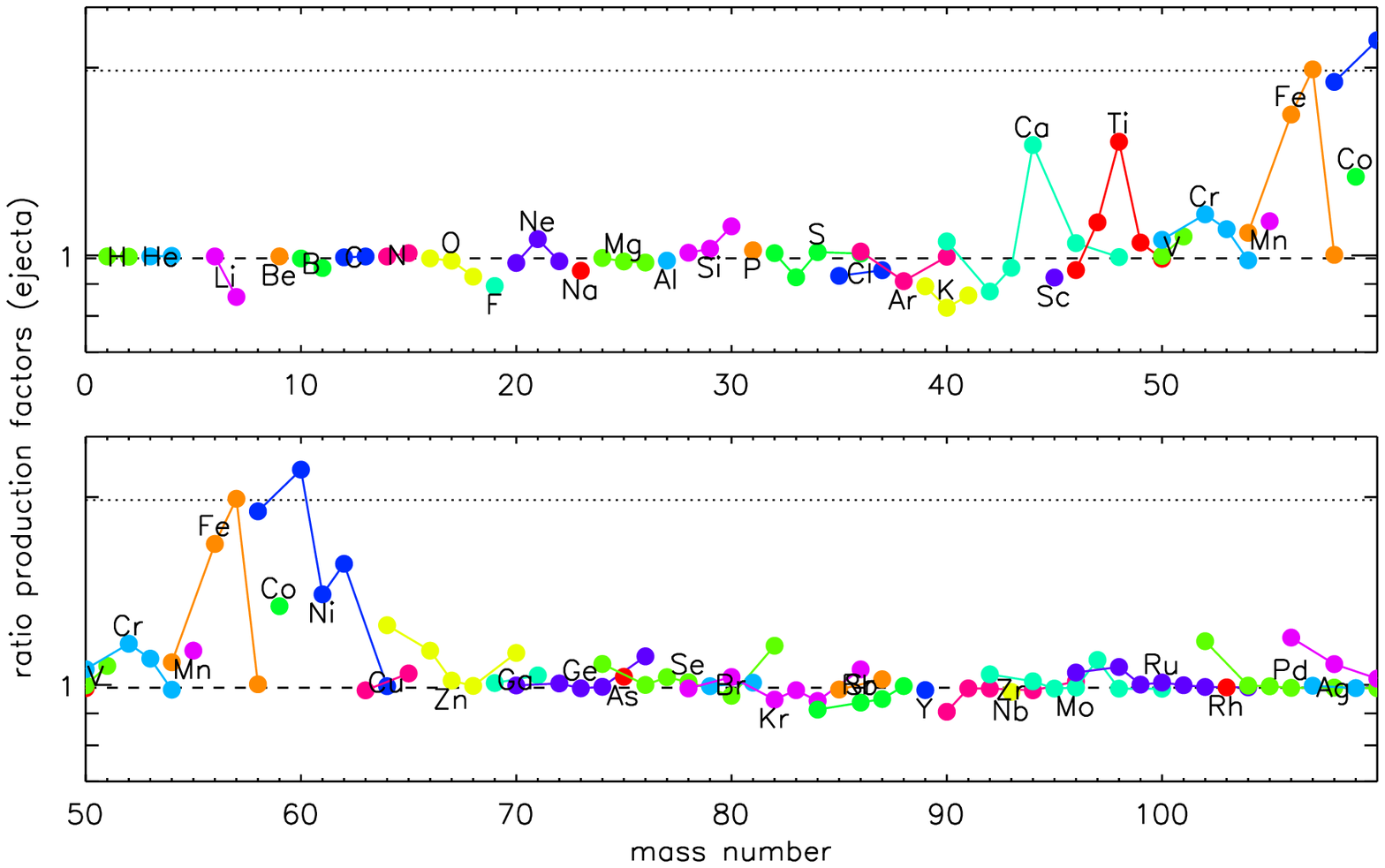} \figcaption[f13.eps]{\lFig{SP25pf} Decayed
post-explosive production factors of Model S25P (high explosion
energy) relative to Model S25.  Shown are only isotopes up to mass
number 110; beyond this the changes are only minor.}
\end{figure}

\clearpage
\begin{figure}
\noindent
{\bf \Large $<$f14.gif$>$\\[1.0cm]
This figure is available as a GIF graphic from the arXiv server}
\figcaption[f14.eps]{\lFig{S15cnv} History of the convective
structure and energy generation in the inner 5 \Msun of Model S15
starting after helium has been depleted in the center of the star.
Time is given on a logarithmic scale measured backwards, in years,
from the instant of iron core collapse (\textsl{x-axis}), The
vertical axis shows the interior mass in solar masses. Nuclear and
neutrino losses are also given on a logarithmic color-coded scale with
pink corresponding to energy loss and blue energy gain (in
\Ep{-1}\,\erggs; each level of more intense (darker) coloring
indicates an increase by one order of magnitude.)  Convective regions
are indicated by \textsl{green hatching} and semiconvective layers by
\textsl{red cross hatching}. Note the convective red supergiant
envelope outside about 4 \Msun. Going along the x-axis at e.g., an
interior mass of 1 \Msun, one encounters sequential episodes of
convective carbon, neon, oxygen, and silicon burning (neon burning is
comparatively brief). Unlike stars above about 20 \Msun, carbon
burning occurs convectively in the center of this star. Note the
existence of a convective helium shell at death reaching from 3.0 to
3.8 \Msun and a merged carbon, neon, and oxygen convective shell from
1.8 to 2.6 \Msun.}
\end{figure}

\clearpage

\begin{figure}
\noindent
{\bf \Large $<$f15.gif$>$\\[1.0cm]
This figure is available as a GIF graphic from the arXiv server}
\figcaption[f15.eps]{\lFig{S20cnv} Similar to \Fig{S15cnv} but for
Model S20, a 20\,\Msun star.  Carbon burning ignites in the middle of
the star, but barely so.  Note the merging of the oxygen-burning shell
with the first carbon-burning region about a day before the death of
the star.  This leaves enough time for hydrodynamic adjustment of the
CO core and extended merging of the carbon, neon, and oxygen
burning regions before death.  The result is the very peculiar
nucleosynthesis pattern we observe only in this model.}
\end{figure}

\clearpage

\begin{figure}
\noindent
{\bf \Large $<$f16.gif$>$\\[1.0cm]
This figure is available as a GIF graphic from the arXiv server}
\figcaption[f16.eps]{\lFig{S25cnv} Similar to \Fig{S15cnv} but for
Model S25, our standard 25 \Msun star.  Note that carbon burning
starts out radiatively (not convectively) in the center of the star
and is not exoergic when neutrino losses are included.  Shell carbon
burning becomes exoergic only later, off center.  A narrow
semiconvective region, poorly resolved in the figure, separates the
oxygen burning and carbon-neon burning shell until about 5 s before
core collapse (as compared to a day in the 20\,\Msun model).  Mixing
does start, but the remaining life-time of the star is too short to
significantly alter its structure or nucleosynthesis.}
\end{figure}

\clearpage
\begin{figure}
\epsscale{0.8} 
\plotone{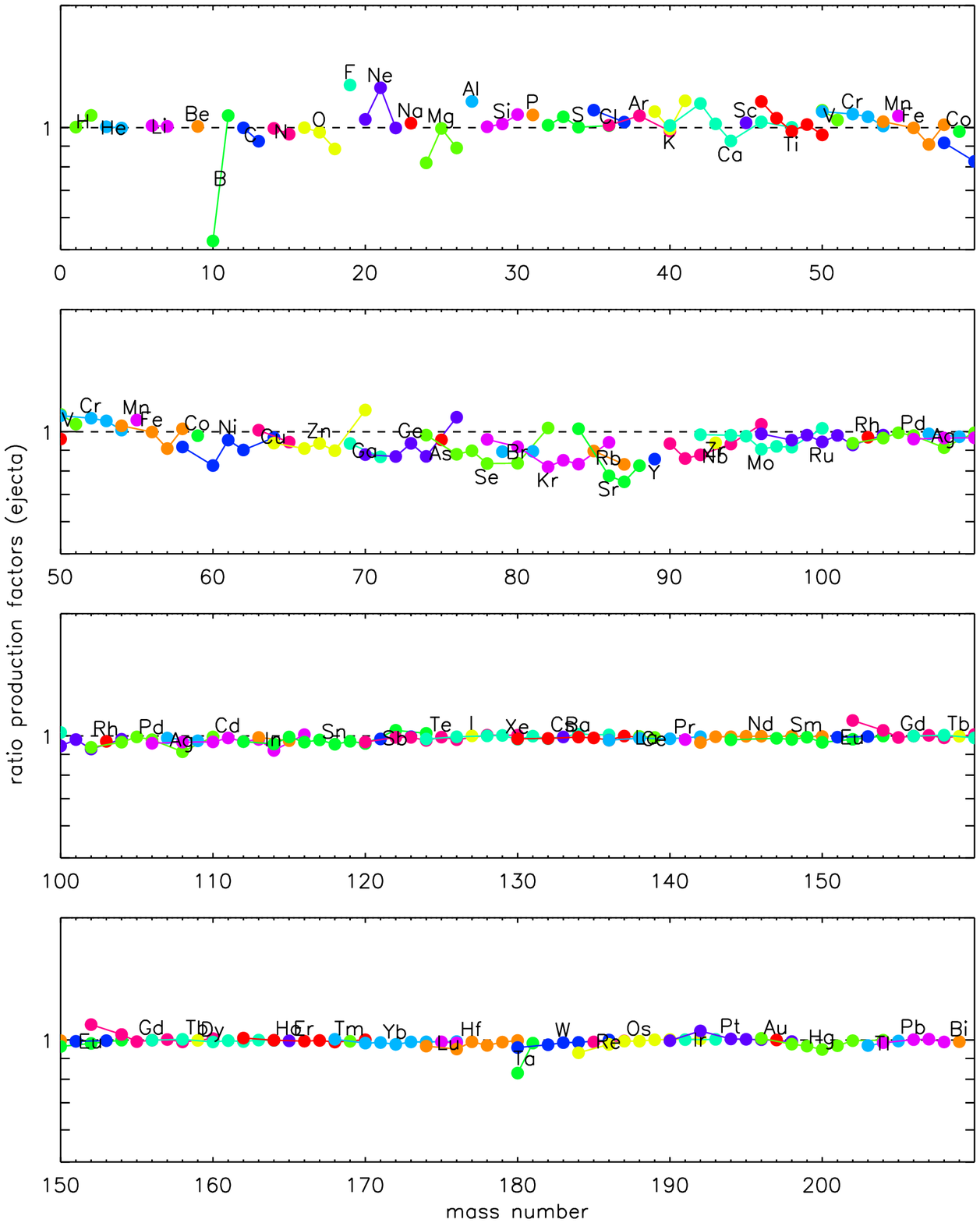}
\figcaption[f17.eps]{\lFig{SM25pf} Ratio of the production factors
of the 25\,\Msun Model M25 divided by S25 (standard).  Model M25 was a
test calculation using the NACRE rate set except for the
\I{22}{Ne}\Ran and \I{22}{Ne}\Rag rates for which we implemented our
``standard'' rates \citep{HWW01}.}
\end{figure}

%\clearpage

%% Tables should be submitted one per page, so put a \clearpage before

%% Two options are available to the author for producing tables:  the
%% deluxetable environment provided by the AASTeX package or the LaTeX
%% table environment.  Use of deluxetable is preferred.  
%%
%% NOT BY ME --a
%%

\clearpage

%%%%%%%%%%%%%%%%%%%%%%%%%%%%%%%%%%%%%%%%%%%%%%%%%%%%%%%%%%%%%%%%%%%%%%%%
% Reaction reference table

%/export/home/hoffman/tommy/rtbl00/aug00/rtbl00/data/sep/exp_rates/smtrx921b.tex

\begin{deluxetable}{lllllll}
\tabletypesize{\footnotesize}
\tablewidth{0pt}
\tablecaption{\label{tab:rrrm}Experimental reaction rate reference matrix.
Identification labels are explained in Table \protect{\ref{tab:rrrm_expl}}.
Only those targets are given for which there is experimental data beyond
\protect{\citet{bao00}}. The full rate set is constructed by combination of
theoretical rates \protect{\citep{RATH}} with the data given in this
table and Table \protect{\ref{tab:macs}}.}
\tablehead{
\colhead{$^AZ$} &
\colhead{(${\rm n}, \gamma$)$^{A+1}Z$} &
\colhead{(${\rm p},{\rm n}$)$^{A}{Z+1}$} &
\colhead{(${\rm p}, \gamma$)$^{A+1}{Z+1}$} &
\colhead{($\alpha, {\rm p}$)$^{A+3}{Z+1}$} &
\colhead{($\alpha, {\rm n}$)$^{A+3}{Z+2}$} &
\colhead{($\alpha,  \gamma$)$^{A+4}{Z+2}$} 
}
\startdata
 $^{ 1}$H& * WIES  &  0.        &  0.        &  0.        &  0.        &  0.        \\
 $^{ 2}$H& * WIES  &  0.        &  * CF88\ud &  0.        &  0.        &  * CF88\ud \\
 $^{ 3}$H&  0.     &  * CF88    &  * CF88    &  0.        &  * CF88    &  * CF88\ud \\
 $^{ 3}$He& * WIES  &  0.        &  0.        &  * CF88\ud &  0.        &  * CF88\ud \\
 $^{ 6}$Li&  0.     &  0.        &  * CF88\ud &  * CF88\ud &  0.        &  * CF88    \\
 $^{ 7}$Li& * RA94  &  * CF88    &  * CF88\ud &  0.        &  * CF88\ud &  * CF88\ud \\
 $^{ 8}$Li& * WIES  &  0.        &  0.        &  0.        &  * RA94    &    FKTH    \\
 $^{ 7}$Be&  0.     &  0.        &  * CF88\ud &  * CF88\ud &  0.        &  * CF88\ud \\
 $^{ 9}$Be& * WIES  &  * CF88\ud &  * CF88\ud &  0.        &  * WB94\ud &  0.        \\
 $^{ 8}$B&  0.     &  0.        &  * WIES    &  * CF88    &  0.        &  0.        \\
 $^{10}$B&  * WIES &  0.        &  * CF88\ud &  0.        &  * CF88    &  0.        \\
 $^{11}$B&  * RA94 &  * CF88\ud &  * CF88\ud &  * CF88    &  * CF88    &  0.        \\
 $^{11}$C&  * RA94 &  0.        &  * CF88    &  * CF88\ud &  0.        &  0.        \\
 $^{12}$C&  * BAAL &  0.        &  * FCZ2\ud &  * HFCZ\ud &  * HFCZ    &  * BU96    \\
 $^{13}$C&  * RA94 &  * FCZ2\ud &  * CF88\ud &  0.        &  * FCZ2\ud &    RATH    \\
 $^{14}$C&  * RA94 &  * CF88    &  * CF88    &  0.        &  * RA94    &  * CF88    \\
 $^{13}$N&  * WIES &  0.        &  * KL93\ud &  * HFCZ    &  0.        &  0.        \\
 $^{14}$N&  * WIES &  * FCZ2\ud &  * CF88\ud &  * LA90\ud &  * CF88\ud &  * FCZ2\ud \\
 $^{15}$N&  * WIES &  * CF88\ud &  * FCZ2\ud &  * CF88\ud &  * CF88    &  * CF88\ud \\
 $^{14}$O&  0.     &  0.        &  0.        &  * CF88    &    RATH    &  * CF88    \\
 $^{15}$O&  0.     &  0.        &  0.        &  * WK82    &  0.        &  * CF88    \\
 $^{16}$O&  * BAAL &  0.        &  * FCZ2\ud &  * HFCZ\ud &    RATH    &  * CF88\ud \\
 $^{17}$O&  * RA94 &  FKTH      &  * LA90\ud &  0.        &  * FCZ2\ud &  * FCZ2    \\
 $^{18}$O&  * RA94 &  FKTH      &  * CF88\ud &  0.        &  * CF88\ud &  * GB94\ud \\
 $^{17}$F&    FKTH &  0.        &  * WK82    &  * CF88\ud &  0.        &  0.        \\
 $^{18}$F&    FKTH &  0.        &  * WK82    &            &  0.        &  0.        \\
 $^{19}$F&  * BAAL &  * CF88\ud &  * FCZ2\ud &  * CF88    &  * CF88    &  0.        \\
 $^{19}$Ne&    RATH &    RATH    &  * CF88    &    RATH    &    RATH    &    RATH    \\
 $^{20}$Ne&  * WM88 &    RATH    &  * ID01\ud &  * ID01\ud &    RATH    &    RTGW\ud \\
 $^{21}$Ne&  * BAAL &    RATH    &  * ID01\ud &    RATH    &  * FCZ2\ud &  * HFCZ    \\
 $^{22}$Ne&  * BAAL &  * CF88    &  * ID01\ud &    RATH    &  * KA94\ud &  * KA94\ud \\
 $^{20}$Na&    RATH &    RATH    &  * ID01    &    RATH    &    RATH    &    RATH    \\
 $^{21}$Na&    RATH &    RATH    &  * ID01\ud &  * CF88\ud &    RATH    &    RATH    \\
 $^{22}$Na&    RATH &    RATH    &  * ID01\ud &    RATH    &    RATH    &    RATH    \\
 $^{23}$Na&  * BAAL &  * CF88\ud &  * ID01\ud &    RATH    &    RATH\ud &    RATH    \\
 $^{22}$Mg&    RATH &    RATH    &  * ID01    &    RATH    &    RATH    &    RATH    \\
 $^{23}$Mg&    RATH &    RATH    &  * ID01    &    RATH    &    RATH    &    RATH    \\
 $^{24}$Mg&  * BAAL &    RATH    &  * ID01\ud &  * ID01\ud &    RATH    &    RTGW    \\
 $^{25}$Mg&  * BAAL &    RATH    &  * ID01\ud &  * CF88    &  * FCZ2\ud &  * HFCZ    \\
 $^{26}$Mg&  * BAAL &    RATH    &  * ID01\ud &    RATH    &  * FCZ2\ud &  * HFCZ    \\
 $^{23}$Al&    RATH &    RATH    &  * ID01    &    RATH    &    RATH    &    RATH    \\
 $^{24}$Al&    RATH &    RATH    &  * ID01    &    RATH    &    RATH    &    RATH    \\
 $^{25}$Al&    RATH &    RATH    &  * ID01\ud &    RATH    &    RATH    &    RATH    \\
 $^{26}$Al&    RATH &    RATH    &  * ID01\ud &    RATH    &    RATH    &    RATH    \\
 $^{27}$Al&  * BAAL &    RATH    &  * ID01\ud &    RATH    &  * CF88\ud &    RATH    \\
 $^{26}$Si&    RATH &    RATH    &  * ID01    &    RATH    &    RATH    &    RATH    \\
 $^{27}$Si&    RATH &    RATH    &  * ID01\ud &    RATH    &    RATH    &    RATH    \\
 $^{28}$Si&  * BAAL &    RATH    &  * ID01\ud &  * ID01    &    RATH    &    RTGW    \\
 $^{29}$Si&  * BAAL &    RATH    &  * ID01\ud &    RATH    &    RATH    &    RATH    \\
 $^{30}$Si&  * BAAL &    RATH    &  * ID01\ud &    RATH    &    RATH    &    RATH    \\
 $^{27}$P&    RATH &    RATH    &  * ID01    &    RATH    &    RATH    &    RATH    \\ 
 $^{28}$P&    RATH &    RATH    &    ID01    &    RATH    &    RATH    &    RATH    \\
 $^{29}$P&    RATH &    RATH    &  * ID01    &    RATH    &    RATH    &    RATH    \\
 $^{30}$P&    RATH &    RATH    &    ID01    &    RATH    &    RATH    &    RATH    \\
 $^{31}$P&  * BAAL &    RATH    &  * ID01    &    RATH &    RATH &    RATH \\
 $^{30}$S&    RATH &    RATH    &  * ID01    &    RATH &    RATH &    RATH \\
 $^{31}$S&    RATH &    RATH    &  * ID01    &    RATH &    RATH &    RATH \\
 $^{32}$S&  * BAAL &    RATH    &  * ID01    &    RATH &    RATH &    RTGW \\
 $^{33}$S&  * BAAL &    RATH    &    ID01    &    RATH &    RATH &    RATH \\
 $^{34}$S&  * BAAL &    RATH    &    ID01    &  * TS92 &  * TS92 &  * TS92 \\
 $^{31}$Cl&    RATH &    RATH    &  * ID01    &    RATH &    RATH &    RATH \\
 $^{32}$Cl&    RATH &    RATH    &  * ID01    &    RATH &    RATH &    RATH \\
 $^{33}$Cl&    RATH &    RATH    &    ID01    &    RATH &    RATH &    RATH \\
 $^{34}$Cl&    RATH &    RATH    &    ID01    &    RATH &    RATH &    RATH \\
 $^{35}$Cl&  * BAAL &    RATH    &  * ID01    &    RATH &    RATH &    RATH \\
 $^{34}$Ar&    RATH &    RATH    &  * ID01 &    RATH &    RATH &    RATH \\
 $^{35}$Ar&    RATH &    RATH    &  * ID01 &    RATH &    RATH &    RATH \\
 $^{36}$Ar&  BAAL &    RATH    &  * ID01 &    ID01 &    RATH &    RTGW \\
 $^{38}$Ar&  BAAL &    RATH    &    RATH &  * SM86 &  * SM86 &    RATH \\
 $^{35}$K&    RATH &    RATH    &  * ID01 &    RATH &    RATH &    RATH \\
 $^{36}$K&    RATH &    RATH    &    ID01 &    RATH &    RATH &    RATH \\
 $^{37}$K&    RATH &    RATH    &    ID01 &    RATH &    RATH &    RATH \\
 $^{38}$K&    RATH &    RATH    &    ID01 &    RATH &    RATH &    RATH \\
 $^{39}$K&  * BAAL &    RATH    &    ID01 &    RATH &    RATH &    RATH \\
 $^{40}$K&  BAAL &    RATH    &    RATH &    RATH &    RATH &    RATH \\
 $^{41}$K&  * BAAL &    RATH    &    RATH &  * SM91 &    RATH &    RATH \\
 $^{39}$Ca&    RATH &    RATH    &  * ID01 &    RATH &    RATH &    RATH \\
 $^{40}$Ca&  * BAAL &    RATH    &  * ID01 &    RATH &    RATH &    RTGW \\
 $^{42}$Ca&  * BAAL &    RATH    &    RATH &  * MK85 &    RATH &  * MK85 \\
 $^{45}$Sc&  * BAAL &    RATH    &    RATH &  * HT89 &  * HT89 &    RATH \\
 $^{48}$Ti&  * BAAL &    RATH    &    RATH &  * MT92 &    RATH &    RATH \\
&&&&&&\\
 $^{70}$Ge&  * BAAL &    RATH    &    RATH &    THIS &    THIS &  * FU96 \\
$^{144}$Sm&  * BAAL &    RATH    &    RATH &    THIS &    THIS &  * SO98 \\
\enddata
\tablecomments{\ud Indicates the reaction rate was varied in the calculations
(see the text). A "*" indicates the reaction rate is based on experiment
. A "0." indicates no rate is available for the given channel.}

\end{deluxetable}

\clearpage

\begin{deluxetable}{lll}
\tablecaption{\label{tab:rrrm_expl}Reference list for the reaction rate
reference matrix (Table \protect{\ref{tab:rrrm}}).}
\tablehead{
\colhead{Label} & \colhead{Reference} & \colhead{Comment}
}
\startdata
BAAL & \protect{\citet{bao00}} & see Table \protect{\ref{tab:macs}}\\
BU96 & \protect{\citet{Buc96}} &\\
CF88 & \protect{\citet{CF88}} &\\
EC95 & \protect{\citet{ec95}} &\\
FCZ2 & \protect{\citet{fcz2}} &\\
FKTH & \protect{\citet{fkth,REACLIB95}} &\\
FU96 & \protect{\citet{fue96}} &\\
GB94 & \protect{\citet{gb94}} &\\
GW89 & \protect{\citet{gw89}} &\\
HFCZ & \protect{\citet{hfcz}} &\\
HT89 & \protect{\citet{ht89}} &\\
ID01 & \protect{\citet{id01}} &\\
KA94 & \protect{\citet{kaepp94}} & modified (see text)\\
KL93 & \protect{\citet{kl93}} &\\
LA90 & \protect{\citet{la90}} &\\
MK85 & \protect{\citet{mk85}} &\\
MT92 & \protect{\citet{mt92}} &\\
RA94 & \protect{\citet{ra94}} &\\
RTGW & \protect{\citet{rtgw00}} & for implementation, see Table 
\protect{\ref{tab:iso}}\\
SO98 & \protect{\citet{som98}} &\\
THIS & this work & see Table \protect{\ref{tab:alpha}}\\
TS92 & \protect{\citet{ts92}} &\\
SM86 & \protect{\citet{sm86}} &\\
SM91 & \protect{\citet{sm91}} &\\
WB94 & \protect{\citet{wb94}} &\\
WIES & \protect{\citet{wies}} & see also \protect{\citet{REACLIB95}}\\
WFHZ & \protect{\citet{wfhz}} &\\
WM88 & \protect{\citet{wm88}} &\\
\enddata

\end{deluxetable}

\clearpage

\begin{deluxetable}{lrrrr}
%\tabletypesize{\scriptsize}
\tablecaption{\label{tab:alpha}Fit parameters for reactions including the
$\alpha$+$^{70}$Ge and $\alpha$+$^{144}$Sm channels; see \protect{\citet{RATH}}
for the definition of the coefficients.}
\tablewidth{0pt}
\tablehead{
\colhead{Reaction} & \colhead{$a_0$} & \colhead{$a_1$}  & \colhead{$a_2$}
& \colhead{$a_3$} \\
&& \colhead{$a_4$} & \colhead{$a_5$} & \colhead{$a_6$}}
\startdata
$^{70}$Ge($\alpha$,$\gamma$)$^{74}$Se & -9.051749E+02&2.096414E+01&-2.210416E+03
&3.222456E+03\\
&&-1.757347E+02&9.295969E+00&-1.599497E+03\\
%\multicolumn{1}{r}{reverse rate}&&&&&&&\\
$^{144}$Sm($\alpha$,$\gamma$)$^{144}$Gd & -9.547989E+02&5.832285E+00
&-4.444873E+03&5.555737E+03\\
&&-2.560688E+02&1.178841E+01&-3.015931E+03\\
%\multicolumn{1}{r}{reverse rate}&&&&&&&\\
\tableline
$^{73}$As(p,$\alpha$)$^{70}$Ge & 1.513141E+02&-9.065257E+00&3.503307E+02
&-5.256190E+02\\
&&2.489691E+01&-1.153807E+00&2.762916E+02\\
%\multicolumn{1}{r}{reverse rate}&&&&&&&\\
$^{147}$Eu(p,$\alpha$)$^{144}$Sm & -3.239141E+02&-7.576453E+00&-3.349700E+02
&6.806995E+02\\
&&-4.521575E+01&2.685919E+00&-2.840773E+02\\
%\multicolumn{1}{r}{reverse rate}&&&&&&&\\
\tableline
$^{73}$Se(n,$\alpha$)$^{70}$Ge & 2.750996E+01&-1.475808E-01&1.093870E+01
&-2.401239E+01\\
&&1.486014E+00&-3.967125E-02&1.028549E+01\\
%\multicolumn{1}{r}{reverse rate}&&&&&&&\\
$^{147}$Gd(n,$\alpha$)$^{144}$Sm & -1.203181E+01&1.696984E-01&-1.675541E+01
&4.310292E+01\\
&&-3.403980E+00&3.122087E-01&-1.666362E+01\\
%\multicolumn{1}{r}{reverse rate}&&&&&&&\\
\enddata

\tablecomments{The reverse rates can be derived as explained in
\protect{\citet{RATH}} and \protect{\citet{rt01}}.}
\end{deluxetable}

\clearpage

\begin{deluxetable}{ccr}
%\tabletypesize{\scriptsize}
\tablewidth{0pt}
\tablecaption{\label{tab:iso}Rates for $\alpha$ capture reactions
on self-conjugated targets.}
\tablehead{
\colhead{Reaction} & \colhead{$T_{\rm match}$} & \colhead{$a_0^{\rm renorm}$}}
\startdata
$^{20}$Ne($\alpha$,$\gamma$)$^{24}$Mg&3.0&$1.333837363\times 10^{2}$\\
$^{24}$Mg($\alpha$,$\gamma$)$^{28}$Si&2.0&$1.428649975\times 10^{2}$\\
$^{28}$Si($\alpha$,$\gamma$)$^{32}$S &3.0&$9.4108\times 10^{1}$\\
$^{32}$S($\alpha$,$\gamma$)$^{36}$Ar &0.0&$-1.915768\times 10^{2}$\\
$^{36}$Ar($\alpha$,$\gamma$)$^{40}$Ca&0.0&$-1.289706\times 10^{2}$\\
$^{40}$Ca($\alpha$,$\gamma$)$^{44}$Ti&0.0&$-7.490256\times 10^{2}$
\enddata
\tablecomments{For $T_9<T_{\rm match}$ a sum of resonances
\protect{\citep[taken from][]{rtgw00}} was used; for $T_9\geq T_{\rm match}$ a 
renormalized RATH rate was used with the new parameter
$a_0=a_0^{\rm renorm}$ (the
reverse rate has to be renormalized by the same factor).}

\end{deluxetable}

\clearpage

\begin{deluxetable}{rrrrrrrrrrrrrr}
\tabletypesize{\scriptsize}
\tablecaption{\label{tab:macs}Renormalization factors $f=r_{\rm
exp}/r_{\rm theory}$ of the theoretical
(n,$\gamma$) rates of
\protect{\citet{RATH}} in order to yield
a 30 keV  Maxwell-averaged cross section  consistent with \protect{\citet{bao00}}.}
\tablewidth{0pt}
\tablehead{
\colhead{Target} & \colhead{$f$} & \colhead{Target} & \colhead{$f$}
& \colhead{Target} & \colhead{$f$} & \colhead{Target} &
\colhead{$f$} & \colhead{Target} & \colhead{$f$}
& \colhead{Target} & \colhead{$f$}
& \colhead{Target} & \colhead{$f$}}
\startdata
$^{20}$Ne & 0.072 &$^{54}$Fe & 0.594 &$^{86}$Rb & 0.428 &$^{112}$Cd & 1.053 &$^{135}$Ba & 0.918 &$^{156}$Gd & 1.183 &$^{182}$W & 0.680 \\
$^{21}$Ne & 0.332 &$^{55}$Fe & 0.895 &$^{87}$Rb & 0.302 &$^{113}$Cd & 1.232 &$^{136}$Ba & 0.596 &$^{157}$Gd & 1.122 &$^{183}$W & 0.746 \\
$^{22}$Ne & 0.132 &$^{56}$Fe & 0.447 &$^{84}$Sr & 0.924 &$^{114}$Cd & 0.894 &$^{137}$Ba & 0.647 &$^{158}$Gd & 1.145 &$^{184}$W & 0.894 \\
$^{23}$Na & 0.697 &$^{57}$Fe & 1.186 &$^{86}$Sr & 0.305 &$^{115}$Cd & 0.667 &$^{138}$Ba & 0.696 &$^{160}$Gd & 0.758 &$^{185}$W & 1.369 \\
$^{24}$Mg & 0.965 &$^{58}$Fe & 0.944 &$^{87}$Sr & 0.278 &$^{116}$Cd & 0.656 &$^{139}$La & 0.514 &$^{159}$Tb & 1.091 &$^{186}$W & 1.221 \\
$^{25}$Mg & 0.777 &$^{59}$Co & 0.740 &$^{88}$Sr & 0.402 &$^{113}$In & 0.659 &$^{132}$Ce & 0.913 &$^{160}$Tb & 1.204 &$^{185}$Re & 1.432 \\
$^{26}$Mg & 0.075 &$^{58}$Ni & 0.811 &$^{89}$Sr & 0.406 &$^{114}$In & 0.800 &$^{133}$Ce & 0.616 &$^{156}$Dy & 1.342 &$^{186}$Re & 1.080 \\
$^{27}$Al & 0.630 &$^{59}$Ni & 0.940 &$^{89}$Y & 0.214 &$^{115}$In & 0.709 &$^{134}$Ce & 0.787 &$^{158}$Dy & 0.899 &$^{187}$Re & 1.443 \\
$^{28}$Si & 0.529 &$^{60}$Ni & 0.902 &$^{90}$Zr & 0.420 &$^{112}$Sn & 0.543 &$^{135}$Ce & 0.621 &$^{160}$Dy & 1.386 &$^{184}$Os & 0.868 \\
$^{29}$Si & 0.895 &$^{61}$Ni & 1.136 &$^{91}$Zr & 0.386 &$^{114}$Sn & 0.490 &$^{136}$Ce & 0.547 &$^{161}$Dy & 1.200 &$^{186}$Os & 0.676 \\
$^{30}$Si & 3.218 &$^{62}$Ni & 0.650 &$^{92}$Zr & 0.477 &$^{115}$Sn & 0.648 &$^{137}$Ce & 0.685 &$^{162}$Dy & 0.991 &$^{187}$Os & 0.874 \\
$^{31}$P & 0.156 &$^{63}$Ni & 0.873 &$^{93}$Zr & 0.501 &$^{116}$Sn & 0.531 &$^{138}$Ce & 0.614 &$^{163}$Dy & 1.042 &$^{188}$Os & 1.131 \\
$^{32}$S & 0.367 &$^{64}$Ni & 0.858 &$^{94}$Zr & 0.391 &$^{117}$Sn & 0.750 &$^{139}$Ce & 0.611 &$^{164}$Dy & 1.035 &$^{189}$Os & 1.023 \\
$^{33}$S & 0.489 &$^{63}$Cu & 0.925 &$^{95}$Zr & 0.624 &$^{118}$Sn & 0.597 &$^{140}$Ce & 0.487 &$^{163}$Ho & 1.025 &$^{190}$Os & 1.374 \\
$^{34}$S & 0.064 &$^{65}$Cu & 0.731 &$^{96}$Zr & 0.742 &$^{119}$Sn & 0.594 &$^{141}$Ce & 0.597 &$^{165}$Ho & 1.038 &$^{191}$Os & 1.684 \\
$^{36}$S & 0.599 &$^{64}$Zn & 0.773 &$^{93}$Nb & 0.620 &$^{120}$Sn & 0.572 &$^{142}$Ce & 0.744 &$^{162}$Er & 1.456 &$^{192}$Os & 2.537 \\
$^{35}$Cl & 0.627 &$^{65}$Zn & 0.799 &$^{94}$Nb & 0.596 &$^{121}$Sn & 0.714 &$^{141}$Pr & 0.425 &$^{164}$Er & 1.083 &$^{191}$Ir & 1.366 \\
$^{36}$Cl & 0.659 &$^{66}$Zn & 0.735 &$^{92}$Mo & 0.546 &$^{122}$Sn & 0.556 &$^{142}$Pr & 0.531 &$^{166}$Er & 1.156 &$^{192}$Ir & 1.291 \\
$^{37}$Cl & 0.992 &$^{67}$Zn & 1.041 &$^{94}$Mo & 0.674 &$^{124}$Sn & 0.559 &$^{143}$Pr & 0.811 &$^{167}$Er & 1.063 &$^{193}$Ir & 1.416 \\
$^{36}$Ar & 0.615 &$^{68}$Zn & 0.671 &$^{95}$Mo & 0.605 &$^{121}$Sb & 0.752 &$^{142}$Nd & 0.423 &$^{168}$Er & 1.200 &$^{190}$Pt & 0.762 \\
$^{38}$Ar & 0.789 &$^{70}$Zn & 1.285 &$^{96}$Mo & 0.712 &$^{122}$Sb & 0.410 &$^{143}$Nd & 0.526 &$^{169}$Er & 1.488 &$^{192}$Pt & 0.954 \\
$^{39}$Ar & 0.900 &$^{69}$Ga & 0.768 &$^{97}$Mo & 0.830 &$^{123}$Sb & 0.553 &$^{144}$Nd & 0.715 &$^{170}$Er & 0.983 &$^{193}$Pt & 0.948 \\
$^{40}$Ar & 0.709 &$^{71}$Ga & 1.318 &$^{98}$Mo & 1.050 &$^{120}$Te & 0.753 &$^{145}$Nd & 0.809 &$^{169}$Tm & 1.183 &$^{194}$Pt & 1.120 \\
$^{39}$K & 0.855 &$^{70}$Ge & 0.820 &$^{100}$Mo & 1.762 &$^{122}$Te & 0.870 &$^{146}$Nd & 0.895 &$^{170}$Tm & 1.333 &$^{195}$Pt & 1.504 \\
$^{40}$K & 1.047 &$^{72}$Ge & 1.272 &$^{99}$Tc & 1.141 &$^{123}$Te & 1.113 &$^{147}$Nd & 1.471 &$^{171}$Tm & 0.862 &$^{196}$Pt & 1.179 \\
$^{41}$K & 0.968 &$^{73}$Ge & 1.246 &$^{96}$Ru & 0.842 &$^{124}$Te & 0.772 &$^{148}$Nd & 1.630 &$^{168}$Yb & 1.207 &$^{198}$Pt & 1.211 \\
$^{40}$Ca & 0.544 &$^{74}$Ge & 1.269 &$^{98}$Ru & 0.659 &$^{125}$Te & 0.945 &$^{150}$Nd & 1.512 &$^{170}$Yb & 1.020 &$^{197}$Au & 1.136 \\
$^{41}$Ca & 0.973 &$^{76}$Ge & 1.442 &$^{99}$Ru & 0.836 &$^{126}$Te & 0.717 &$^{147}$Pm & 1.038 &$^{171}$Yb & 1.177 &$^{198}$Au & 1.089 \\
$^{42}$Ca & 0.932 &$^{75}$As & 1.893 &$^{100}$Ru & 0.965 &$^{128}$Te & 0.829 &$^{148}$Pm & 1.466 &$^{172}$Yb & 0.851 &$^{196}$Hg & 1.556 \\
$^{43}$Ca & 1.533 &$^{74}$Se & 1.290 &$^{101}$Ru & 1.504 &$^{130}$Te & 0.609 &$^{149}$Pm & 2.179 &$^{173}$Yb & 0.873 &$^{198}$Hg & 0.729 \\
$^{44}$Ca & 1.186 &$^{76}$Se & 1.380 &$^{102}$Ru & 1.262 &$^{127}$I & 0.866 &$^{144}$Sm & 0.426 &$^{174}$Yb & 0.834 &$^{199}$Hg & 1.208 \\
$^{45}$Ca & 1.045 &$^{77}$Se & 1.356 &$^{103}$Ru & 0.935 &$^{129}$I & 1.114 &$^{147}$Sm & 0.655 &$^{175}$Yb & 1.112 &$^{200}$Hg & 1.040 \\
$^{46}$Ca & 1.694 &$^{78}$Se & 1.477 &$^{104}$Ru & 1.595 &$^{124}$Xe & 0.803 &$^{148}$Sm & 0.929 &$^{176}$Yb & 0.977 &$^{201}$Hg & 2.333 \\
$^{48}$Ca & 1.315 &$^{79}$Se & 1.193 &$^{103}$Rh & 1.001 &$^{126}$Xe & 0.666 &$^{149}$Sm & 1.596 &$^{175}$Lu & 1.077 &$^{202}$Hg & 1.533 \\
$^{45}$Sc & 1.189 &$^{80}$Se & 1.097 &$^{102}$Pd & 0.998 &$^{128}$Xe & 0.799 &$^{150}$Sm & 1.754 &$^{176}$Lu & 1.036 &$^{203}$Hg & 1.450 \\
$^{46}$Ti & 0.770 &$^{82}$Se & 0.482 &$^{104}$Pd & 0.955 &$^{129}$Xe & 0.734 &$^{151}$Sm & 1.987 &$^{174}$Hf & 1.182 &$^{204}$Hg & 2.796 \\
$^{47}$Ti & 0.803 &$^{79}$Br & 1.422 &$^{105}$Pd & 1.236 &$^{130}$Xe & 0.880 &$^{152}$Sm & 1.644 &$^{176}$Hf & 0.802 &$^{203}$Tl & 0.975 \\
$^{48}$Ti & 2.083 &$^{81}$Br & 1.170 &$^{106}$Pd & 1.137 &$^{131}$Xe & 0.744 &$^{153}$Sm & 1.176 &$^{177}$Hf & 1.286 &$^{204}$Tl & 1.274 \\
$^{49}$Ti & 0.966 &$^{78}$Kr & 0.916 &$^{107}$Pd & 1.604 &$^{132}$Xe & 0.884 &$^{154}$Sm & 0.955 &$^{178}$Hf & 0.900 &$^{205}$Tl & 1.033 \\
$^{50}$Ti & 1.092 &$^{79}$Kr & 1.179 &$^{108}$Pd & 1.292 &$^{133}$Xe & 0.807 &$^{151}$Eu & 1.604 &$^{179}$Hf & 0.991 &$^{204}$Pb & 1.345 \\
$^{50}$V & 0.974 &$^{80}$Kr & 1.313 &$^{110}$Pd & 1.395 &$^{134}$Xe & 0.576 &$^{152}$Eu & 1.312 &$^{180}$Hf & 0.955 &$^{205}$Pb & 1.505 \\
$^{51}$V & 1.959 &$^{81}$Kr & 1.142 &$^{107}$Ag & 0.849 &$^{136}$Xe & 0.612 &$^{153}$Eu & 1.073 &$^{181}$Hf & 0.840 &$^{206}$Pb & 0.588 \\
$^{50}$Cr & 1.091 &$^{82}$Kr & 0.764 &$^{109}$Ag & 0.891 &$^{133}$Cs & 0.752 &$^{154}$Eu & 1.145 &$^{182}$Hf & 1.101 &$^{207}$Pb & 0.637 \\
$^{51}$Cr & 1.015 &$^{83}$Kr & 0.786 &$^{110}$Ag & 0.768 &$^{134}$Cs & 0.697 &$^{155}$Eu & 1.004 &$^{179}$Ta & 0.899 &$^{208}$Pb & 0.918 \\
$^{52}$Cr & 0.523 &$^{84}$Kr & 0.609 &$^{106}$Cd & 0.665 &$^{135}$Cs & 0.690 &$^{152}$Gd & 1.802 &$^{180}$Ta & ---\tablenotemark{a} &$^{209}$Bi & 0.334 \\
$^{53}$Cr & 2.237 &$^{85}$Kr & 0.448 &$^{108}$Cd & 0.536 &$^{130}$Ba & 1.038 &$^{153}$Gd & 1.519 &$^{181}$Ta & 0.854 &$^{210}$Bi & 0.465 \\
$^{54}$Cr & 0.715 &$^{86}$Kr & 0.419 &$^{110}$Cd & 0.814 &$^{132}$Ba & 0.803 &$^{154}$Gd & 1.631 &$^{182}$Ta & 0.854 & & \\
$^{55}$Mn & 1.135 &$^{85}$Rb & 0.529 &$^{111}$Cd & 1.318 &$^{134}$Ba & 0.767 &$^{155}$Gd & 1.198 &$^{180}$W & 0.793 & & \\
\enddata
\tablenotetext{a}{The RATH rate is for $^{180\rm g}$Ta whereas in 
\protect{\citet{bao00}} the MACS for $^{180\rm m}$Ta is given. The proper
renormalized rate coefficients for $^{180\rm m}$Ta(n,$\gamma$) are 
$a_0=-9.899046$, $a_1=0.2659302$, $a_2=-27.6891$, $a_3=59.43646$,
$a_4=-3.639849$, $a_5=0.03887359$, and $a_6=-25.80838$.}
\tablecomments{Forward as well as reverse rate have to be multiplied by
the given factor.}
\end{deluxetable}

\clearpage

{% BEGINNING OF TABLE ENVIRONMENT

%-----------------------------------------------------------------------
% These commands are required in for the tables to work properly
\renewcommand{\E}[1]{&{\ensuremath{(#1)}}}
\newcommand{\EE}{&}
\renewcommand{\I}[2]{{\ensuremath{^{#1}}}&{\ensuremath{\mathrm{#2}}}}
\newcommand{\NoData}{\multicolumn{2}{c}{\nodata}}
\newcommand{\PPI}{\I{\phantom{99}}{\phantom{Mm}}}
\newcommand{\PPE}{\phantom{$9.99$}&\phantom{$(-99)$}}
%-----------------------------------------------------------------------

%%%%%%%%%%%%%%%%%%%%%%%%%%%%%%%%%%%%%%%%%%%%%%%%%%%%%%%%%%%%%%%%%%%%%%%%
% Parameters

\begin{table}
\setlength{\tabcolsep}{1ex}
%\centering
\caption{\label{tab:struc_models}\scshape Properties of stellar models at the onset of core collapse and explosion parameters\lTab{p}\vspace{1ex}}
\begin{tabular}{rrrrrrr}
\hline
\hline
\multicolumn{1}{c}{model(s)} &
\multicolumn{1}{c}{   S15} &
\multicolumn{1}{c}{   S19} &
\multicolumn{1}{c}{   S20} &
\multicolumn{1}{c}{   S21} &
\multicolumn{1}{c}{   S25} &
\multicolumn{1}{c}{  S25P} \\
\multicolumn{1}{c}{} &
\multicolumn{1}{c}{   N15} &
\multicolumn{1}{c}{      } &
\multicolumn{1}{c}{   N20} &
\multicolumn{1}{c}{      } &
\multicolumn{1}{c}{   N25} &
\multicolumn{1}{c}{      } \\
\multicolumn{1}{c}{} &
\multicolumn{1}{c}{      } &
\multicolumn{1}{c}{      } &
\multicolumn{1}{c}{      } &
\multicolumn{1}{c}{      } &
\multicolumn{1}{c}{   H25} &
\multicolumn{1}{c}{      } \\
\hline
initial mass (\Msun)         &15.081 &19.103 &20.109 &21.114 &\multicolumn{2}{c}{25.136} \\
wind mass loss (\Msun)       & 2.469 & 4.268 & 5.369 & 6.403 &\multicolumn{2}{c}{12.057} \\
final mass (\Msun)           &12.612 &14.835 &14.740 &14.711 &\multicolumn{2}{c}{13.079} \\
helium core (\Msun)          & 4.163 & 5.646 & 6.131 & 6.540 &\multicolumn{2}{c}{\phantom{0}8.317} \\
C/O core (\Msun)             & 2.819 & 4.103 & 4.508 & 4.849 &\multicolumn{2}{c}{\phantom{0}6.498} \\
Ne/Mg/O core (\Msun)         & 1.858 & 2.172 & 1.695 & 2.021 &\multicolumn{2}{c}{\phantom{0}2.443} \\
Si core (\Msun)              & 1.808 & 1.699 & 1.601 & 1.739 &\multicolumn{2}{c}{\phantom{0}2.121} \\
``iron'' core (\Msun)        & 1.452 & 1.458 & 1.461 & 1.548 &\multicolumn{2}{c}{\phantom{0}1.619} \\
\Mpist / ``\Ye'' core (\Msun)& 1.315 & 1.458 & 1.461 & 1.357 &\multicolumn{2}{c}{\phantom{0}1.619} \\
\gPist (G\Mpist/r)           & 0.460 & 0.710 & 2.320 & 0.770 & 0.670 & 0.930 \\
\Eexp (\foe)                 & 1.205 & 1.204 & 2.203 & 1.765 & 1.735 & 2.293 \\
fallback (\Msun)             & 0.368 & 0.221 & 0.087 & 0.403 & 0.422 & 0.342 \\
remnant (baryonic, \Msun)    & 1.683 & 1.679 & 1.548 & 1.760 & 2.041 & 1.961 \\
\hline			     
                             &       &       &       &       &\phantom{00.000}&\phantom{00.000}\\
\end{tabular}
\vspace{-1.5\baselineskip}
\end{table}

}% END OF TABLE ENVIRONMENT

\clearpage

{% BEGINNING OF TABLE ENVIRONMENT

%-----------------------------------------------------------------------
% These commands are required in for the tables to work properly
\renewcommand{\E}[1]{&{\ensuremath{(#1)}}}
\newcommand{\EE}{&}
\renewcommand{\I}[2]{{\ensuremath{^{#1}}}&{\ensuremath{\mathrm{#2}}}}
\newcommand{\NoData}{\multicolumn{2}{c}{\nodata}}
\newcommand{\PPI}{\I{\phantom{999}}{\phantom{Mg}}}
\newcommand{\PPE}{\phantom{$9.999$}&\phantom{$(-99)$}}
%-----------------------------------------------------------------------

%%%%%%%%%%%%%%%%%%%%%%%%%%%%%%%%%%%%%%%%%%%%%%%%%%%%%%%%%%%%%%%%%%%%%%%%
% Solar abundances
\newcommand{\TableName}{Initial Mass Fractions}

\begin{table}
\setlength{\tabcolsep}{1ex}
\centering
\caption{\scshape \TableName\lTab{s}}
\scalebox{1.00}{
\begin{tabular}{r@{}lr@{}lr@{}lr@{}lr@{}lr@{}lr@{}lr@{}lr@{}lr@{}l}
\hline
\hline
\multicolumn{2}{c}{ion} &
\multicolumn{2}{c}{X} &
\multicolumn{2}{c}{ion} &
\multicolumn{2}{c}{X} &
\multicolumn{2}{c}{ion} &
\multicolumn{2}{c}{X} &
\multicolumn{2}{c}{ion} &
\multicolumn{2}{c}{X} &
\multicolumn{2}{c}{ion} &
\multicolumn{2}{c}{X} \\
\hline
\I{  1}{ H} &    7.057\E{- 1} & \I{ 48}{Ca} &    1.384\E{- 7} & \I{ 81}{Br} &    1.190\E{- 8} & \I{111}{Cd} &    5.780\E{-10} & \I{139}{La} &    1.570\E{- 9} \\
\I{  2}{ H} &    4.801\E{- 5} & \I{ 45}{Sc} &    3.893\E{- 8} & \I{ 78}{Kr} &    3.020\E{-10} & \I{112}{Cd} &    1.100\E{- 9} & \I{136}{Ce} &    7.430\E{-12} \\
\I{  3}{He} &    2.929\E{- 5} & \I{ 46}{Ti} &    2.234\E{- 7} & \I{ 80}{Kr} &    2.020\E{- 9} & \I{113}{Cd} &    5.630\E{-10} & \I{138}{Ce} &    9.880\E{-12} \\
\I{  4}{He} &    2.752\E{- 1} & \I{ 47}{Ti} &    2.081\E{- 7} & \I{ 82}{Kr} &    1.070\E{- 8} & \I{114}{Cd} &    1.340\E{- 9} & \I{140}{Ce} &    3.580\E{- 9} \\
\I{  6}{Li} &    6.496\E{-10} & \I{ 48}{Ti} &    2.149\E{- 6} & \I{ 83}{Kr} &    1.080\E{- 8} & \I{116}{Cd} &    3.550\E{-10} & \I{142}{Ce} &    4.530\E{-10} \\
\I{  7}{Li} &    9.349\E{- 9} & \I{ 49}{Ti} &    1.636\E{- 7} & \I{ 84}{Kr} &    5.460\E{- 8} & \I{113}{In} &    2.260\E{-11} & \I{141}{Pr} &    5.960\E{-10} \\
\I{  9}{Be} &    1.662\E{-10} & \I{ 50}{Ti} &    1.644\E{- 7} & \I{ 86}{Kr} &    1.710\E{- 8} & \I{115}{In} &    5.120\E{-10} & \I{142}{Nd} &    8.080\E{-10} \\
\I{ 10}{ B} &    1.067\E{- 9} & \I{ 50}{ V} &    9.258\E{-10} & \I{ 85}{Rb} &    1.100\E{- 8} & \I{112}{Sn} &    1.050\E{-10} & \I{143}{Nd} &    3.620\E{-10} \\
\I{ 11}{ B} &    4.730\E{- 9} & \I{ 51}{ V} &    3.767\E{- 7} & \I{ 87}{Rb} &    4.640\E{- 9} & \I{114}{Sn} &    7.180\E{-11} & \I{144}{Nd} &    7.180\E{-10} \\
\I{ 12}{ C} &    3.032\E{- 3} & \I{ 50}{Cr} &    7.424\E{- 7} & \I{ 84}{Sr} &    2.800\E{-10} & \I{115}{Sn} &    3.750\E{-11} & \I{145}{Nd} &    2.520\E{-10} \\
\I{ 13}{ C} &    3.650\E{- 5} & \I{ 52}{Cr} &    1.486\E{- 5} & \I{ 86}{Sr} &    5.050\E{- 9} & \I{116}{Sn} &    1.630\E{- 9} & \I{146}{Nd} &    5.240\E{-10} \\
\I{ 14}{ N} &    1.105\E{- 3} & \I{ 53}{Cr} &    1.716\E{- 6} & \I{ 87}{Sr} &    3.320\E{- 9} & \I{117}{Sn} &    8.670\E{-10} & \I{148}{Nd} &    1.790\E{-10} \\
\I{ 15}{ N} &    4.363\E{- 6} & \I{ 54}{Cr} &    4.357\E{- 7} & \I{ 88}{Sr} &    4.320\E{- 8} & \I{118}{Sn} &    2.760\E{- 9} & \I{150}{Nd} &    1.770\E{-10} \\
\I{ 16}{ O} &    9.592\E{- 3} & \I{ 55}{Mn} &    1.329\E{- 5} & \I{ 89}{ Y} &    1.040\E{- 8} & \I{119}{Sn} &    9.870\E{-10} & \I{144}{Sm} &    2.910\E{-11} \\
\I{ 17}{ O} &    3.887\E{- 6} & \I{ 54}{Fe} &    7.130\E{- 5} & \I{ 90}{Zr} &    1.340\E{- 8} & \I{120}{Sn} &    3.790\E{- 9} & \I{147}{Sm} &    1.480\E{-10} \\
\I{ 18}{ O} &    2.167\E{- 5} & \I{ 56}{Fe} &    1.169\E{- 3} & \I{ 91}{Zr} &    2.950\E{- 9} & \I{122}{Sn} &    5.460\E{-10} & \I{148}{Sm} &    1.090\E{-10} \\
\I{ 19}{ F} &    4.051\E{- 7} & \I{ 57}{Fe} &    2.855\E{- 5} & \I{ 92}{Zr} &    4.560\E{- 9} & \I{124}{Sn} &    6.930\E{-10} & \I{149}{Sm} &    1.340\E{-10} \\
\I{ 20}{Ne} &    1.619\E{- 3} & \I{ 58}{Fe} &    3.697\E{- 6} & \I{ 94}{Zr} &    4.710\E{- 9} & \I{121}{Sb} &    5.420\E{-10} & \I{150}{Sm} &    7.250\E{-11} \\
\I{ 21}{Ne} &    4.127\E{- 6} & \I{ 59}{Co} &    3.358\E{- 6} & \I{ 96}{Zr} &    7.770\E{-10} & \I{123}{Sb} &    4.110\E{-10} & \I{152}{Sm} &    2.650\E{-10} \\
\I{ 22}{Ne} &    1.302\E{- 4} & \I{ 58}{Ni} &    4.944\E{- 5} & \I{ 93}{Nb} &    1.640\E{- 9} & \I{120}{Te} &    1.310\E{-11} & \I{154}{Sm} &    2.280\E{-10} \\
\I{ 23}{Na} &    3.339\E{- 5} & \I{ 60}{Ni} &    1.958\E{- 5} & \I{ 92}{Mo} &    8.800\E{-10} & \I{122}{Te} &    3.830\E{-10} & \I{151}{Eu} &    1.780\E{-10} \\
\I{ 24}{Mg} &    5.148\E{- 4} & \I{ 61}{Ni} &    8.594\E{- 7} & \I{ 94}{Mo} &    5.610\E{-10} & \I{123}{Te} &    1.330\E{-10} & \I{153}{Eu} &    1.970\E{-10} \\
\I{ 25}{Mg} &    6.766\E{- 5} & \I{ 62}{Ni} &    2.776\E{- 6} & \I{ 95}{Mo} &    9.760\E{-10} & \I{124}{Te} &    7.180\E{-10} & \I{152}{Gd} &    2.540\E{-12} \\
\I{ 26}{Mg} &    7.760\E{- 5} & \I{ 64}{Ni} &    7.269\E{- 7} & \I{ 96}{Mo} &    1.030\E{- 9} & \I{125}{Te} &    1.080\E{- 9} & \I{154}{Gd} &    2.800\E{-11} \\
\I{ 27}{Al} &    5.805\E{- 5} & \I{ 63}{Cu} &    5.753\E{- 7} & \I{ 97}{Mo} &    5.990\E{-10} & \I{126}{Te} &    2.900\E{- 9} & \I{155}{Gd} &    1.910\E{-10} \\
\I{ 28}{Si} &    6.530\E{- 4} & \I{ 65}{Cu} &    2.647\E{- 7} & \I{ 98}{Mo} &    1.520\E{- 9} & \I{128}{Te} &    4.950\E{- 9} & \I{156}{Gd} &    2.670\E{-10} \\
\I{ 29}{Si} &    3.426\E{- 5} & \I{ 64}{Zn} &    9.924\E{- 7} & \I{100}{Mo} &    6.220\E{-10} & \I{130}{Te} &    5.360\E{- 9} & \I{157}{Gd} &    2.050\E{-10} \\
\I{ 30}{Si} &    2.352\E{- 5} & \I{ 66}{Zn} &    5.877\E{- 7} & \I{ 96}{Ru} &    2.500\E{-10} & \I{127}{ I} &    2.890\E{- 9} & \I{158}{Gd} &    3.280\E{-10} \\
\I{ 31}{ P} &    8.155\E{- 6} & \I{ 67}{Zn} &    8.762\E{- 8} & \I{ 98}{Ru} &    8.680\E{-11} & \I{124}{Xe} &    1.790\E{-11} & \I{160}{Gd} &    2.920\E{-10} \\
\I{ 32}{ S} &    3.958\E{- 4} & \I{ 68}{Zn} &    4.059\E{- 7} & \I{ 99}{Ru} &    5.910\E{-10} & \I{126}{Xe} &    1.620\E{-11} & \I{159}{Tb} &    2.430\E{-10} \\
\I{ 33}{ S} &    3.222\E{- 6} & \I{ 70}{Zn} &    1.339\E{- 8} & \I{100}{Ru} &    5.920\E{-10} & \I{128}{Xe} &    3.330\E{-10} & \I{156}{Dy} &    8.720\E{-13} \\
\I{ 34}{ S} &    1.866\E{- 5} & \I{ 69}{Ga} &    3.962\E{- 8} & \I{101}{Ru} &    8.070\E{-10} & \I{129}{Xe} &    4.180\E{- 9} & \I{158}{Dy} &    1.510\E{-12} \\
\I{ 36}{ S} &    9.379\E{- 8} & \I{ 71}{Ga} &    2.630\E{- 8} & \I{102}{Ru} &    1.520\E{- 9} & \I{130}{Xe} &    6.740\E{-10} & \I{160}{Dy} &    3.730\E{-11} \\
\I{ 35}{Cl} &    2.532\E{- 6} & \I{ 70}{Ge} &    4.320\E{- 8} & \I{104}{Ru} &    9.150\E{-10} & \I{131}{Xe} &    3.380\E{- 9} & \I{161}{Dy} &    3.030\E{-10} \\
\I{ 37}{Cl} &    8.545\E{- 7} & \I{ 72}{Ge} &    5.940\E{- 8} & \I{103}{Rh} &    8.960\E{-10} & \I{132}{Xe} &    4.140\E{- 9} & \I{162}{Dy} &    4.140\E{-10} \\
\I{ 36}{Ar} &    7.740\E{- 5} & \I{ 73}{Ge} &    1.710\E{- 8} & \I{102}{Pd} &    3.660\E{-11} & \I{134}{Xe} &    1.560\E{- 9} & \I{163}{Dy} &    4.050\E{-10} \\
\I{ 38}{Ar} &    1.538\E{- 5} & \I{ 74}{Ge} &    8.120\E{- 8} & \I{104}{Pd} &    4.080\E{-10} & \I{136}{Xe} &    1.280\E{- 9} & \I{164}{Dy} &    4.600\E{-10} \\
\I{ 40}{Ar} &    2.529\E{- 8} & \I{ 76}{Ge} &    1.780\E{- 8} & \I{105}{Pd} &    8.230\E{-10} & \I{133}{Cs} &    1.250\E{- 9} & \I{165}{Ho} &    3.710\E{-10} \\
\I{ 39}{ K} &    3.472\E{- 6} & \I{ 75}{As} &    1.240\E{- 8} & \I{106}{Pd} &    1.020\E{- 9} & \I{130}{Ba} &    1.570\E{-11} & \I{162}{Er} &    1.440\E{-12} \\
\I{ 40}{ K} &    5.545\E{- 9} & \I{ 74}{Se} &    1.030\E{- 9} & \I{108}{Pd} &    1.010\E{- 9} & \I{132}{Ba} &    1.510\E{-11} & \I{164}{Er} &    1.680\E{-11} \\
\I{ 41}{ K} &    2.634\E{- 7} & \I{ 76}{Se} &    1.080\E{- 8} & \I{110}{Pd} &    4.540\E{-10} & \I{134}{Ba} &    3.690\E{-10} & \I{166}{Er} &    3.540\E{-10} \\
\I{ 40}{Ca} &    5.990\E{- 5} & \I{ 77}{Se} &    9.150\E{- 9} & \I{107}{Ag} &    6.820\E{-10} & \I{135}{Ba} &    1.010\E{- 9} & \I{167}{Er} &    2.430\E{-10} \\
\I{ 42}{Ca} &    4.196\E{- 7} & \I{ 78}{Se} &    2.900\E{- 8} & \I{109}{Ag} &    6.450\E{-10} & \I{136}{Ba} &    1.210\E{- 9} & \I{168}{Er} &    2.860\E{-10} \\
\I{ 43}{Ca} &    8.973\E{- 8} & \I{ 80}{Se} &    6.250\E{- 8} & \I{106}{Cd} &    5.390\E{-11} & \I{137}{Ba} &    1.750\E{- 9} & \I{170}{Er} &    1.610\E{-10} \\
\I{ 44}{Ca} &    1.425\E{- 6} & \I{ 82}{Se} &    1.180\E{- 8} & \I{108}{Cd} &    3.910\E{-11} & \I{138}{Ba} &    1.120\E{- 8} & \I{169}{Tm} &    1.620\E{-10} \\
\I{ 46}{Ca} &    2.793\E{- 9} & \I{ 79}{Br} &    1.190\E{- 8} & \I{110}{Cd} &    5.590\E{-10} & \I{138}{La} &    1.430\E{-12} & \I{168}{Yb} &    1.370\E{-12} \\
\hline
\PPI&\PPE&\PPI&\PPE&\PPI&\PPE&\PPI&\PPE&\PPI&\PPE \\
\end{tabular}
}
\vspace{-1.5\baselineskip}
\begin{flushright}\textsc{(continued on next page)}\end{flushright}
\end{table}

\clearpage

\addtocounter{table}{-1}

\begin{table}
\setlength{\tabcolsep}{1ex}
\centering
\caption{\scshape (continued) \TableName}
\scalebox{1.00}{
\begin{tabular}{r@{}lr@{}lr@{}lr@{}lr@{}lr@{}lr@{}lr@{}lr@{}lr@{}l}
\hline
\hline
\multicolumn{2}{c}{ion} &
\multicolumn{2}{c}{X} &
\multicolumn{2}{c}{ion} &
\multicolumn{2}{c}{X} &
\multicolumn{2}{c}{ion} &
\multicolumn{2}{c}{X} &
\multicolumn{2}{c}{ion} &
\multicolumn{2}{c}{X} &
\multicolumn{2}{c}{ion} &
\multicolumn{2}{c}{X} \\
\hline
\I{170}{Yb} &    3.250\E{-11} & \I{178}{Hf} &    1.890\E{-10} & \I{187}{Re} &    1.660\E{-10} & \I{192}{Pt} &    5.100\E{-11} & \I{201}{Hg} &    2.280\E{-10} \\
\I{171}{Yb} &    1.530\E{-10} & \I{179}{Hf} &    9.510\E{-11} & \I{184}{Os} &    5.680\E{-13} & \I{194}{Pt} &    2.160\E{- 9} & \I{202}{Hg} &    5.160\E{-10} \\
\I{172}{Yb} &    2.630\E{-10} & \I{180}{Hf} &    2.460\E{-10} & \I{186}{Os} &    5.030\E{-11} & \I{195}{Pt} &    2.230\E{- 9} & \I{204}{Hg} &    1.200\E{-10} \\
\I{173}{Yb} &    1.750\E{-10} & \I{180}{Ta} &    1.130\E{-14} & \I{187}{Os} &    3.820\E{-11} & \I{196}{Pt} &    1.680\E{- 9} & \I{203}{Tl} &    2.790\E{-10} \\
\I{174}{Yb} &    3.470\E{-10} & \I{181}{Ta} &    9.480\E{-11} & \I{188}{Os} &    4.270\E{-10} & \I{198}{Pt} &    4.820\E{-10} & \I{205}{Tl} &    6.740\E{-10} \\
\I{176}{Yb} &    1.400\E{-10} & \I{180}{ W} &    7.880\E{-13} & \I{189}{Os} &    5.210\E{-10} & \I{197}{Au} &    9.320\E{-10} & \I{204}{Pb} &    3.150\E{-10} \\
\I{175}{Lu} &    1.580\E{-10} & \I{182}{ W} &    1.610\E{-10} & \I{190}{Os} &    8.550\E{-10} & \I{196}{Hg} &    2.380\E{-12} & \I{206}{Pb} &    3.090\E{- 9} \\
\I{176}{Lu} &    4.630\E{-12} & \I{183}{ W} &    8.800\E{-11} & \I{192}{Os} &    1.350\E{- 9} & \I{198}{Hg} &    1.710\E{-10} & \I{207}{Pb} &    3.370\E{- 9} \\
\I{174}{Hf} &    1.100\E{-12} & \I{184}{ W} &    1.900\E{-10} & \I{191}{Ir} &    1.190\E{- 9} & \I{199}{Hg} &    2.880\E{-10} & \I{208}{Pb} &    9.630\E{- 9} \\
\I{176}{Hf} &    3.530\E{-11} & \I{186}{ W} &    1.790\E{-10} & \I{193}{Ir} &    2.020\E{- 9} & \I{200}{Hg} &    3.980\E{-10} & \I{209}{Bi} &    7.610\E{-10} \\
\I{177}{Hf} &    1.280\E{-10} & \I{185}{Re} &    9.030\E{-11} & \I{190}{Pt} &    8.170\E{-13} \\
\hline
\PPI&\PPE&\PPI&\PPE&\PPI&\PPE&\PPI&\PPE&\PPI&\PPE \\
\end{tabular}
}
\vspace{-1.5\baselineskip}
\begin{flushright}\textsc{(End of \TableName\ Table)}\end{flushright}
\end{table}

}% END OF TABLE ENVIRONMENT

\clearpage

{% BEGINNING OF TABLE ENVIRONMENT

%-----------------------------------------------------------------------
% These commands are required in for the tables to work properly
\renewcommand{\E}[1]{&{\ensuremath{(#1)}}}
\newcommand{\EE}{&}
\renewcommand{\I}[2]{{\ensuremath{^{#1}}}&{\ensuremath{\mathrm{#2}}}}
\newcommand{\NoData}{\multicolumn{2}{c}{\nodata}}
\newcommand{\PPI}{\I{\phantom{99}}{\phantom{Mm}}}
\newcommand{\PPE}{\phantom{$9.99$}&\phantom{$(-99)$}}
%-----------------------------------------------------------------------

%%%%%%%%%%%%%%%%%%%%%%%%%%%%%%%%%%%%%%%%%%%%%%%%%%%%%%%%%%%%%%%%%%%%%%%%
% YIELDS
\newcommand{\Scale}{     0.760000}

\begin{table}
\setlength{\tabcolsep}{1ex}
\centering
\caption{Yields (in solar masses)\lTab{y}}
\scalebox{\Scale}{
\begin{tabular}{r@{}lr@{}lr@{}lr@{}lr@{}lr@{}lr@{}lr@{}lr@{}lr@{}lr@{}l}
\hline
\hline
\multicolumn{2}{c}{ion} &
\multicolumn{2}{c}{   S15} &
\multicolumn{2}{c}{   S19} &
\multicolumn{2}{c}{   S20} &
\multicolumn{2}{c}{   S21} &
\multicolumn{2}{c}{   S25} &
\multicolumn{2}{c}{  S25P} &
\multicolumn{2}{c}{   N15} &
\multicolumn{2}{c}{   N20} &
\multicolumn{2}{c}{   N25} &
\multicolumn{2}{c}{   H25} \\
\hline
\I{  1}{ H} &    7.136\EE     &    8.446\EE     &    8.700\EE     &    8.994\EE     &    1.012\E{  1} &    1.012\E{  1} &    7.146\EE     &    8.717\EE     &    1.014\E{  1} &    1.012\E{  1} \\
\I{  2}{ H} &    2.735\E{- 7} &    2.021\E{- 7} &    2.094\E{- 7} &    2.156\E{- 7} &    2.429\E{- 7} &    2.427\E{- 7} &    2.927\E{- 7} &    2.321\E{- 7} &    2.604\E{- 7} &    2.425\E{- 7} \\
\I{  3}{He} &    5.695\E{- 4} &    6.612\E{- 4} &    6.836\E{- 4} &    7.065\E{- 4} &    7.994\E{- 4} &    7.994\E{- 4} &    5.716\E{- 4} &    6.866\E{- 4} &    8.029\E{- 4} &    7.994\E{- 4} \\
\I{  4}{He} &    4.684\EE     &    5.939\EE     &    6.229\EE     &    6.514\EE     &    7.597\EE     &    7.607\EE     &    4.673\EE     &    6.209\EE     &    7.574\EE     &    7.596\EE     \\
\I{  6}{Li} &    5.476\E{-11} &    6.837\E{-11} &    7.161\E{-11} &    7.457\E{-11} &    8.782\E{-11} &    8.781\E{-11} &    5.535\E{-11} &    7.238\E{-11} &    8.877\E{-11} &    8.782\E{-11} \\
\I{  7}{Li} &    1.820\E{- 7} &    1.824\E{- 7} &    1.588\E{- 7} &    1.315\E{- 7} &    1.327\E{- 7} &    1.142\E{- 7} &    1.814\E{- 7} &    1.598\E{- 7} &    1.335\E{- 7} &    1.321\E{- 7} \\
\I{  9}{Be} &    5.457\E{-11} &    6.768\E{-11} &    6.873\E{-11} &    6.857\E{-11} &    8.024\E{-11} &    8.022\E{-11} &    5.461\E{-11} &    6.891\E{-11} &    8.071\E{-11} &    8.024\E{-11} \\
\I{ 10}{ B} &    2.003\E{- 9} &    2.080\E{- 9} &    1.974\E{- 9} &    2.284\E{- 9} &    2.337\E{- 9} &    2.319\E{- 9} &    1.385\E{- 9} &    1.072\E{- 9} &    1.223\E{- 9} &    2.307\E{- 9} \\
\I{ 11}{ B} &    6.905\E{- 7} &    9.790\E{- 7} &    4.046\E{- 7} &    5.633\E{- 7} &    1.542\E{- 6} &    1.477\E{- 6} &    7.387\E{- 7} &    4.477\E{- 7} &    1.654\E{- 6} &    1.010\E{- 6} \\
\I{ 12}{ C} &    1.555\E{- 1} &    2.520\E{- 1} &    2.233\E{- 1} &    2.764\E{- 1} &    4.093\E{- 1} &    4.080\E{- 1} &    1.554\E{- 1} &    2.231\E{- 1} &    4.089\E{- 1} &    4.098\E{- 1} \\
\I{ 13}{ C} &    1.264\E{- 3} &    1.408\E{- 3} &    1.412\E{- 3} &    1.417\E{- 3} &    1.570\E{- 3} &    1.570\E{- 3} &    1.165\E{- 3} &    1.311\E{- 3} &    1.452\E{- 3} &    1.570\E{- 3} \\
\I{ 14}{ N} &    4.662\E{- 2} &    6.128\E{- 2} &    6.440\E{- 2} &    6.857\E{- 2} &    8.101\E{- 2} &    8.100\E{- 2} &    4.675\E{- 2} &    6.413\E{- 2} &    8.063\E{- 2} &    8.098\E{- 2} \\
\I{ 15}{ N} &    1.775\E{- 4} &    7.392\E{- 5} &    5.191\E{- 5} &    9.192\E{- 5} &    1.391\E{- 4} &    1.409\E{- 4} &    1.091\E{- 4} &    5.329\E{- 5} &    1.347\E{- 4} &    1.373\E{- 4} \\
\I{ 16}{ O} &    8.495\E{- 1} &    1.760\EE     &    2.205\EE     &    2.395\EE     &    3.316\EE     &    3.292\EE     &    8.472\E{- 1} &    2.210\EE     &    3.312\EE     &    3.336\EE     \\
\I{ 17}{ O} &    9.941\E{- 5} &    9.941\E{- 5} &    9.820\E{- 5} &    1.361\E{- 4} &    1.262\E{- 4} &    1.243\E{- 4} &    9.784\E{- 5} &    9.417\E{- 5} &    1.206\E{- 4} &    8.338\E{- 5} \\
\I{ 18}{ O} &    3.304\E{- 3} &    4.025\E{- 3} &    3.122\E{- 3} &    2.560\E{- 3} &    1.205\E{- 3} &    1.118\E{- 3} &    3.092\E{- 3} &    2.963\E{- 3} &    1.068\E{- 3} &    1.204\E{- 3} \\
\I{ 19}{ F} &    2.989\E{- 5} &    5.886\E{- 5} &    1.081\E{- 5} &    6.950\E{- 5} &    7.820\E{- 5} &    7.012\E{- 5} &    7.363\E{- 5} &    7.820\E{- 6} &    9.949\E{- 5} &    7.517\E{- 5} \\
\I{ 20}{Ne} &    1.267\E{- 1} &    3.724\E{- 1} &    6.971\E{- 2} &    4.692\E{- 1} &    5.356\E{- 1} &    5.231\E{- 1} &    1.336\E{- 1} &    7.272\E{- 2} &    5.617\E{- 1} &    5.243\E{- 1} \\
\I{ 21}{Ne} &    8.376\E{- 4} &    1.170\E{- 3} &    2.965\E{- 4} &    1.429\E{- 3} &    1.517\E{- 3} &    1.617\E{- 3} &    8.364\E{- 4} &    3.401\E{- 4} &    1.629\E{- 3} &    9.987\E{- 4} \\
\I{ 22}{Ne} &    1.089\E{- 2} &    1.818\E{- 2} &    2.028\E{- 2} &    2.509\E{- 2} &    2.796\E{- 2} &    2.746\E{- 2} &    1.021\E{- 2} &    1.921\E{- 2} &    2.530\E{- 2} &    2.752\E{- 2} \\
\I{ 23}{Na} &    2.625\E{- 3} &    1.387\E{- 2} &    2.193\E{- 3} &    1.765\E{- 2} &    1.281\E{- 2} &    1.216\E{- 2} &    2.535\E{- 3} &    2.243\E{- 3} &    1.298\E{- 2} &    1.253\E{- 2} \\
\I{ 24}{Mg} &    3.999\E{- 2} &    6.070\E{- 2} &    7.260\E{- 2} &    1.018\E{- 1} &    1.444\E{- 1} &    1.437\E{- 1} &    3.495\E{- 2} &    6.173\E{- 2} &    1.192\E{- 1} &    1.454\E{- 1} \\
\I{ 25}{Mg} &    8.809\E{- 3} &    1.997\E{- 2} &    5.627\E{- 3} &    2.217\E{- 2} &    3.146\E{- 2} &    3.090\E{- 2} &    7.827\E{- 3} &    5.064\E{- 3} &    2.568\E{- 2} &    3.891\E{- 2} \\
\I{ 26}{Mg} &    8.469\E{- 3} &    2.083\E{- 2} &    8.622\E{- 3} &    2.323\E{- 2} &    3.901\E{- 2} &    3.815\E{- 2} &    8.693\E{- 3} &    9.542\E{- 3} &    3.977\E{- 2} &    3.134\E{- 2} \\
\I{ 27}{Al} &    4.682\E{- 3} &    1.016\E{- 2} &    1.205\E{- 2} &    1.685\E{- 2} &    2.206\E{- 2} &    2.172\E{- 2} &    5.679\E{- 3} &    1.178\E{- 2} &    2.877\E{- 2} &    1.713\E{- 2} \\
\I{ 28}{Si} &    9.684\E{- 2} &    1.585\E{- 1} &    4.416\E{- 1} &    1.613\E{- 1} &    3.540\E{- 1} &    3.589\E{- 1} &    1.005\E{- 1} &    4.277\E{- 1} &    3.557\E{- 1} &    3.503\E{- 1} \\
\I{ 29}{Si} &    3.166\E{- 3} &    3.062\E{- 3} &    1.574\E{- 2} &    6.624\E{- 3} &    1.042\E{- 2} &    1.072\E{- 2} &    3.207\E{- 3} &    1.358\E{- 2} &    1.074\E{- 2} &    7.823\E{- 3} \\
\I{ 30}{Si} &    4.504\E{- 3} &    2.496\E{- 3} &    1.672\E{- 2} &    9.206\E{- 3} &    1.020\E{- 2} &    1.139\E{- 2} &    4.588\E{- 3} &    1.673\E{- 2} &    1.128\E{- 2} &    8.130\E{- 3} \\
\I{ 31}{ P} &    1.096\E{- 3} &    1.368\E{- 3} &    9.892\E{- 3} &    2.458\E{- 3} &    3.993\E{- 3} &    4.084\E{- 3} &    1.135\E{- 3} &    1.007\E{- 2} &    4.177\E{- 3} &    2.741\E{- 3} \\
\I{ 32}{ S} &    4.165\E{- 2} &    8.395\E{- 2} &    1.922\E{- 1} &    6.442\E{- 2} &    1.475\E{- 1} &    1.493\E{- 1} &    4.156\E{- 2} &    1.983\E{- 1} &    1.496\E{- 1} &    1.555\E{- 1} \\
\I{ 33}{ S} &    2.084\E{- 4} &    3.833\E{- 4} &    2.453\E{- 3} &    3.265\E{- 4} &    9.121\E{- 4} &    8.445\E{- 4} &    2.229\E{- 4} &    2.601\E{- 3} &    9.721\E{- 4} &    9.136\E{- 4} \\
\I{ 34}{ S} &    2.211\E{- 3} &    3.461\E{- 3} &    1.578\E{- 2} &    3.589\E{- 3} &    9.039\E{- 3} &    9.183\E{- 3} &    2.289\E{- 3} &    1.533\E{- 2} &    9.008\E{- 3} &    7.623\E{- 3} \\
\I{ 36}{ S} &    4.898\E{- 6} &    9.400\E{- 6} &    6.185\E{- 5} &    1.149\E{- 5} &    1.740\E{- 5} &    1.759\E{- 5} &    4.596\E{- 6} &    4.898\E{- 5} &    1.559\E{- 5} &    4.082\E{- 5} \\
\I{ 35}{Cl} &    1.509\E{- 4} &    4.174\E{- 4} &    8.342\E{- 3} &    2.461\E{- 4} &    6.545\E{- 4} &    6.091\E{- 4} &    1.671\E{- 4} &    8.852\E{- 3} &    7.237\E{- 4} &    5.982\E{- 4} \\
\I{ 37}{Cl} &    6.046\E{- 5} &    1.869\E{- 4} &    7.696\E{- 4} &    1.622\E{- 4} &    2.830\E{- 4} &    2.689\E{- 4} &    5.920\E{- 5} &    8.358\E{- 4} &    2.882\E{- 4} &    2.740\E{- 4} \\
\I{ 36}{Ar} &    7.403\E{- 3} &    1.640\E{- 2} &    4.493\E{- 2} &    1.115\E{- 2} &    2.315\E{- 2} &    2.357\E{- 2} &    7.095\E{- 3} &    5.289\E{- 2} &    2.348\E{- 2} &    2.508\E{- 2} \\
\I{ 38}{Ar} &    1.016\E{- 3} &    5.471\E{- 3} &    1.649\E{- 2} &    1.621\E{- 3} &    7.505\E{- 3} &    6.850\E{- 3} &    1.025\E{- 3} &    1.684\E{- 2} &    7.922\E{- 3} &    5.620\E{- 3} \\
\I{ 40}{Ar} &    3.400\E{- 6} &    8.330\E{- 6} &    2.161\E{- 5} &    8.264\E{- 6} &    1.569\E{- 5} &    1.567\E{- 5} &    2.863\E{- 6} &    1.762\E{- 5} &    1.186\E{- 5} &    1.760\E{- 5} \\
\I{ 39}{ K} &    1.256\E{- 4} &    5.868\E{- 4} &    1.039\E{- 2} &    2.012\E{- 4} &    5.551\E{- 4} &    4.972\E{- 4} &    1.234\E{- 4} &    1.196\E{- 2} &    5.988\E{- 4} &    4.669\E{- 4} \\
\I{ 40}{ K} &    7.592\E{- 7} &    1.970\E{- 6} &    2.503\E{- 4} &    2.745\E{- 6} &    5.100\E{- 6} &    4.219\E{- 6} &    7.122\E{- 7} &    2.496\E{- 4} &    4.891\E{- 6} &    3.388\E{- 6} \\
\I{ 41}{ K} &    8.807\E{- 6} &    3.392\E{- 5} &    4.573\E{- 4} &    1.574\E{- 5} &    4.374\E{- 5} &    3.788\E{- 5} &    8.566\E{- 6} &    5.850\E{- 4} &    5.031\E{- 5} &    3.817\E{- 5} \\
\I{ 40}{Ca} &    6.284\E{- 3} &    1.278\E{- 2} &    2.391\E{- 2} &    9.273\E{- 3} &    1.716\E{- 2} &    1.814\E{- 2} &    5.915\E{- 3} &    2.951\E{- 2} &    1.737\E{- 2} &    1.843\E{- 2} \\
\I{ 42}{Ca} &    3.087\E{- 5} &    2.469\E{- 4} &    1.195\E{- 3} &    5.149\E{- 5} &    2.225\E{- 4} &    1.954\E{- 4} &    3.028\E{- 5} &    1.462\E{- 3} &    2.516\E{- 4} &    1.532\E{- 4} \\
\I{ 43}{Ca} &    2.688\E{- 6} &    5.698\E{- 6} &    1.390\E{- 4} &    5.314\E{- 6} &    8.935\E{- 6} &    8.569\E{- 6} &    2.543\E{- 6} &    1.485\E{- 4} &    8.999\E{- 6} &    5.924\E{- 6} \\
\I{ 44}{Ca} &    3.421\E{- 5} &    5.539\E{- 5} &    1.802\E{- 4} &    5.140\E{- 5} &    6.164\E{- 5} &    9.301\E{- 5} &    2.986\E{- 5} &    1.811\E{- 4} &    5.607\E{- 5} &    6.152\E{- 5} \\
\I{ 46}{Ca} &    5.837\E{- 7} &    1.658\E{- 6} &    1.292\E{- 6} &    6.488\E{- 7} &    1.979\E{- 6} &    2.079\E{- 6} &    5.201\E{- 7} &    1.063\E{- 6} &    1.648\E{- 6} &    1.158\E{- 6} \\
\I{ 48}{Ca} &    1.763\E{- 6} &    2.266\E{- 6} &    2.207\E{- 6} &    2.481\E{- 6} &    2.895\E{- 6} &    2.888\E{- 6} &    1.764\E{- 6} &    2.207\E{- 6} &    2.905\E{- 6} &    2.901\E{- 6} \\
\I{ 45}{Sc} &    2.048\E{- 6} &    4.896\E{- 6} &    5.518\E{- 5} &    3.984\E{- 6} &    8.322\E{- 6} &    7.697\E{- 6} &    1.908\E{- 6} &    6.110\E{- 5} &    7.927\E{- 6} &    5.065\E{- 6} \\
\I{ 46}{Ti} &    1.392\E{- 5} &    7.153\E{- 5} &    2.072\E{- 4} &    2.373\E{- 5} &    9.140\E{- 5} &    8.693\E{- 5} &    1.333\E{- 5} &    2.574\E{- 4} &    1.033\E{- 4} &    6.252\E{- 5} \\
\I{ 47}{Ti} &    5.235\E{- 6} &    8.564\E{- 6} &    7.551\E{- 5} &    7.949\E{- 6} &    1.505\E{- 5} &    1.706\E{- 5} &    4.590\E{- 6} &    8.393\E{- 5} &    1.506\E{- 5} &    9.722\E{- 6} \\
\I{ 48}{Ti} &    1.276\E{- 4} &    1.691\E{- 4} &    2.390\E{- 4} &    1.664\E{- 4} &    2.050\E{- 4} &    3.130\E{- 4} &    1.201\E{- 4} &    2.214\E{- 4} &    2.002\E{- 4} &    2.118\E{- 4} \\
\I{ 49}{Ti} &    9.988\E{- 6} &    1.428\E{- 5} &    3.613\E{- 5} &    1.606\E{- 5} &    2.525\E{- 5} &    2.657\E{- 5} &    1.005\E{- 5} &    3.965\E{- 5} &    2.460\E{- 5} &    2.359\E{- 5} \\
\I{ 50}{Ti} &    4.577\E{- 6} &    9.499\E{- 6} &    9.085\E{- 6} &    1.218\E{- 5} &    1.803\E{- 5} &    1.788\E{- 5} &    4.276\E{- 6} &    8.087\E{- 6} &    1.509\E{- 5} &    1.561\E{- 5} \\
\I{ 50}{ V} &    1.270\E{- 7} &    1.475\E{- 7} &    9.219\E{- 6} &    2.686\E{- 7} &    6.913\E{- 7} &    6.924\E{- 7} &    1.223\E{- 7} &    9.931\E{- 6} &    7.040\E{- 7} &    3.286\E{- 7} \\
\I{ 51}{ V} &    3.146\E{- 5} &    4.253\E{- 5} &    5.739\E{- 5} &    4.149\E{- 5} &    6.878\E{- 5} &    7.403\E{- 5} &    3.256\E{- 5} &    6.222\E{- 5} &    7.088\E{- 5} &    4.645\E{- 5} \\
\I{ 50}{Cr} &    8.179\E{- 5} &    1.547\E{- 4} &    1.054\E{- 4} &    1.185\E{- 4} &    2.452\E{- 4} &    2.607\E{- 4} &    8.566\E{- 5} &    1.228\E{- 4} &    2.649\E{- 4} &    1.710\E{- 4} \\
\I{ 52}{Cr} &    1.597\E{- 3} &    1.798\E{- 3} &    1.291\E{- 3} &    2.096\E{- 3} &    2.947\E{- 3} &    3.443\E{- 3} &    1.754\E{- 3} &    1.264\E{- 3} &    3.188\E{- 3} &    3.194\E{- 3} \\
\I{ 53}{Cr} &    2.007\E{- 4} &    2.396\E{- 4} &    1.580\E{- 4} &    2.584\E{- 4} &    3.960\E{- 4} &    4.378\E{- 4} &    2.193\E{- 4} &    1.572\E{- 4} &    4.197\E{- 4} &    3.494\E{- 4} \\
\I{ 54}{Cr} &    1.520\E{- 5} &    2.641\E{- 5} &    2.506\E{- 5} &    3.133\E{- 5} &    4.283\E{- 5} &    4.222\E{- 5} &    1.543\E{- 5} &    2.556\E{- 5} &    4.522\E{- 5} &    4.088\E{- 5} \\
\hline
\PPI&\PPE&\PPE&\PPE&\PPE&\PPE&\PPE&\PPE&\PPE&\PPE&\PPE \\
\end{tabular}
}% end scale box
\vspace{-1.5\baselineskip}
\begin{flushright}\textsc{(continued on next page)}\end{flushright}
\end{table}

\clearpage

\addtocounter{table}{-1}

\begin{table}
\setlength{\tabcolsep}{1ex}
\centering
\caption{(continued) yields}
\scalebox{\Scale}{
\begin{tabular}{r@{}lr@{}lr@{}lr@{}lr@{}lr@{}lr@{}lr@{}lr@{}lr@{}lr@{}l}
\hline
\hline
\multicolumn{2}{c}{ion} &
\multicolumn{2}{c}{   S15} &
\multicolumn{2}{c}{   S19} &
\multicolumn{2}{c}{   S20} &
\multicolumn{2}{c}{   S21} &
\multicolumn{2}{c}{   S25} &
\multicolumn{2}{c}{  S25P} &
\multicolumn{2}{c}{   N15} &
\multicolumn{2}{c}{   N20} &
\multicolumn{2}{c}{   N25} &
\multicolumn{2}{c}{   H25} \\
\hline
\I{ 55}{Mn} &    1.271\E{- 3} &    1.496\E{- 3} &    9.887\E{- 4} &    1.540\E{- 3} &    2.321\E{- 3} &    2.643\E{- 3} &    1.390\E{- 3} &    9.616\E{- 4} &    2.467\E{- 3} &    1.511\E{- 3} \\
\I{ 54}{Fe} &    7.118\E{- 3} &    9.521\E{- 3} &    5.634\E{- 3} &    9.748\E{- 3} &    1.703\E{- 2} &    1.858\E{- 2} &    7.476\E{- 3} &    5.578\E{- 3} &    1.742\E{- 2} &    1.312\E{- 2} \\
\I{ 56}{Fe} &    1.261\E{- 1} &    1.233\E{- 1} &    1.096\E{- 1} &    1.239\E{- 1} &    1.294\E{- 1} &    2.184\E{- 1} &    1.263\E{- 1} &    1.109\E{- 1} &    1.294\E{- 1} &    1.330\E{- 1} \\
\I{ 57}{Fe} &    4.211\E{- 3} &    4.093\E{- 3} &    5.016\E{- 3} &    3.783\E{- 3} &    3.414\E{- 3} &    6.805\E{- 3} &    3.977\E{- 3} &    4.990\E{- 3} &    3.171\E{- 3} &    3.341\E{- 3} \\
\I{ 58}{Fe} &    3.939\E{- 4} &    7.141\E{- 4} &    6.953\E{- 4} &    9.639\E{- 4} &    1.161\E{- 3} &    1.167\E{- 3} &    4.132\E{- 4} &    7.663\E{- 4} &    1.432\E{- 3} &    1.066\E{- 3} \\
\I{ 59}{Co} &    4.542\E{- 4} &    5.830\E{- 4} &    7.622\E{- 4} &    5.939\E{- 4} &    6.682\E{- 4} &    8.965\E{- 4} &    4.317\E{- 4} &    8.019\E{- 4} &    7.355\E{- 4} &    6.631\E{- 4} \\
\I{ 58}{Ni} &    7.326\E{- 3} &    5.464\E{- 3} &    7.982\E{- 3} &    4.863\E{- 3} &    4.840\E{- 3} &    9.218\E{- 3} &    6.958\E{- 3} &    8.041\E{- 3} &    4.407\E{- 3} &    3.727\E{- 3} \\
\I{ 60}{Ni} &    2.539\E{- 3} &    2.813\E{- 3} &    3.642\E{- 3} &    2.372\E{- 3} &    2.092\E{- 3} &    4.646\E{- 3} &    2.119\E{- 3} &    3.522\E{- 3} &    1.769\E{- 3} &    2.237\E{- 3} \\
\I{ 61}{Ni} &    2.148\E{- 4} &    3.078\E{- 4} &    5.084\E{- 4} &    3.398\E{- 4} &    3.672\E{- 4} &    5.147\E{- 4} &    1.847\E{- 4} &    4.843\E{- 4} &    3.754\E{- 4} &    3.332\E{- 4} \\
\I{ 62}{Ni} &    1.405\E{- 3} &    1.464\E{- 3} &    2.437\E{- 3} &    1.582\E{- 3} &    1.744\E{- 3} &    2.737\E{- 3} &    1.144\E{- 3} &    2.140\E{- 3} &    1.453\E{- 3} &    9.272\E{- 4} \\
\I{ 64}{Ni} &    1.169\E{- 4} &    3.259\E{- 4} &    2.152\E{- 4} &    4.184\E{- 4} &    7.453\E{- 4} &    7.444\E{- 4} &    9.166\E{- 5} &    1.431\E{- 4} &    5.184\E{- 4} &    9.306\E{- 4} \\
\I{ 63}{Cu} &    8.870\E{- 5} &    2.638\E{- 4} &    1.121\E{- 4} &    2.929\E{- 4} &    4.340\E{- 4} &    4.267\E{- 4} &    7.740\E{- 5} &    9.578\E{- 5} &    3.598\E{- 4} &    4.681\E{- 4} \\
\I{ 65}{Cu} &    3.667\E{- 5} &    7.554\E{- 5} &    9.614\E{- 5} &    1.149\E{- 4} &    2.013\E{- 4} &    2.107\E{- 4} &    2.905\E{- 5} &    6.714\E{- 5} &    1.412\E{- 4} &    2.445\E{- 4} \\
\I{ 64}{Zn} &    2.772\E{- 5} &    3.200\E{- 5} &    5.240\E{- 5} &    3.938\E{- 5} &    5.525\E{- 5} &    6.912\E{- 5} &    2.361\E{- 5} &    4.784\E{- 5} &    4.767\E{- 5} &    7.542\E{- 5} \\
\I{ 66}{Zn} &    5.667\E{- 5} &    8.345\E{- 5} &    1.502\E{- 4} &    1.216\E{- 4} &    1.966\E{- 4} &    2.239\E{- 4} &    4.321\E{- 5} &    1.067\E{- 4} &    1.287\E{- 4} &    3.147\E{- 4} \\
\I{ 67}{Zn} &    8.346\E{- 6} &    2.345\E{- 5} &    1.061\E{- 5} &    3.342\E{- 5} &    6.660\E{- 5} &    6.790\E{- 5} &    6.118\E{- 6} &    6.835\E{- 6} &    4.256\E{- 5} &    1.004\E{- 4} \\
\I{ 68}{Zn} &    2.518\E{- 5} &    8.078\E{- 5} &    6.653\E{- 5} &    1.329\E{- 4} &    2.521\E{- 4} &    2.520\E{- 4} &    1.782\E{- 5} &    3.765\E{- 5} &    1.388\E{- 4} &    4.367\E{- 4} \\
\I{ 70}{Zn} &    2.637\E{- 6} &    4.671\E{- 6} &    5.420\E{- 7} &    5.507\E{- 6} &    1.996\E{- 5} &    2.253\E{- 5} &    1.904\E{- 6} &    3.788\E{- 7} &    1.152\E{- 5} &    1.967\E{- 5} \\
\I{ 69}{Ga} &    4.071\E{- 6} &    1.354\E{- 5} &    1.063\E{- 5} &    1.953\E{- 5} &    4.111\E{- 5} &    4.153\E{- 5} &    2.848\E{- 6} &    6.236\E{- 6} &    2.242\E{- 5} &    6.872\E{- 5} \\
\I{ 71}{Ga} &    2.945\E{- 6} &    8.446\E{- 6} &    9.592\E{- 6} &    1.765\E{- 5} &    3.040\E{- 5} &    3.160\E{- 5} &    1.913\E{- 6} &    4.910\E{- 6} &    1.493\E{- 5} &    4.851\E{- 5} \\
\I{ 70}{Ge} &    3.209\E{- 6} &    1.231\E{- 5} &    1.553\E{- 5} &    1.858\E{- 5} &    2.856\E{- 5} &    2.859\E{- 5} &    2.199\E{- 6} &    9.135\E{- 6} &    1.458\E{- 5} &    5.449\E{- 5} \\
\I{ 72}{Ge} &    3.848\E{- 6} &    1.389\E{- 5} &    2.204\E{- 5} &    2.076\E{- 5} &    3.691\E{- 5} &    3.723\E{- 5} &    2.601\E{- 6} &    1.089\E{- 5} &    1.731\E{- 5} &    8.387\E{- 5} \\
\I{ 73}{Ge} &    1.757\E{- 6} &    7.394\E{- 6} &    1.938\E{- 6} &    9.719\E{- 6} &    2.232\E{- 5} &    2.212\E{- 5} &    1.171\E{- 6} &    1.049\E{- 6} &    1.038\E{- 5} &    3.121\E{- 5} \\
\I{ 74}{Ge} &    4.611\E{- 6} &    1.733\E{- 5} &    1.115\E{- 5} &    2.579\E{- 5} &    5.030\E{- 5} &    5.011\E{- 5} &    3.041\E{- 6} &    5.510\E{- 6} &    2.180\E{- 5} &    8.813\E{- 5} \\
\I{ 76}{Ge} &    1.274\E{- 6} &    2.719\E{- 6} &    4.070\E{- 7} &    3.418\E{- 6} &    1.036\E{- 5} &    1.156\E{- 5} &    9.136\E{- 7} &    3.376\E{- 7} &    4.872\E{- 6} &    2.779\E{- 5} \\
\I{ 75}{As} &    1.376\E{- 6} &    3.344\E{- 6} &    2.548\E{- 6} &    4.844\E{- 6} &    1.408\E{- 5} &    1.456\E{- 5} &    9.093\E{- 7} &    1.292\E{- 6} &    6.199\E{- 6} &    2.328\E{- 5} \\
\I{ 74}{Se} &    1.039\E{- 7} &    3.318\E{- 7} &    2.788\E{- 7} &    3.396\E{- 7} &    1.036\E{- 6} &    1.123\E{- 6} &    7.241\E{- 8} &    1.588\E{- 7} &    4.500\E{- 7} &    2.188\E{- 6} \\
\I{ 76}{Se} &    8.126\E{- 7} &    4.065\E{- 6} &    4.269\E{- 6} &    5.540\E{- 6} &    8.389\E{- 6} &    8.414\E{- 6} &    5.236\E{- 7} &    2.124\E{- 6} &    3.348\E{- 6} &    1.458\E{- 5} \\
\I{ 77}{Se} &    8.420\E{- 7} &    3.512\E{- 6} &    2.149\E{- 6} &    3.638\E{- 6} &    9.806\E{- 6} &    1.012\E{- 5} &    5.282\E{- 7} &    1.017\E{- 6} &    3.957\E{- 6} &    1.297\E{- 5} \\
\I{ 78}{Se} &    1.328\E{- 6} &    5.368\E{- 6} &    8.425\E{- 6} &    7.151\E{- 6} &    1.072\E{- 5} &    1.087\E{- 5} &    8.718\E{- 7} &    3.724\E{- 6} &    4.327\E{- 6} &    2.448\E{- 5} \\
\I{ 80}{Se} &    2.570\E{- 6} &    1.026\E{- 5} &    7.081\E{- 6} &    1.392\E{- 5} &    2.527\E{- 5} &    2.436\E{- 5} &    1.692\E{- 6} &    3.376\E{- 6} &    9.580\E{- 6} &    4.970\E{- 5} \\
\I{ 82}{Se} &    8.656\E{- 7} &    1.689\E{- 6} &    2.310\E{- 7} &    2.386\E{- 6} &    5.419\E{- 6} &    6.280\E{- 6} &    5.813\E{- 7} &    2.043\E{- 7} &    2.194\E{- 6} &    9.179\E{- 6} \\
\I{ 79}{Br} &    7.744\E{- 7} &    3.493\E{- 6} &    1.132\E{- 6} &    4.109\E{- 6} &    8.674\E{- 6} &    8.663\E{- 6} &    4.992\E{- 7} &    5.747\E{- 7} &    3.427\E{- 6} &    1.793\E{- 5} \\
\I{ 81}{Br} &    6.830\E{- 7} &    2.039\E{- 6} &    1.261\E{- 6} &    2.709\E{- 6} &    6.393\E{- 6} &    6.468\E{- 6} &    4.437\E{- 7} &    6.585\E{- 7} &    2.454\E{- 6} &    1.468\E{- 5} \\
\I{ 78}{Kr} &    9.545\E{- 9} &    9.648\E{- 8} &    3.274\E{- 8} &    2.474\E{- 8} &    1.382\E{- 7} &    1.368\E{- 7} &    7.196\E{- 9} &    1.834\E{- 8} &    5.230\E{- 8} &    4.924\E{- 7} \\
\I{ 80}{Kr} &    1.291\E{- 7} &    6.763\E{- 7} &    4.501\E{- 7} &    4.057\E{- 7} &    9.503\E{- 7} &    9.819\E{- 7} &    8.782\E{- 8} &    2.392\E{- 7} &    3.445\E{- 7} &    2.056\E{- 6} \\
\I{ 82}{Kr} &    4.846\E{- 7} &    2.402\E{- 6} &    3.352\E{- 6} &    3.679\E{- 6} &    5.228\E{- 6} &    4.970\E{- 6} &    3.100\E{- 7} &    1.552\E{- 6} &    1.860\E{- 6} &    1.156\E{- 5} \\
\I{ 83}{Kr} &    5.326\E{- 7} &    1.904\E{- 6} &    1.727\E{- 6} &    2.510\E{- 6} &    6.170\E{- 6} &    6.071\E{- 6} &    3.463\E{- 7} &    7.977\E{- 7} &    2.177\E{- 6} &    1.282\E{- 5} \\
\I{ 84}{Kr} &    1.546\E{- 6} &    5.998\E{- 6} &    6.501\E{- 6} &    7.186\E{- 6} &    1.466\E{- 5} &    1.385\E{- 5} &    1.112\E{- 6} &    3.010\E{- 6} &    5.449\E{- 6} &    4.323\E{- 5} \\
\I{ 86}{Kr} &    1.662\E{- 6} &    5.653\E{- 6} &    1.506\E{- 6} &    4.642\E{- 6} &    1.261\E{- 5} &    1.340\E{- 5} &    1.004\E{- 6} &    7.318\E{- 7} &    4.431\E{- 6} &    2.886\E{- 5} \\
\I{ 85}{Rb} &    7.364\E{- 7} &    3.160\E{- 6} &    1.104\E{- 6} &    2.559\E{- 6} &    7.717\E{- 6} &    7.612\E{- 6} &    4.601\E{- 7} &    5.443\E{- 7} &    2.714\E{- 6} &    7.653\E{- 6} \\
\I{ 87}{Rb} &    2.659\E{- 7} &    9.607\E{- 7} &    1.121\E{- 6} &    1.611\E{- 6} &    3.012\E{- 6} &    3.086\E{- 6} &    1.735\E{- 7} &    4.549\E{- 7} &    9.064\E{- 7} &    1.990\E{- 6} \\
\I{ 84}{Sr} &    5.227\E{- 9} &    1.950\E{- 7} &    6.621\E{- 8} &    1.549\E{- 8} &    1.912\E{- 7} &    1.752\E{- 7} &    4.709\E{- 9} &    3.811\E{- 8} &    7.560\E{- 8} &    3.036\E{- 7} \\
\I{ 86}{Sr} &    1.286\E{- 7} &    4.127\E{- 7} &    1.227\E{- 6} &    1.525\E{- 6} &    1.917\E{- 6} &    1.802\E{- 6} &    1.019\E{- 7} &    6.370\E{- 7} &    7.182\E{- 7} &    9.693\E{- 7} \\
\I{ 87}{Sr} &    6.853\E{- 8} &    1.335\E{- 7} &    8.588\E{- 7} &    8.668\E{- 7} &    1.281\E{- 6} &    1.221\E{- 6} &    5.700\E{- 8} &    4.298\E{- 7} &    4.696\E{- 7} &    3.062\E{- 7} \\
\I{ 88}{Sr} &    1.070\E{- 6} &    2.345\E{- 6} &    4.688\E{- 6} &    5.869\E{- 6} &    1.145\E{- 5} &    1.143\E{- 5} &    8.873\E{- 7} &    2.497\E{- 6} &    4.586\E{- 6} &    7.797\E{- 6} \\
\I{ 89}{ Y} &    2.384\E{- 7} &    5.985\E{- 7} &    8.811\E{- 7} &    1.007\E{- 6} &    1.991\E{- 6} &    1.960\E{- 6} &    2.027\E{- 7} &    5.257\E{- 7} &    8.896\E{- 7} &    1.525\E{- 6} \\
\I{ 90}{Zr} &    2.460\E{- 7} &    8.488\E{- 7} &    2.739\E{- 6} &    6.807\E{- 7} &    1.880\E{- 6} &    1.709\E{- 6} &    2.225\E{- 7} &    1.505\E{- 6} &    9.298\E{- 7} &    2.132\E{- 6} \\
\I{ 91}{Zr} &    5.846\E{- 8} &    1.302\E{- 7} &    2.137\E{- 7} &    2.078\E{- 7} &    4.120\E{- 7} &    4.081\E{- 7} &    5.157\E{- 8} &    1.269\E{- 7} &    2.051\E{- 7} &    4.494\E{- 7} \\
\I{ 92}{Zr} &    8.629\E{- 8} &    1.827\E{- 7} &    2.473\E{- 7} &    2.534\E{- 7} &    4.426\E{- 7} &    4.381\E{- 7} &    7.697\E{- 8} &    1.599\E{- 7} &    2.433\E{- 7} &    7.119\E{- 7} \\
\I{ 94}{Zr} &    8.407\E{- 8} &    1.639\E{- 7} &    1.040\E{- 7} &    1.953\E{- 7} &    3.428\E{- 7} &    3.371\E{- 7} &    7.641\E{- 8} &    9.188\E{- 8} &    2.100\E{- 7} &    5.278\E{- 7} \\
\I{ 96}{Zr} &    2.363\E{- 8} &    6.184\E{- 8} &    1.905\E{- 8} &    3.090\E{- 8} &    8.282\E{- 8} &    8.417\E{- 8} &    2.023\E{- 8} &    1.669\E{- 8} &    5.093\E{- 8} &    1.253\E{- 7} \\
\I{ 93}{Nb} &    3.224\E{- 8} &    7.076\E{- 8} &    3.873\E{- 8} &    8.403\E{- 8} &    1.742\E{- 7} &    1.704\E{- 7} &    2.884\E{- 8} &    3.312\E{- 8} &    9.275\E{- 8} &    2.994\E{- 7} \\
\I{ 92}{Mo} &    1.122\E{- 8} &    1.309\E{- 8} &    1.618\E{- 8} &    1.666\E{- 8} &    1.964\E{- 8} &    2.049\E{- 8} &    1.118\E{- 8} &    1.565\E{- 8} &    1.804\E{- 8} &    3.090\E{- 8} \\
\I{ 94}{Mo} &    7.221\E{- 9} &    8.621\E{- 9} &    1.511\E{- 8} &    1.062\E{- 8} &    1.278\E{- 8} &    1.300\E{- 8} &    7.205\E{- 9} &    1.312\E{- 8} &    1.160\E{- 8} &    2.350\E{- 8} \\
\I{ 95}{Mo} &    1.842\E{- 8} &    3.929\E{- 8} &    2.075\E{- 8} &    3.117\E{- 8} &    6.705\E{- 8} &    6.639\E{- 8} &    1.675\E{- 8} &    1.861\E{- 8} &    4.263\E{- 8} &    1.337\E{- 7} \\
\I{ 96}{Mo} &    1.383\E{- 8} &    1.890\E{- 8} &    3.127\E{- 8} &    3.312\E{- 8} &    3.635\E{- 8} &    3.612\E{- 8} &    1.346\E{- 8} &    2.542\E{- 8} &    2.808\E{- 8} &    9.055\E{- 8} \\
\I{ 97}{Mo} &    1.039\E{- 8} &    1.216\E{- 8} &    1.059\E{- 8} &    1.614\E{- 8} &    2.365\E{- 8} &    2.603\E{- 8} &    9.630\E{- 9} &    1.007\E{- 8} &    1.835\E{- 8} &    6.849\E{- 8} \\
\I{ 98}{Mo} &    2.153\E{- 8} &    3.153\E{- 8} &    2.994\E{- 8} &    4.807\E{- 8} &    7.918\E{- 8} &    7.836\E{- 8} &    2.072\E{- 8} &    2.746\E{- 8} &    5.391\E{- 8} &    2.630\E{- 7} \\
\I{100}{Mo} &    8.223\E{- 9} &    1.246\E{- 8} &    9.997\E{- 9} &    1.137\E{- 8} &    1.734\E{- 8} &    1.715\E{- 8} &    8.043\E{- 9} &    9.806\E{- 9} &    1.418\E{- 8} &    3.973\E{- 8} \\
\hline
\PPI&\PPE&\PPE&\PPE&\PPE&\PPE&\PPE&\PPE&\PPE&\PPE&\PPE \\
\end{tabular}
}% end scale box
\vspace{-1.5\baselineskip}
\begin{flushright}\textsc{(continued on next page)}\end{flushright}
\end{table}

\clearpage

\addtocounter{table}{-1}

\begin{table}
\setlength{\tabcolsep}{1ex}
\centering
\caption{(continued) yields}
\scalebox{\Scale}{
\begin{tabular}{r@{}lr@{}lr@{}lr@{}lr@{}lr@{}lr@{}lr@{}lr@{}lr@{}lr@{}l}
\hline
\hline
\multicolumn{2}{c}{ion} &
\multicolumn{2}{c}{   S15} &
\multicolumn{2}{c}{   S19} &
\multicolumn{2}{c}{   S20} &
\multicolumn{2}{c}{   S21} &
\multicolumn{2}{c}{   S25} &
\multicolumn{2}{c}{  S25P} &
\multicolumn{2}{c}{   N15} &
\multicolumn{2}{c}{   N20} &
\multicolumn{2}{c}{   N25} &
\multicolumn{2}{c}{   H25} \\
\hline
\I{ 96}{Ru} &    3.222\E{- 9} &    3.596\E{- 9} &    4.099\E{- 9} &    4.656\E{- 9} &    5.392\E{- 9} &    5.664\E{- 9} &    3.219\E{- 9} &    4.053\E{- 9} &    5.074\E{- 9} &    1.145\E{- 8} \\
\I{ 98}{Ru} &    1.273\E{- 9} &    1.277\E{- 9} &    2.327\E{- 9} &    2.250\E{- 9} &    2.730\E{- 9} &    2.928\E{- 9} &    1.249\E{- 9} &    2.120\E{- 9} &    2.231\E{- 9} &    2.135\E{- 8} \\
\I{ 99}{Ru} &    7.682\E{- 9} &    1.086\E{- 8} &    1.005\E{- 8} &    1.241\E{- 8} &    1.399\E{- 8} &    1.405\E{- 8} &    7.507\E{- 9} &    9.680\E{- 9} &    1.241\E{- 8} &    2.113\E{- 8} \\
\I{100}{Ru} &    8.160\E{- 9} &    1.120\E{- 8} &    1.608\E{- 8} &    1.666\E{- 8} &    1.705\E{- 8} &    1.726\E{- 8} &    8.005\E{- 9} &    1.385\E{- 8} &    1.452\E{- 8} &    5.959\E{- 8} \\
\I{101}{Ru} &    9.795\E{- 9} &    1.239\E{- 8} &    1.226\E{- 8} &    1.470\E{- 8} &    1.605\E{- 8} &    1.609\E{- 8} &    9.671\E{- 9} &    1.213\E{- 8} &    1.528\E{- 8} &    2.170\E{- 8} \\
\I{102}{Ru} &    1.984\E{- 8} &    2.630\E{- 8} &    2.858\E{- 8} &    3.371\E{- 8} &    4.119\E{- 8} &    4.102\E{- 8} &    1.954\E{- 8} &    2.671\E{- 8} &    3.505\E{- 8} &    1.156\E{- 7} \\
\I{104}{Ru} &    1.227\E{- 8} &    1.724\E{- 8} &    1.495\E{- 8} &    1.901\E{- 8} &    3.033\E{- 8} &    3.015\E{- 8} &    1.210\E{- 8} &    1.451\E{- 8} &    2.353\E{- 8} &    1.291\E{- 7} \\
\I{103}{Rh} &    1.100\E{- 8} &    1.475\E{- 8} &    1.407\E{- 8} &    1.683\E{- 8} &    1.983\E{- 8} &    1.972\E{- 8} &    1.093\E{- 8} &    1.392\E{- 8} &    1.802\E{- 8} &    1.662\E{- 8} \\
\I{102}{Pd} &    9.811\E{-10} &    5.679\E{-10} &    1.364\E{- 9} &    2.110\E{- 9} &    2.381\E{- 9} &    2.807\E{- 9} &    9.424\E{-10} &    1.280\E{- 9} &    1.840\E{- 9} &         \NoData \\
\I{104}{Pd} &    5.657\E{- 9} &    7.069\E{- 9} &    1.016\E{- 8} &    9.491\E{- 9} &    1.051\E{- 8} &    1.052\E{- 8} &    5.593\E{- 9} &    9.483\E{- 9} &    9.430\E{- 9} &         \NoData \\
\I{105}{Pd} &    9.972\E{- 9} &    1.285\E{- 8} &    1.279\E{- 8} &    1.416\E{- 8} &    1.639\E{- 8} &    1.636\E{- 8} &    9.916\E{- 9} &    1.262\E{- 8} &    1.556\E{- 8} &    5.963\E{- 9} \\
\I{106}{Pd} &    1.371\E{- 8} &    1.907\E{- 8} &    2.264\E{- 8} &    2.154\E{- 8} &    2.586\E{- 8} &    2.567\E{- 8} &    1.349\E{- 8} &    2.056\E{- 8} &    2.298\E{- 8} &    5.147\E{- 7} \\
\I{108}{Pd} &    1.343\E{- 8} &    1.816\E{- 8} &    1.883\E{- 8} &    2.369\E{- 8} &    3.070\E{- 8} &    3.049\E{- 8} &    1.311\E{- 8} &    1.771\E{- 8} &    2.542\E{- 8} &         \NoData \\
\I{110}{Pd} &    6.380\E{- 9} &    8.488\E{- 9} &    7.406\E{- 9} &    9.492\E{- 9} &    1.838\E{- 8} &    1.822\E{- 8} &    6.280\E{- 9} &    7.256\E{- 9} &    1.377\E{- 8} &         \NoData \\
\I{107}{Ag} &    8.271\E{- 9} &    1.038\E{- 8} &    1.048\E{- 8} &    1.182\E{- 8} &    1.321\E{- 8} &    1.322\E{- 8} &    8.253\E{- 9} &    1.043\E{- 8} &    1.276\E{- 8} &         \NoData \\
\I{109}{Ag} &    7.865\E{- 9} &    1.030\E{- 8} &    1.042\E{- 8} &    1.168\E{- 8} &    1.414\E{- 8} &    1.403\E{- 8} &    7.820\E{- 9} &    1.023\E{- 8} &    1.299\E{- 8} &         \NoData \\
\I{106}{Cd} &    1.506\E{- 9} &    8.046\E{-10} &    1.653\E{- 9} &    2.849\E{- 9} &    3.353\E{- 9} &    4.007\E{- 9} &    1.470\E{- 9} &    1.570\E{- 9} &    2.788\E{- 9} &         \NoData \\
\I{108}{Cd} &    1.076\E{- 9} &    8.946\E{-10} &    2.241\E{- 9} &    1.778\E{- 9} &    2.299\E{- 9} &    2.491\E{- 9} &    1.056\E{- 9} &    2.120\E{- 9} &    1.856\E{- 9} &         \NoData \\
\I{110}{Cd} &    7.934\E{- 9} &    1.015\E{- 8} &    1.480\E{- 8} &    1.385\E{- 8} &    1.551\E{- 8} &    1.594\E{- 8} &    7.852\E{- 9} &    1.348\E{- 8} &    1.386\E{- 8} &         \NoData \\
\I{111}{Cd} &    7.051\E{- 9} &    9.032\E{- 9} &    9.063\E{- 9} &    1.030\E{- 8} &    1.171\E{- 8} &    1.171\E{- 8} &    7.005\E{- 9} &    8.918\E{- 9} &    1.111\E{- 8} &         \NoData \\
\I{112}{Cd} &    1.428\E{- 8} &    1.879\E{- 8} &    2.278\E{- 8} &    2.262\E{- 8} &    2.613\E{- 8} &    2.606\E{- 8} &    1.409\E{- 8} &    2.099\E{- 8} &    2.378\E{- 8} &         \NoData \\
\I{113}{Cd} &    6.950\E{- 9} &    9.033\E{- 9} &    8.802\E{- 9} &    1.031\E{- 8} &    1.173\E{- 8} &    1.168\E{- 8} &    6.898\E{- 9} &    8.709\E{- 9} &    1.109\E{- 8} &         \NoData \\
\I{114}{Cd} &    1.833\E{- 8} &    2.498\E{- 8} &    2.550\E{- 8} &    3.209\E{- 8} &    4.459\E{- 8} &    4.396\E{- 8} &    1.793\E{- 8} &    2.416\E{- 8} &    3.644\E{- 8} &         \NoData \\
\I{116}{Cd} &    6.211\E{- 9} &    1.040\E{- 8} &    7.215\E{- 9} &    9.294\E{- 9} &    1.978\E{- 8} &    1.942\E{- 8} &    5.978\E{- 9} &    6.818\E{- 9} &    1.417\E{- 8} &         \NoData \\
\I{113}{In} &    3.410\E{-10} &    3.331\E{-10} &    5.510\E{-10} &    5.081\E{-10} &    4.767\E{-10} &    5.020\E{-10} &    3.376\E{-10} &    5.562\E{-10} &    4.600\E{-10} &         \NoData \\
\I{115}{In} &    6.401\E{- 9} &    8.827\E{- 9} &    8.408\E{- 9} &    9.918\E{- 9} &    1.224\E{- 8} &    1.210\E{- 8} &    6.340\E{- 9} &    8.252\E{- 9} &    1.101\E{- 8} &         \NoData \\
\I{112}{Sn} &    3.180\E{- 9} &    1.518\E{- 9} &    3.332\E{- 9} &    5.905\E{- 9} &    6.257\E{- 9} &    7.466\E{- 9} &    3.075\E{- 9} &    3.244\E{- 9} &    5.436\E{- 9} &         \NoData \\
\I{114}{Sn} &    1.824\E{- 9} &    1.069\E{- 9} &    5.919\E{- 9} &    3.291\E{- 9} &    4.316\E{- 9} &    4.807\E{- 9} &    1.782\E{- 9} &    5.675\E{- 9} &    3.594\E{- 9} &         \NoData \\
\I{115}{Sn} &    4.675\E{-10} &    5.499\E{-10} &    1.205\E{- 9} &    6.436\E{-10} &    7.201\E{-10} &    7.243\E{-10} &    4.674\E{-10} &    1.141\E{- 9} &    7.078\E{-10} &         \NoData \\
\I{116}{Sn} &    2.109\E{- 8} &    2.748\E{- 8} &    3.949\E{- 8} &    3.577\E{- 8} &    3.935\E{- 8} &    3.916\E{- 8} &    2.095\E{- 8} &    3.668\E{- 8} &    3.607\E{- 8} &         \NoData \\
\I{117}{Sn} &    1.081\E{- 8} &    1.441\E{- 8} &    1.404\E{- 8} &    1.672\E{- 8} &    1.882\E{- 8} &    1.869\E{- 8} &    1.075\E{- 8} &    1.383\E{- 8} &    1.770\E{- 8} &         \NoData \\
\I{118}{Sn} &    3.644\E{- 8} &    5.059\E{- 8} &    5.709\E{- 8} &    6.142\E{- 8} &    7.478\E{- 8} &    7.401\E{- 8} &    3.604\E{- 8} &    5.429\E{- 8} &    6.654\E{- 8} &         \NoData \\
\I{119}{Sn} &    1.258\E{- 8} &    1.709\E{- 8} &    1.602\E{- 8} &    1.977\E{- 8} &    2.370\E{- 8} &    2.343\E{- 8} &    1.249\E{- 8} &    1.585\E{- 8} &    2.178\E{- 8} &         \NoData \\
\I{120}{Sn} &    5.244\E{- 8} &    7.390\E{- 8} &    7.272\E{- 8} &    8.907\E{- 8} &    1.168\E{- 7} &    1.156\E{- 7} &    5.190\E{- 8} &    7.103\E{- 8} &    1.029\E{- 7} &         \NoData \\
\I{122}{Sn} &    1.145\E{- 8} &    1.835\E{- 8} &    1.166\E{- 8} &    1.432\E{- 8} &    2.908\E{- 8} &    2.934\E{- 8} &    1.131\E{- 8} &    1.132\E{- 8} &    2.409\E{- 8} &         \NoData \\
\I{124}{Sn} &    9.903\E{- 9} &    1.102\E{- 8} &    1.098\E{- 8} &    1.139\E{- 8} &    1.555\E{- 8} &    1.635\E{- 8} &    9.913\E{- 9} &    1.100\E{- 8} &    1.513\E{- 8} &         \NoData \\
\I{121}{Sb} &    7.310\E{- 9} &    1.054\E{- 8} &    9.482\E{- 9} &    1.102\E{- 8} &    1.551\E{- 8} &    1.537\E{- 8} &    7.260\E{- 9} &    9.360\E{- 9} &    1.386\E{- 8} &         \NoData \\
\I{123}{Sb} &    5.489\E{- 9} &    7.264\E{- 9} &    6.724\E{- 9} &    7.582\E{- 9} &    9.190\E{- 9} &    9.327\E{- 9} &    5.455\E{- 9} &    6.640\E{- 9} &    8.757\E{- 9} &         \NoData \\
\I{120}{Te} &    3.792\E{-10} &    1.909\E{-10} &    5.168\E{-10} &    6.916\E{-10} &    9.574\E{-10} &    1.051\E{- 9} &    3.775\E{-10} &    5.321\E{-10} &    8.471\E{-10} &         \NoData \\
\I{122}{Te} &    5.263\E{- 9} &    6.423\E{- 9} &    8.331\E{- 9} &    8.299\E{- 9} &    9.163\E{- 9} &    9.342\E{- 9} &    5.267\E{- 9} &    8.331\E{- 9} &    8.896\E{- 9} &         \NoData \\
\I{123}{Te} &    1.773\E{- 9} &    2.149\E{- 9} &    2.254\E{- 9} &    2.712\E{- 9} &    2.880\E{- 9} &    2.900\E{- 9} &    1.770\E{- 9} &    2.231\E{- 9} &    2.820\E{- 9} &         \NoData \\
\I{124}{Te} &    9.497\E{- 9} &    1.314\E{- 8} &    1.607\E{- 8} &    1.605\E{- 8} &    1.695\E{- 8} &    1.681\E{- 8} &    9.457\E{- 9} &    1.573\E{- 8} &    1.635\E{- 8} &         \NoData \\
\I{125}{Te} &    1.347\E{- 8} &    1.646\E{- 8} &    1.649\E{- 8} &    1.850\E{- 8} &    2.108\E{- 8} &    2.121\E{- 8} &    1.346\E{- 8} &    1.651\E{- 8} &    2.071\E{- 8} &         \NoData \\
\I{126}{Te} &    3.716\E{- 8} &    4.833\E{- 8} &    4.967\E{- 8} &    5.590\E{- 8} &    6.456\E{- 8} &    6.409\E{- 8} &    3.708\E{- 8} &    4.947\E{- 8} &    6.271\E{- 8} &         \NoData \\
\I{128}{Te} &    6.106\E{- 8} &    7.654\E{- 8} &    7.644\E{- 8} &    8.184\E{- 8} &    9.915\E{- 8} &    9.849\E{- 8} &    6.112\E{- 8} &    7.651\E{- 8} &    9.738\E{- 8} &         \NoData \\
\I{130}{Te} &    6.594\E{- 8} &    7.983\E{- 8} &    8.209\E{- 8} &    8.620\E{- 8} &    9.919\E{- 8} &    9.950\E{- 8} &    6.604\E{- 8} &    8.231\E{- 8} &    9.945\E{- 8} &         \NoData \\
\I{127}{ I} &    3.395\E{- 8} &    4.195\E{- 8} &    4.332\E{- 8} &    4.581\E{- 8} &    5.298\E{- 8} &    5.285\E{- 8} &    3.405\E{- 8} &    4.341\E{- 8} &    5.261\E{- 8} &         \NoData \\
\I{124}{Xe} &    1.081\E{- 9} &    2.510\E{-10} &    8.870\E{-10} &    2.010\E{- 9} &    2.304\E{- 9} &    2.703\E{- 9} &    1.032\E{- 9} &    8.678\E{-10} &    2.114\E{- 9} &         \NoData \\
\I{126}{Xe} &    7.958\E{-10} &    2.697\E{-10} &    2.416\E{- 9} &    1.411\E{- 9} &    2.299\E{- 9} &    2.535\E{- 9} &    7.930\E{-10} &    2.577\E{- 9} &    2.138\E{- 9} &         \NoData \\
\I{128}{Xe} &    6.540\E{- 9} &    7.382\E{- 9} &    1.385\E{- 8} &    1.100\E{- 8} &    1.221\E{- 8} &    1.267\E{- 8} &    6.519\E{- 9} &    1.353\E{- 8} &    1.191\E{- 8} &         \NoData \\
\I{129}{Xe} &    4.946\E{- 8} &    6.086\E{- 8} &    6.251\E{- 8} &    6.625\E{- 8} &    7.571\E{- 8} &    7.565\E{- 8} &    4.952\E{- 8} &    6.256\E{- 8} &    7.551\E{- 8} &         \NoData \\
\I{130}{Xe} &    1.090\E{- 8} &    1.538\E{- 8} &    1.944\E{- 8} &    1.792\E{- 8} &    1.912\E{- 8} &    1.876\E{- 8} &    1.082\E{- 8} &    1.893\E{- 8} &    1.888\E{- 8} &         \NoData \\
\I{131}{Xe} &    4.083\E{- 8} &    5.052\E{- 8} &    5.165\E{- 8} &    5.548\E{- 8} &    6.324\E{- 8} &    6.330\E{- 8} &    4.085\E{- 8} &    5.169\E{- 8} &    6.300\E{- 8} &         \NoData \\
\I{132}{Xe} &    5.431\E{- 8} &    7.074\E{- 8} &    7.211\E{- 8} &    8.101\E{- 8} &    9.371\E{- 8} &    9.304\E{- 8} &    5.412\E{- 8} &    7.191\E{- 8} &    9.277\E{- 8} &         \NoData \\
\I{134}{Xe} &    2.280\E{- 8} &    3.221\E{- 8} &    2.716\E{- 8} &    2.964\E{- 8} &    4.210\E{- 8} &    4.212\E{- 8} &    2.263\E{- 8} &    2.716\E{- 8} &    4.082\E{- 8} &         \NoData \\
\I{136}{Xe} &    1.799\E{- 8} &    2.168\E{- 8} &    2.092\E{- 8} &    2.347\E{- 8} &    2.931\E{- 8} &    3.009\E{- 8} &    1.805\E{- 8} &    2.096\E{- 8} &    2.922\E{- 8} &         \NoData \\
\I{133}{Cs} &    1.591\E{- 8} &    2.113\E{- 8} &    2.187\E{- 8} &    2.308\E{- 8} &    2.823\E{- 8} &    2.801\E{- 8} &    1.583\E{- 8} &    2.178\E{- 8} &    2.780\E{- 8} &         \NoData \\
\I{130}{Ba} &    1.355\E{- 9} &    2.440\E{-10} &    1.630\E{- 9} &    3.014\E{- 9} &    4.114\E{- 9} &    4.494\E{- 9} &    1.279\E{- 9} &    1.639\E{- 9} &    3.792\E{- 9} &         \NoData \\
\I{132}{Ba} &    9.946\E{-10} &    4.917\E{-10} &    6.432\E{- 9} &    2.207\E{- 9} &    3.232\E{- 9} &    3.675\E{- 9} &    9.544\E{-10} &    6.781\E{- 9} &    2.980\E{- 9} &         \NoData \\
\I{134}{Ba} &    6.077\E{- 9} &    7.662\E{- 9} &    2.370\E{- 8} &    1.020\E{- 8} &    1.155\E{- 8} &    1.187\E{- 8} &    6.018\E{- 9} &    2.231\E{- 8} &    1.139\E{- 8} &         \NoData \\
\hline
\PPI&\PPE&\PPE&\PPE&\PPE&\PPE&\PPE&\PPE&\PPE&\PPE&\PPE \\
\end{tabular}
}% end scale box
\vspace{-1.5\baselineskip}
\begin{flushright}\textsc{(continued on next page)}\end{flushright}
\end{table}

\clearpage

\addtocounter{table}{-1}

\begin{table}
\setlength{\tabcolsep}{1ex}
\centering
\caption{(continued) yields}
\scalebox{\Scale}{
\begin{tabular}{r@{}lr@{}lr@{}lr@{}lr@{}lr@{}lr@{}lr@{}lr@{}lr@{}lr@{}l}
\hline
\hline
\multicolumn{2}{c}{ion} &
\multicolumn{2}{c}{   S15} &
\multicolumn{2}{c}{   S19} &
\multicolumn{2}{c}{   S20} &
\multicolumn{2}{c}{   S21} &
\multicolumn{2}{c}{   S25} &
\multicolumn{2}{c}{  S25P} &
\multicolumn{2}{c}{   N15} &
\multicolumn{2}{c}{   N20} &
\multicolumn{2}{c}{   N25} &
\multicolumn{2}{c}{   H25} \\
\hline
\I{135}{Ba} &    1.366\E{- 8} &    1.763\E{- 8} &    1.801\E{- 8} &    2.043\E{- 8} &    2.337\E{- 8} &    2.380\E{- 8} &    1.359\E{- 8} &    1.777\E{- 8} &    2.297\E{- 8} &         \NoData \\
\I{136}{Ba} &    1.742\E{- 8} &    2.424\E{- 8} &    4.168\E{- 8} &    3.350\E{- 8} &    3.606\E{- 8} &    3.588\E{- 8} &    1.735\E{- 8} &    4.029\E{- 8} &    3.615\E{- 8} &         \NoData \\
\I{137}{Ba} &    2.629\E{- 8} &    4.747\E{- 8} &    3.251\E{- 8} &    4.475\E{- 8} &    5.407\E{- 8} &    5.320\E{- 8} &    2.622\E{- 8} &    3.222\E{- 8} &    5.279\E{- 8} &         \NoData \\
\I{138}{Ba} &    1.743\E{- 7} &    2.721\E{- 7} &    2.461\E{- 7} &    3.274\E{- 7} &    4.288\E{- 7} &    4.223\E{- 7} &    1.719\E{- 7} &    2.398\E{- 7} &    4.095\E{- 7} &         \NoData \\
\I{138}{La} &    2.918\E{-11} &    3.916\E{-11} &    2.331\E{-10} &    5.064\E{-11} &    6.772\E{-11} &    7.208\E{-11} &    2.793\E{-11} &    2.023\E{-10} &    6.233\E{-11} &         \NoData \\
\I{139}{La} &    2.376\E{- 8} &    3.695\E{- 8} &    3.315\E{- 8} &    4.263\E{- 8} &    5.409\E{- 8} &    5.370\E{- 8} &    2.331\E{- 8} &    3.197\E{- 8} &    5.076\E{- 8} &         \NoData \\
\I{136}{Ce} &    2.297\E{-10} &    1.131\E{-10} &    4.131\E{-10} &    5.207\E{-10} &    7.351\E{-10} &    7.913\E{-10} &    2.134\E{-10} &    3.937\E{-10} &    6.195\E{-10} &         \NoData \\
\I{138}{Ce} &    3.194\E{-10} &    2.457\E{-10} &    1.975\E{- 9} &    7.203\E{-10} &    1.084\E{- 9} &    1.201\E{- 9} &    3.026\E{-10} &    1.769\E{- 9} &    9.268\E{-10} &         \NoData \\
\I{140}{Ce} &    5.170\E{- 8} &    7.589\E{- 8} &    7.723\E{- 8} &    9.218\E{- 8} &    1.190\E{- 7} &    1.180\E{- 7} &    5.065\E{- 8} &    7.316\E{- 8} &    1.088\E{- 7} &         \NoData \\
\I{142}{Ce} &    6.475\E{- 9} &    1.218\E{- 8} &    8.216\E{- 9} &    8.150\E{- 9} &    1.199\E{- 8} &    1.181\E{- 8} &    6.321\E{- 9} &    8.050\E{- 9} &    1.137\E{- 8} &         \NoData \\
\I{141}{Pr} &    8.019\E{- 9} &    1.087\E{- 8} &    1.194\E{- 8} &    1.302\E{- 8} &    1.704\E{- 8} &    1.693\E{- 8} &    7.913\E{- 9} &    1.149\E{- 8} &    1.597\E{- 8} &         \NoData \\
\I{142}{Nd} &    1.151\E{- 8} &    1.519\E{- 8} &    2.762\E{- 8} &    2.285\E{- 8} &    2.811\E{- 8} &    2.811\E{- 8} &    1.134\E{- 8} &    2.497\E{- 8} &    2.543\E{- 8} &         \NoData \\
\I{143}{Nd} &    4.696\E{- 9} &    6.052\E{- 9} &    6.197\E{- 9} &    7.062\E{- 9} &    9.231\E{- 9} &    9.280\E{- 9} &    4.654\E{- 9} &    6.098\E{- 9} &    8.737\E{- 9} &         \NoData \\
\I{144}{Nd} &    9.601\E{- 9} &    1.335\E{- 8} &    1.200\E{- 8} &    1.517\E{- 8} &    2.054\E{- 8} &    2.040\E{- 8} &    9.494\E{- 9} &    1.183\E{- 8} &    1.893\E{- 8} &         \NoData \\
\I{145}{Nd} &    3.140\E{- 9} &    4.174\E{- 9} &    4.033\E{- 9} &    4.527\E{- 9} &    5.272\E{- 9} &    5.333\E{- 9} &    3.121\E{- 9} &    3.994\E{- 9} &    5.114\E{- 9} &         \NoData \\
\I{146}{Nd} &    7.366\E{- 9} &    1.135\E{- 8} &    8.730\E{- 9} &    1.164\E{- 8} &    1.458\E{- 8} &    1.459\E{- 8} &    7.243\E{- 9} &    8.609\E{- 9} &    1.326\E{- 8} &         \NoData \\
\I{148}{Nd} &    2.428\E{- 9} &    4.322\E{- 9} &    2.923\E{- 9} &    3.202\E{- 9} &    3.981\E{- 9} &    4.081\E{- 9} &    2.383\E{- 9} &    2.895\E{- 9} &    3.793\E{- 9} &         \NoData \\
\I{150}{Nd} &    2.221\E{- 9} &    2.613\E{- 9} &    2.708\E{- 9} &    2.811\E{- 9} &    3.465\E{- 9} &    3.510\E{- 9} &    2.202\E{- 9} &    2.710\E{- 9} &    3.417\E{- 9} &         \NoData \\
\I{144}{Sm} &    2.368\E{- 9} &    4.769\E{-10} &    1.004\E{- 8} &    4.996\E{- 9} &    5.987\E{- 9} &    6.795\E{- 9} &    2.271\E{- 9} &    9.618\E{- 9} &    5.340\E{- 9} &         \NoData \\
\I{147}{Sm} &    1.922\E{- 9} &    2.302\E{- 9} &    2.275\E{- 9} &    2.564\E{- 9} &    3.154\E{- 9} &    3.309\E{- 9} &    1.895\E{- 9} &    2.257\E{- 9} &    3.038\E{- 9} &         \NoData \\
\I{148}{Sm} &    1.441\E{- 9} &    1.811\E{- 9} &    1.847\E{- 9} &    2.135\E{- 9} &    2.510\E{- 9} &    2.488\E{- 9} &    1.438\E{- 9} &    1.827\E{- 9} &    2.415\E{- 9} &         \NoData \\
\I{149}{Sm} &    1.628\E{- 9} &    2.008\E{- 9} &    1.991\E{- 9} &    2.164\E{- 9} &    2.609\E{- 9} &    2.642\E{- 9} &    1.609\E{- 9} &    1.980\E{- 9} &    2.554\E{- 9} &         \NoData \\
\I{150}{Sm} &    1.003\E{- 9} &    1.369\E{- 9} &    1.269\E{- 9} &    1.772\E{- 9} &    2.017\E{- 9} &    1.976\E{- 9} &    1.002\E{- 9} &    1.258\E{- 9} &    1.901\E{- 9} &         \NoData \\
\I{152}{Sm} &    3.387\E{- 9} &    4.018\E{- 9} &    4.074\E{- 9} &    4.531\E{- 9} &    5.696\E{- 9} &    5.819\E{- 9} &    3.356\E{- 9} &    4.064\E{- 9} &    5.469\E{- 9} &         \NoData \\
\I{154}{Sm} &    2.983\E{- 9} &    3.575\E{- 9} &    3.538\E{- 9} &    3.834\E{- 9} &    4.970\E{- 9} &    5.090\E{- 9} &    2.965\E{- 9} &    3.528\E{- 9} &    4.706\E{- 9} &         \NoData \\
\I{151}{Eu} &    2.145\E{- 9} &    2.563\E{- 9} &    2.620\E{- 9} &    2.813\E{- 9} &    3.409\E{- 9} &    3.441\E{- 9} &    2.117\E{- 9} &    2.607\E{- 9} &    3.345\E{- 9} &         \NoData \\
\I{153}{Eu} &    2.340\E{- 9} &    2.828\E{- 9} &    2.918\E{- 9} &    3.123\E{- 9} &    3.609\E{- 9} &    3.632\E{- 9} &    2.327\E{- 9} &    2.906\E{- 9} &    3.565\E{- 9} &         \NoData \\
\I{152}{Gd} &    6.136\E{-11} &    7.365\E{-11} &    9.047\E{-11} &    6.956\E{-11} &    7.891\E{-11} &    7.729\E{-11} &    7.007\E{-11} &    9.720\E{-11} &    8.256\E{-11} &         \NoData \\
\I{154}{Gd} &    4.197\E{-10} &    5.224\E{-10} &    6.522\E{-10} &    5.601\E{-10} &    6.197\E{-10} &    6.144\E{-10} &    4.404\E{-10} &    6.485\E{-10} &    6.225\E{-10} &         \NoData \\
\I{155}{Gd} &    2.324\E{- 9} &    2.878\E{- 9} &    2.897\E{- 9} &    3.250\E{- 9} &    3.804\E{- 9} &    3.822\E{- 9} &    2.308\E{- 9} &    2.881\E{- 9} &    3.700\E{- 9} &         \NoData \\
\I{156}{Gd} &    3.419\E{- 9} &    4.150\E{- 9} &    4.258\E{- 9} &    4.574\E{- 9} &    5.282\E{- 9} &    5.308\E{- 9} &    3.417\E{- 9} &    4.250\E{- 9} &    5.189\E{- 9} &         \NoData \\
\I{157}{Gd} &    2.505\E{- 9} &    3.189\E{- 9} &    3.220\E{- 9} &    3.419\E{- 9} &    4.202\E{- 9} &    4.181\E{- 9} &    2.498\E{- 9} &    3.191\E{- 9} &    4.046\E{- 9} &         \NoData \\
\I{158}{Gd} &    4.285\E{- 9} &    5.486\E{- 9} &    5.337\E{- 9} &    6.113\E{- 9} &    7.489\E{- 9} &    7.435\E{- 9} &    4.260\E{- 9} &    5.294\E{- 9} &    7.138\E{- 9} &         \NoData \\
\I{160}{Gd} &    3.551\E{- 9} &    4.628\E{- 9} &    4.500\E{- 9} &    4.762\E{- 9} &    5.811\E{- 9} &    5.779\E{- 9} &    3.553\E{- 9} &    4.494\E{- 9} &    5.639\E{- 9} &         \NoData \\
\I{159}{Tb} &    2.903\E{- 9} &    3.666\E{- 9} &    3.712\E{- 9} &    4.038\E{- 9} &    4.555\E{- 9} &    4.551\E{- 9} &    2.889\E{- 9} &    3.679\E{- 9} &    4.490\E{- 9} &         \NoData \\
\I{156}{Dy} &    1.957\E{-11} &    2.144\E{-11} &    3.476\E{-10} &    4.754\E{-11} &    6.451\E{-11} &    7.153\E{-11} &    1.916\E{-11} &    3.450\E{-10} &    5.478\E{-11} &         \NoData \\
\I{158}{Dy} &    2.715\E{-11} &    3.794\E{-11} &    7.162\E{-10} &    6.831\E{-11} &    8.994\E{-11} &    9.621\E{-11} &    2.657\E{-11} &    6.223\E{-10} &    7.544\E{-11} &         \NoData \\
\I{160}{Dy} &    6.672\E{-10} &    7.604\E{-10} &    1.343\E{- 9} &    1.103\E{- 9} &    1.321\E{- 9} &    1.393\E{- 9} &    6.711\E{-10} &    1.281\E{- 9} &    1.224\E{- 9} &         \NoData \\
\I{161}{Dy} &    3.592\E{- 9} &    4.516\E{- 9} &    4.611\E{- 9} &    4.939\E{- 9} &    5.677\E{- 9} &    5.680\E{- 9} &    3.577\E{- 9} &    4.583\E{- 9} &    5.597\E{- 9} &         \NoData \\
\I{162}{Dy} &    5.187\E{- 9} &    6.482\E{- 9} &    6.500\E{- 9} &    7.252\E{- 9} &    8.285\E{- 9} &    8.281\E{- 9} &    5.176\E{- 9} &    6.495\E{- 9} &    8.111\E{- 9} &         \NoData \\
\I{163}{Dy} &    4.790\E{- 9} &    6.012\E{- 9} &    6.304\E{- 9} &    6.518\E{- 9} &    7.392\E{- 9} &    7.383\E{- 9} &    4.786\E{- 9} &    6.262\E{- 9} &    7.336\E{- 9} &         \NoData \\
\I{164}{Dy} &    6.031\E{- 9} &    7.984\E{- 9} &    7.560\E{- 9} &    8.725\E{- 9} &    1.066\E{- 8} &    1.051\E{- 8} &    6.009\E{- 9} &    7.505\E{- 9} &    1.010\E{- 8} &         \NoData \\
\I{165}{Ho} &    4.418\E{- 9} &    5.503\E{- 9} &    5.747\E{- 9} &    6.125\E{- 9} &    7.002\E{- 9} &    7.003\E{- 9} &    4.400\E{- 9} &    5.702\E{- 9} &    6.896\E{- 9} &         \NoData \\
\I{162}{Er} &    4.070\E{-11} &    4.884\E{-11} &    1.245\E{- 9} &    7.384\E{-11} &    1.095\E{-10} &    1.242\E{-10} &    3.937\E{-11} &    1.186\E{- 9} &    9.111\E{-11} &         \NoData \\
\I{164}{Er} &    3.048\E{-10} &    2.796\E{-10} &    1.201\E{- 9} &    5.053\E{-10} &    6.628\E{-10} &    7.027\E{-10} &    3.037\E{-10} &    1.021\E{- 9} &    6.179\E{-10} &         \NoData \\
\I{166}{Er} &    4.495\E{- 9} &    5.583\E{- 9} &    6.144\E{- 9} &    6.212\E{- 9} &    7.364\E{- 9} &    7.449\E{- 9} &    4.488\E{- 9} &    6.108\E{- 9} &    7.148\E{- 9} &         \NoData \\
\I{167}{Er} &    2.945\E{- 9} &    3.751\E{- 9} &    3.989\E{- 9} &    4.024\E{- 9} &    4.677\E{- 9} &    4.694\E{- 9} &    2.940\E{- 9} &    3.938\E{- 9} &    4.590\E{- 9} &         \NoData \\
\I{168}{Er} &    3.894\E{- 9} &    5.019\E{- 9} &    4.739\E{- 9} &    5.552\E{- 9} &    6.998\E{- 9} &    6.935\E{- 9} &    3.852\E{- 9} &    4.703\E{- 9} &    6.569\E{- 9} &         \NoData \\
\I{170}{Er} &    2.064\E{- 9} &    3.187\E{- 9} &    2.559\E{- 9} &    3.009\E{- 9} &    3.790\E{- 9} &    3.792\E{- 9} &    2.043\E{- 9} &    2.544\E{- 9} &    3.512\E{- 9} &         \NoData \\
\I{169}{Tm} &    2.028\E{- 9} &    2.658\E{- 9} &    2.730\E{- 9} &    2.912\E{- 9} &    3.317\E{- 9} &    3.316\E{- 9} &    2.007\E{- 9} &    2.684\E{- 9} &    3.220\E{- 9} &         \NoData \\
\I{168}{Yb} &    7.253\E{-11} &    6.792\E{-11} &    1.737\E{- 9} &    1.582\E{-10} &    2.272\E{-10} &    2.478\E{-10} &    7.240\E{-11} &    1.550\E{- 9} &    1.946\E{-10} &         \NoData \\
\I{170}{Yb} &    6.438\E{-10} &    7.639\E{-10} &    1.154\E{- 9} &    1.032\E{- 9} &    1.255\E{- 9} &    1.328\E{- 9} &    6.417\E{-10} &    1.053\E{- 9} &    1.158\E{- 9} &         \NoData \\
\I{171}{Yb} &    1.974\E{- 9} &    2.612\E{- 9} &    2.590\E{- 9} &    2.831\E{- 9} &    3.329\E{- 9} &    3.328\E{- 9} &    1.951\E{- 9} &    2.544\E{- 9} &    3.200\E{- 9} &         \NoData \\
\I{172}{Yb} &    3.518\E{- 9} &    4.538\E{- 9} &    4.672\E{- 9} &    5.150\E{- 9} &    6.318\E{- 9} &    6.345\E{- 9} &    3.473\E{- 9} &    4.609\E{- 9} &    5.972\E{- 9} &         \NoData \\
\I{173}{Yb} &    2.180\E{- 9} &    2.926\E{- 9} &    2.787\E{- 9} &    3.140\E{- 9} &    3.617\E{- 9} &    3.617\E{- 9} &    2.168\E{- 9} &    2.760\E{- 9} &    3.497\E{- 9} &         \NoData \\
\I{174}{Yb} &    4.646\E{- 9} &    6.303\E{- 9} &    5.713\E{- 9} &    7.240\E{- 9} &    9.348\E{- 9} &    9.207\E{- 9} &    4.609\E{- 9} &    5.651\E{- 9} &    8.527\E{- 9} &         \NoData \\
\I{176}{Yb} &    1.771\E{- 9} &    2.851\E{- 9} &    2.309\E{- 9} &    2.638\E{- 9} &    3.010\E{- 9} &    3.048\E{- 9} &    1.760\E{- 9} &    2.294\E{- 9} &    2.884\E{- 9} &         \NoData \\
\I{175}{Lu} &    1.946\E{- 9} &    2.509\E{- 9} &    2.515\E{- 9} &    2.795\E{- 9} &    3.227\E{- 9} &    3.237\E{- 9} &    1.933\E{- 9} &    2.495\E{- 9} &    3.119\E{- 9} &         \NoData \\
\I{176}{Lu} &    7.749\E{-11} &    1.221\E{-10} &    8.595\E{-11} &    1.052\E{-10} &    1.260\E{-10} &    1.226\E{-10} &    7.376\E{-11} &    8.446\E{-11} &    1.186\E{-10} &         \NoData \\
\I{174}{Hf} &    7.755\E{-11} &    8.733\E{-11} &    7.201\E{-10} &    1.666\E{-10} &    2.104\E{-10} &    2.306\E{-10} &    7.604\E{-11} &    6.357\E{-10} &    1.676\E{-10} &         \NoData \\
\hline
\PPI&\PPE&\PPE&\PPE&\PPE&\PPE&\PPE&\PPE&\PPE&\PPE&\PPE \\
\end{tabular}
}% end scale box
\vspace{-1.5\baselineskip}
\begin{flushright}\textsc{(continued on next page)}\end{flushright}
\end{table}

\clearpage

\addtocounter{table}{-1}

\begin{table}
\setlength{\tabcolsep}{1ex}
\centering
\caption{(continued) yields}
\scalebox{\Scale}{
\begin{tabular}{r@{}lr@{}lr@{}lr@{}lr@{}lr@{}lr@{}lr@{}lr@{}lr@{}lr@{}l}
\hline
\hline
\multicolumn{2}{c}{ion} &
\multicolumn{2}{c}{   S15} &
\multicolumn{2}{c}{   S19} &
\multicolumn{2}{c}{   S20} &
\multicolumn{2}{c}{   S21} &
\multicolumn{2}{c}{   S25} &
\multicolumn{2}{c}{  S25P} &
\multicolumn{2}{c}{   N15} &
\multicolumn{2}{c}{   N20} &
\multicolumn{2}{c}{   N25} &
\multicolumn{2}{c}{   H25} \\
\hline
\I{176}{Hf} &    6.674\E{-10} &    7.909\E{-10} &    7.830\E{-10} &    1.282\E{- 9} &    1.402\E{- 9} &    1.453\E{- 9} &    6.557\E{-10} &    7.642\E{-10} &    1.231\E{- 9} &         \NoData \\
\I{177}{Hf} &    1.582\E{- 9} &    2.084\E{- 9} &    2.002\E{- 9} &    2.258\E{- 9} &    2.638\E{- 9} &    2.643\E{- 9} &    1.570\E{- 9} &    1.989\E{- 9} &    2.548\E{- 9} &         \NoData \\
\I{178}{Hf} &    2.732\E{- 9} &    3.384\E{- 9} &    3.386\E{- 9} &    4.009\E{- 9} &    5.131\E{- 9} &    5.153\E{- 9} &    2.697\E{- 9} &    3.360\E{- 9} &    4.733\E{- 9} &         \NoData \\
\I{179}{Hf} &    1.213\E{- 9} &    1.611\E{- 9} &    1.592\E{- 9} &    1.781\E{- 9} &    2.062\E{- 9} &    2.066\E{- 9} &    1.207\E{- 9} &    1.571\E{- 9} &    1.972\E{- 9} &         \NoData \\
\I{180}{Hf} &    3.422\E{- 9} &    4.977\E{- 9} &    4.074\E{- 9} &    5.708\E{- 9} &    6.819\E{- 9} &    6.722\E{- 9} &    3.402\E{- 9} &    4.033\E{- 9} &    6.146\E{- 9} &         \NoData \\
\I{180}{Ta} &    9.794\E{-13} &    2.155\E{-12} &    2.524\E{-12} &    4.036\E{-12} &    8.862\E{-12} &    8.846\E{-12} &    8.633\E{-13} &    2.051\E{-12} &    6.210\E{-12} &         \NoData \\
\I{181}{Ta} &    1.236\E{- 9} &    1.781\E{- 9} &    1.639\E{- 9} &    1.929\E{- 9} &    2.412\E{- 9} &    2.392\E{- 9} &    1.227\E{- 9} &    1.621\E{- 9} &    2.245\E{- 9} &         \NoData \\
\I{180}{ W} &    9.414\E{-11} &    9.019\E{-11} &    2.290\E{-10} &    1.900\E{-10} &    2.531\E{-10} &    2.831\E{-10} &    9.395\E{-11} &    2.091\E{-10} &    2.167\E{-10} &         \NoData \\
\I{182}{ W} &    2.338\E{- 9} &    4.140\E{- 9} &    2.914\E{- 9} &    3.783\E{- 9} &    4.403\E{- 9} &    4.424\E{- 9} &    2.325\E{- 9} &    2.887\E{- 9} &    4.085\E{- 9} &         \NoData \\
\I{183}{ W} &    1.176\E{- 9} &    1.709\E{- 9} &    1.492\E{- 9} &    1.784\E{- 9} &    2.127\E{- 9} &    2.119\E{- 9} &    1.161\E{- 9} &    1.478\E{- 9} &    2.012\E{- 9} &         \NoData \\
\I{184}{ W} &    2.653\E{- 9} &    3.306\E{- 9} &    3.053\E{- 9} &    4.140\E{- 9} &    5.096\E{- 9} &    5.045\E{- 9} &    2.646\E{- 9} &    3.033\E{- 9} &    4.683\E{- 9} &         \NoData \\
\I{186}{ W} &    2.305\E{- 9} &    3.585\E{- 9} &    2.872\E{- 9} &    3.520\E{- 9} &    3.808\E{- 9} &    3.812\E{- 9} &    2.300\E{- 9} &    2.866\E{- 9} &    3.632\E{- 9} &         \NoData \\
\I{185}{Re} &    1.125\E{- 9} &    1.432\E{- 9} &    1.410\E{- 9} &    1.595\E{- 9} &    1.847\E{- 9} &    1.849\E{- 9} &    1.120\E{- 9} &    1.404\E{- 9} &    1.791\E{- 9} &         \NoData \\
\I{187}{Re} &    1.964\E{- 9} &    2.450\E{- 9} &    2.495\E{- 9} &    2.688\E{- 9} &    3.143\E{- 9} &    3.141\E{- 9} &    1.960\E{- 9} &    2.493\E{- 9} &    3.089\E{- 9} &         \NoData \\
\I{184}{Os} &    3.213\E{-11} &    3.892\E{-11} &    2.342\E{-10} &    8.055\E{-11} &    1.365\E{-10} &    1.483\E{-10} &    3.372\E{-11} &    2.185\E{-10} &    1.066\E{-10} &         \NoData \\
\I{186}{Os} &    6.897\E{-10} &    8.528\E{-10} &    9.142\E{-10} &    1.015\E{- 9} &    1.209\E{- 9} &    1.210\E{- 9} &    6.928\E{-10} &    9.091\E{-10} &    1.148\E{- 9} &         \NoData \\
\I{187}{Os} &    4.691\E{-10} &    5.780\E{-10} &    6.027\E{-10} &    6.501\E{-10} &    7.468\E{-10} &    7.441\E{-10} &    4.697\E{-10} &    6.005\E{-10} &    7.349\E{-10} &         \NoData \\
\I{188}{Os} &    5.290\E{- 9} &    7.069\E{- 9} &    6.863\E{- 9} &    7.183\E{- 9} &    8.685\E{- 9} &    8.633\E{- 9} &    5.304\E{- 9} &    6.875\E{- 9} &    8.538\E{- 9} &         \NoData \\
\I{189}{Os} &    6.032\E{- 9} &    7.416\E{- 9} &    7.647\E{- 9} &    8.000\E{- 9} &    9.166\E{- 9} &    9.159\E{- 9} &    6.041\E{- 9} &    7.653\E{- 9} &    9.147\E{- 9} &         \NoData \\
\I{190}{Os} &    1.053\E{- 8} &    1.289\E{- 8} &    1.321\E{- 8} &    1.410\E{- 8} &    1.640\E{- 8} &    1.631\E{- 8} &    1.060\E{- 8} &    1.325\E{- 8} &    1.626\E{- 8} &         \NoData \\
\I{192}{Os} &    1.592\E{- 8} &    1.968\E{- 8} &    2.004\E{- 8} &    2.109\E{- 8} &    2.427\E{- 8} &    2.425\E{- 8} &    1.600\E{- 8} &    2.011\E{- 8} &    2.415\E{- 8} &         \NoData \\
\I{191}{Ir} &    1.370\E{- 8} &    1.675\E{- 8} &    1.745\E{- 8} &    1.813\E{- 8} &    2.070\E{- 8} &    2.069\E{- 8} &    1.373\E{- 8} &    1.748\E{- 8} &    2.070\E{- 8} &         \NoData \\
\I{193}{Ir} &    2.346\E{- 8} &    2.873\E{- 8} &    2.992\E{- 8} &    3.102\E{- 8} &    3.541\E{- 8} &    3.534\E{- 8} &    2.351\E{- 8} &    2.996\E{- 8} &    3.547\E{- 8} &         \NoData \\
\I{190}{Pt} &    1.421\E{-11} &    1.707\E{-11} &    1.611\E{-10} &    2.777\E{-11} &    4.122\E{-11} &    4.374\E{-11} &    1.525\E{-11} &    1.662\E{-10} &    3.745\E{-11} &         \NoData \\
\I{192}{Pt} &    1.063\E{- 9} &    1.295\E{- 9} &    1.793\E{- 9} &    1.249\E{- 9} &    1.443\E{- 9} &    1.417\E{- 9} &    1.132\E{- 9} &    1.890\E{- 9} &    1.490\E{- 9} &         \NoData \\
\I{194}{Pt} &    2.648\E{- 8} &    3.273\E{- 8} &    3.587\E{- 8} &    3.468\E{- 8} &    3.976\E{- 8} &    3.956\E{- 8} &    2.663\E{- 8} &    3.617\E{- 8} &    3.987\E{- 8} &         \NoData \\
\I{195}{Pt} &    2.614\E{- 8} &    3.202\E{- 8} &    3.364\E{- 8} &    3.443\E{- 8} &    3.924\E{- 8} &    3.916\E{- 8} &    2.622\E{- 8} &    3.369\E{- 8} &    3.934\E{- 8} &         \NoData \\
\I{196}{Pt} &    2.174\E{- 8} &    2.691\E{- 8} &    2.745\E{- 8} &    2.882\E{- 8} &    3.339\E{- 8} &    3.308\E{- 8} &    2.176\E{- 8} &    2.729\E{- 8} &    3.315\E{- 8} &         \NoData \\
\I{198}{Pt} &    6.098\E{- 9} &    7.372\E{- 9} &    7.432\E{- 9} &    8.426\E{- 9} &    9.862\E{- 9} &    9.905\E{- 9} &    6.020\E{- 9} &    7.439\E{- 9} &    9.637\E{- 9} &         \NoData \\
\I{197}{Au} &    1.151\E{- 8} &    1.417\E{- 8} &    1.563\E{- 8} &    1.574\E{- 8} &    1.770\E{- 8} &    1.763\E{- 8} &    1.149\E{- 8} &    1.558\E{- 8} &    1.762\E{- 8} &         \NoData \\
\I{196}{Hg} &    3.747\E{-10} &    3.406\E{-10} &    3.364\E{- 9} &    8.151\E{-10} &    1.155\E{- 9} &    1.302\E{- 9} &    3.855\E{-10} &    3.442\E{- 9} &    1.107\E{- 9} &         \NoData \\
\I{198}{Hg} &    3.433\E{- 9} &    4.603\E{- 9} &    7.411\E{- 9} &    5.628\E{- 9} &    6.505\E{- 9} &    6.733\E{- 9} &    3.358\E{- 9} &    7.174\E{- 9} &    6.225\E{- 9} &         \NoData \\
\I{199}{Hg} &    3.971\E{- 9} &    4.999\E{- 9} &    6.990\E{- 9} &    6.059\E{- 9} &    6.937\E{- 9} &    6.964\E{- 9} &    3.860\E{- 9} &    6.817\E{- 9} &    6.663\E{- 9} &         \NoData \\
\I{200}{Hg} &    6.517\E{- 9} &    8.088\E{- 9} &    1.481\E{- 8} &    1.105\E{- 8} &    1.336\E{- 8} &    1.383\E{- 8} &    6.429\E{- 9} &    1.440\E{- 8} &    1.266\E{- 8} &         \NoData \\
\I{201}{Hg} &    3.185\E{- 9} &    4.252\E{- 9} &    5.026\E{- 9} &    5.387\E{- 9} &    5.841\E{- 9} &    5.940\E{- 9} &    3.154\E{- 9} &    4.890\E{- 9} &    5.622\E{- 9} &         \NoData \\
\I{202}{Hg} &    8.202\E{- 9} &    1.080\E{- 8} &    1.989\E{- 8} &    1.318\E{- 8} &    1.649\E{- 8} &    1.658\E{- 8} &    8.312\E{- 9} &    1.895\E{- 8} &    1.609\E{- 8} &         \NoData \\
\I{204}{Hg} &    2.220\E{- 9} &    3.935\E{- 9} &    2.817\E{- 9} &    2.674\E{- 9} &    3.638\E{- 9} &    3.609\E{- 9} &    2.266\E{- 9} &    2.875\E{- 9} &    3.611\E{- 9} &         \NoData \\
\I{203}{Tl} &    4.112\E{- 9} &    6.425\E{- 9} &    5.831\E{- 9} &    7.378\E{- 9} &    7.634\E{- 9} &    7.739\E{- 9} &    4.113\E{- 9} &    5.805\E{- 9} &    7.633\E{- 9} &         \NoData \\
\I{205}{Tl} &    9.913\E{- 9} &    1.335\E{- 8} &    1.189\E{- 8} &    1.551\E{- 8} &    1.727\E{- 8} &    1.730\E{- 8} &    1.005\E{- 8} &    1.195\E{- 8} &    1.775\E{- 8} &         \NoData \\
\I{204}{Pb} &    4.864\E{- 9} &    6.112\E{- 9} &    1.594\E{- 8} &    8.532\E{- 9} &    9.889\E{- 9} &    1.005\E{- 8} &    4.864\E{- 9} &    1.535\E{- 8} &    1.009\E{- 8} &         \NoData \\
\I{206}{Pb} &    4.497\E{- 8} &    6.254\E{- 8} &    6.142\E{- 8} &    7.521\E{- 8} &    9.188\E{- 8} &    9.087\E{- 8} &    4.509\E{- 8} &    6.122\E{- 8} &    9.700\E{- 8} &         \NoData \\
\I{207}{Pb} &    4.978\E{- 8} &    7.910\E{- 8} &    5.915\E{- 8} &    8.798\E{- 8} &    1.081\E{- 7} &    1.066\E{- 7} &    4.937\E{- 8} &    5.879\E{- 8} &    1.079\E{- 7} &         \NoData \\
\I{208}{Pb} &    1.343\E{- 7} &    1.973\E{- 7} &    1.655\E{- 7} &    2.217\E{- 7} &    2.847\E{- 7} &    2.819\E{- 7} &    1.324\E{- 7} &    1.628\E{- 7} &    2.649\E{- 7} &         \NoData \\
\I{209}{Bi} &    1.017\E{- 8} &    1.395\E{- 8} &    1.249\E{- 8} &    1.559\E{- 8} &    1.922\E{- 8} &    1.911\E{- 8} &    1.009\E{- 8} &    1.239\E{- 8} &    1.833\E{- 8} &         \NoData \\
\hline
\PPI&\PPE&\PPE&\PPE&\PPE&\PPE&\PPE&\PPE&\PPE&\PPE&\PPE \\
\end{tabular}
}% end scale box
\vspace{-1.5\baselineskip}
\begin{flushright}\textsc{(end of yield table)}\end{flushright}
\end{table}

}% END OF TABLE ENVIRONMENT

\clearpage

{% BEGINNING OF TABLE ENVIRONMENT

%-----------------------------------------------------------------------
% These commands are required in for the tables to work properly
\renewcommand{\E}[1]{&{\ensuremath{(#1)}}}
\newcommand{\EE}{&}
\renewcommand{\I}[2]{{\ensuremath{^{#1}}}&{\ensuremath{\mathrm{#2}}}}
\newcommand{\NoData}{\multicolumn{2}{c}{\nodata}}
\newcommand{\PPI}{\I{\phantom{99}}{\phantom{Mm}}}
\newcommand{\PPE}{\phantom{$9.99$}&\phantom{$(-99)$}}
%-----------------------------------------------------------------------

%%%%%%%%%%%%%%%%%%%%%%%%%%%%%%%%%%%%%%%%%%%%%%%%%%%%%%%%%%%%%%%%%%%%%%%%
% Radioactive Yields
\newcommand{\Scale}{     0.760000}

\begin{table}
\setlength{\tabcolsep}{1ex}
\centering
\caption{\scshape Radioactive Yields (in solar masses)\lTab{r}}
\scalebox{\Scale}{
\begin{tabular}{r@{}lr@{}lr@{}lr@{}lr@{}lr@{}lr@{}lr@{}lr@{}lr@{}lr@{}l}
\hline
\hline
\multicolumn{2}{c}{ion} &
\multicolumn{2}{c}{   S15} &
\multicolumn{2}{c}{   S19} &
\multicolumn{2}{c}{   S20} &
\multicolumn{2}{c}{   S21} &
\multicolumn{2}{c}{   S25} &
\multicolumn{2}{c}{  S25P} &
\multicolumn{2}{c}{   N15} &
\multicolumn{2}{c}{   N20} &
\multicolumn{2}{c}{   N25} &
\multicolumn{2}{c}{   H25} \\
\hline
\I{  3}{ H} &    2.385\E{-10} &    3.358\E{-10} &    2.141\E{-10} &    2.297\E{-10} &    2.086\E{-11} &    2.051\E{-11} &    2.234\E{-10} &    7.774\E{-10} &    2.037\E{-11} &    2.078\E{-11} \\
\I{ 14}{ C} &    4.744\E{- 5} &    9.458\E{- 6} &    6.054\E{- 6} &    1.822\E{- 5} &    9.611\E{- 6} &    1.070\E{- 5} &    3.917\E{- 5} &    5.155\E{- 6} &    7.408\E{- 6} &    2.458\E{- 6} \\
\I{ 22}{Na} &    8.019\E{- 7} &    4.929\E{- 6} &    4.422\E{- 7} &    4.038\E{- 6} &    3.998\E{- 6} &    3.703\E{- 6} &    9.461\E{- 7} &    5.433\E{- 7} &    4.349\E{- 6} &    3.907\E{- 6} \\
\I{ 26}{Al} &    2.589\E{- 5} &    3.182\E{- 5} &    2.969\E{- 5} &    4.574\E{- 5} &    6.991\E{- 5} &    6.953\E{- 5} &    2.704\E{- 5} &    2.858\E{- 5} &    6.495\E{- 5} &    4.806\E{-13} \\
\I{ 32}{Si} &    1.862\E{- 6} &    5.142\E{- 7} &    2.265\E{- 7} &    4.915\E{- 7} &    2.407\E{- 6} &    3.145\E{- 6} &    1.733\E{- 6} &    1.977\E{- 7} &    2.299\E{- 6} &    2.366\E{- 6} \\
\I{ 36}{Cl} &    1.499\E{- 6} &    2.605\E{- 6} &    1.165\E{- 4} &    2.757\E{- 6} &    6.884\E{- 6} &    6.521\E{- 6} &    1.596\E{- 6} &    1.111\E{- 4} &    7.305\E{- 6} &    6.308\E{- 6} \\
\I{ 39}{Ar} &    5.607\E{- 6} &    1.498\E{- 5} &    4.252\E{- 5} &    1.404\E{- 5} &    2.235\E{- 5} &    2.215\E{- 5} &    5.178\E{- 6} &    3.795\E{- 5} &    1.944\E{- 5} &    1.832\E{- 5} \\
\I{ 42}{Ar} &    2.704\E{- 8} &    2.781\E{- 9} &    1.395\E{- 9} &    2.148\E{- 9} &    2.805\E{- 8} &    4.655\E{- 8} &    2.060\E{- 8} &    1.098\E{- 9} &    2.021\E{- 8} &    3.226\E{- 8} \\
\I{ 41}{Ca} &    4.316\E{- 6} &    2.658\E{- 5} &    4.285\E{- 4} &    6.942\E{- 6} &    3.218\E{- 5} &    2.650\E{- 5} &    4.132\E{- 6} &    5.577\E{- 4} &    3.954\E{- 5} &    2.890\E{- 5} \\
\I{ 45}{Ca} &    9.659\E{- 7} &    2.676\E{- 6} &    4.389\E{- 6} &    9.296\E{- 7} &    3.403\E{- 6} &    3.433\E{- 6} &    9.003\E{- 7} &    3.921\E{- 6} &    3.023\E{- 6} &    1.880\E{- 6} \\
\I{ 44}{Ti} &    1.394\E{- 5} &    2.553\E{- 5} &    4.869\E{- 5} &    1.722\E{- 5} &    1.563\E{- 5} &    4.827\E{- 5} &    9.780\E{- 6} &    4.341\E{- 5} &    1.144\E{- 5} &    2.296\E{- 5} \\
\I{ 49}{ V} &    6.921\E{- 6} &    9.117\E{- 6} &    2.266\E{- 5} &    9.411\E{- 6} &    1.632\E{- 5} &    1.778\E{- 5} &    7.116\E{- 6} &    2.654\E{- 5} &    1.692\E{- 5} &    1.635\E{- 5} \\
\I{ 53}{Mn} &    1.791\E{- 4} &    2.127\E{- 4} &    1.276\E{- 4} &    2.289\E{- 4} &    3.624\E{- 4} &    4.043\E{- 4} &    1.977\E{- 4} &    1.265\E{- 4} &    3.855\E{- 4} &    3.160\E{- 4} \\
\I{ 54}{Mn} &    3.019\E{- 6} &    4.290\E{- 6} &    4.938\E{- 6} &    3.861\E{- 6} &    7.706\E{- 6} &    7.307\E{- 6} &    3.245\E{- 6} &    5.523\E{- 6} &    8.039\E{- 6} &    5.546\E{- 6} \\
\I{ 55}{Fe} &    1.106\E{- 3} &    1.293\E{- 3} &    7.741\E{- 4} &    1.320\E{- 3} &    2.073\E{- 3} &    2.396\E{- 3} &    1.225\E{- 3} &    7.460\E{- 4} &    2.217\E{- 3} &    1.263\E{- 3} \\
\I{ 60}{Fe} &    6.636\E{- 5} &    1.096\E{- 4} &    3.586\E{- 5} &    2.452\E{- 5} &    1.508\E{- 4} &    1.647\E{- 4} &    6.788\E{- 5} &    3.575\E{- 5} &    1.583\E{- 4} &    9.850\E{- 5} \\
\I{ 57}{Co} &    3.754\E{- 3} &    3.502\E{- 3} &    4.394\E{- 3} &    3.082\E{- 3} &    2.658\E{- 3} &    6.044\E{- 3} &    3.515\E{- 3} &    4.337\E{- 3} &    2.346\E{- 3} &    2.555\E{- 3} \\
\I{ 60}{Co} &    8.343\E{- 5} &    1.637\E{- 4} &    6.421\E{- 5} &    5.867\E{- 5} &    2.197\E{- 4} &    2.310\E{- 4} &    8.632\E{- 5} &    6.590\E{- 5} &    2.403\E{- 4} &    1.778\E{- 4} \\
\I{ 56}{Ni} &    1.116\E{- 1} &    1.039\E{- 1} &    9.005\E{- 2} &    1.036\E{- 1} &    1.056\E{- 1} &    1.944\E{- 1} &    1.117\E{- 1} &    9.115\E{- 2} &    1.053\E{- 1} &    1.094\E{- 1} \\
\I{ 57}{Ni} &    3.732\E{- 3} &    3.244\E{- 3} &    4.166\E{- 3} &    2.918\E{- 3} &    2.429\E{- 3} &    5.738\E{- 3} &    3.494\E{- 3} &    4.101\E{- 3} &    2.134\E{- 3} &    2.339\E{- 3} \\
\I{ 59}{Ni} &    2.597\E{- 4} &    2.317\E{- 4} &    3.950\E{- 4} &    1.862\E{- 4} &    1.336\E{- 4} &    3.594\E{- 4} &    2.276\E{- 4} &    3.952\E{- 4} &    1.051\E{- 4} &    9.464\E{- 5} \\
\I{ 63}{Ni} &    7.165\E{- 5} &    2.436\E{- 4} &    5.324\E{- 5} &    2.617\E{- 4} &    3.823\E{- 4} &    3.718\E{- 4} &    6.203\E{- 5} &    4.196\E{- 5} &    3.149\E{- 4} &    3.992\E{- 4} \\
\I{ 65}{Zn} &    2.046\E{- 6} &    2.592\E{- 6} &    7.783\E{- 6} &    3.271\E{- 6} &    5.579\E{- 6} &    7.214\E{- 6} &    1.569\E{- 6} &    6.740\E{- 6} &    4.723\E{- 6} &    8.805\E{- 6} \\
\I{ 68}{Ge} &    2.223\E{- 8} &    1.939\E{- 8} &    4.581\E{- 8} &    2.146\E{- 8} &    5.060\E{- 8} &    5.824\E{- 8} &    1.510\E{- 8} &    3.724\E{- 8} &    3.250\E{- 8} &    3.965\E{- 8} \\
\I{ 79}{Se} &    6.207\E{- 7} &    3.317\E{- 6} &    8.498\E{- 7} &    3.863\E{- 6} &    8.268\E{- 6} &    8.269\E{- 6} &    3.513\E{- 7} &    3.456\E{- 7} &    3.146\E{- 6} &    1.716\E{- 5} \\
\I{ 81}{Kr} &    6.691\E{- 9} &    1.042\E{- 8} &    1.901\E{- 7} &    1.949\E{- 8} &    2.045\E{- 8} &    2.063\E{- 8} &    4.445\E{- 9} &    9.775\E{- 8} &    7.864\E{- 9} &    3.214\E{- 8} \\
\I{ 85}{Kr} &    5.688\E{- 7} &    2.933\E{- 6} &    7.495\E{- 7} &    1.762\E{- 6} &    6.960\E{- 6} &    6.869\E{- 6} &    3.077\E{- 7} &    2.893\E{- 7} &    2.334\E{- 6} &    6.719\E{- 6} \\
\I{ 90}{Sr} &    2.926\E{- 8} &    8.998\E{- 8} &    1.164\E{- 8} &    4.523\E{- 8} &    1.905\E{- 7} &    1.944\E{- 7} &    2.118\E{- 8} &    7.311\E{- 9} &    7.550\E{- 8} &    2.817\E{- 7} \\
\I{ 93}{Zr} &    1.297\E{- 8} &    4.723\E{- 8} &    9.824\E{- 9} &    5.855\E{- 8} &    1.450\E{- 7} &    1.412\E{- 7} &    9.485\E{- 9} &    6.069\E{- 9} &    6.373\E{- 8} &    2.691\E{- 7} \\
\I{ 91}{Nb} &    3.958\E{-11} &    4.955\E{-10} &    1.700\E{- 9} &    1.658\E{-10} &    7.397\E{-10} &    5.237\E{-10} &    3.275\E{-11} &    1.072\E{- 9} &    3.449\E{-10} &    1.030\E{- 9} \\
\I{ 92}{Nb} &    6.407\E{-12} &    1.475\E{-11} &    6.575\E{-10} &    2.706\E{-11} &    6.629\E{-11} &    5.804\E{-11} &    5.185\E{-12} &    3.977\E{-10} &    3.051\E{-11} &    1.438\E{-10} \\
\I{ 93}{Nb} &    3.224\E{- 8} &    7.076\E{- 8} &    3.873\E{- 8} &    8.404\E{- 8} &    1.742\E{- 7} &    1.704\E{- 7} &    2.884\E{- 8} &    3.313\E{- 8} &    9.276\E{- 8} &    2.994\E{- 7} \\
\I{ 93}{Mo} &    5.398\E{-11} &    3.924\E{-11} &    5.170\E{-10} &    2.135\E{-10} &    2.390\E{-10} &    2.599\E{-10} &    4.406\E{-11} &    4.179\E{-10} &    1.511\E{-10} &    7.229\E{-10} \\
\I{ 97}{Tc} &    4.827\E{-11} &    4.183\E{-11} &    1.881\E{-10} &    1.341\E{-10} &    8.305\E{-11} &    9.463\E{-11} &    4.824\E{-11} &    1.544\E{-10} &    6.333\E{-11} &    5.124\E{-10} \\
\I{ 99}{Tc} &    8.805\E{-10} &    2.566\E{- 9} &    1.244\E{- 9} &    3.483\E{- 9} &    3.804\E{- 9} &    3.871\E{- 9} &    6.737\E{-10} &    8.852\E{-10} &    2.201\E{- 9} &    1.070\E{- 8} \\
\I{106}{Ru} &    7.213\E{-10} &    1.466\E{- 9} &    4.728\E{-10} &    6.040\E{-10} &    2.861\E{- 9} &    2.823\E{- 9} &    5.783\E{-10} &    3.952\E{-10} &    1.770\E{- 9} &    5.147\E{- 7} \\
\I{101}{Rh} &    3.569\E{-11} &    2.380\E{-12} &    5.180\E{-11} &    1.087\E{-10} &    5.661\E{-11} &    7.466\E{-11} &    3.386\E{-11} &    4.787\E{-11} &    3.725\E{-11} &         \NoData \\
\I{107}{Pd} &    4.082\E{-10} &    8.424\E{-10} &    4.589\E{-10} &    1.440\E{- 9} &    1.433\E{- 9} &    1.424\E{- 9} &    3.626\E{-10} &    3.779\E{-10} &    9.729\E{-10} &         \NoData \\
\I{109}{Cd} &    2.987\E{-11} &    2.452\E{-11} &    1.514\E{-10} &    4.446\E{-11} &    4.807\E{-11} &    5.284\E{-11} &    2.461\E{-11} &    1.343\E{-10} &    3.868\E{-11} &         \NoData \\
\I{121}{Sn} &    8.071\E{-10} &    2.020\E{- 9} &    5.452\E{-10} &    1.215\E{- 9} &    5.365\E{- 9} &    5.252\E{- 9} &    7.405\E{-10} &    4.886\E{-10} &    3.818\E{- 9} &         \NoData \\
\I{126}{Sn} &    3.773\E{-10} &    8.287\E{-12} &    5.930\E{-12} &    3.204\E{-11} &    2.125\E{-10} &    4.473\E{-10} &    3.580\E{-10} &    6.182\E{-12} &    1.586\E{-10} &         \NoData \\
\I{129}{ I} &    6.748\E{-10} &    1.163\E{- 9} &    5.640\E{-10} &    1.329\E{- 9} &    2.255\E{- 9} &    2.245\E{- 9} &    6.028\E{-10} &    5.151\E{-10} &    1.917\E{- 9} &         \NoData \\
\I{135}{Cs} &    1.677\E{- 9} &    3.007\E{- 9} &    2.118\E{- 9} &    4.324\E{- 9} &    4.884\E{- 9} &    5.305\E{- 9} &    1.601\E{- 9} &    1.985\E{- 9} &    4.473\E{- 9} &         \NoData \\
\I{137}{Cs} &    3.874\E{- 9} &    1.782\E{- 8} &    2.602\E{- 9} &    6.522\E{- 9} &    1.202\E{- 8} &    1.161\E{- 8} &    3.750\E{- 9} &    2.402\E{- 9} &    1.033\E{- 8} &         \NoData \\
\I{137}{La} &    1.052\E{-10} &    1.228\E{-10} &    1.224\E{- 9} &    2.662\E{-10} &    3.810\E{-10} &    4.389\E{-10} &    9.862\E{-11} &    1.146\E{- 9} &    3.382\E{-10} &         \NoData \\
\I{144}{Ce} &    2.793\E{-10} &    4.824\E{-10} &    3.610\E{-10} &    5.584\E{-11} &    1.937\E{- 9} &    2.101\E{- 9} &    2.361\E{-10} &    3.133\E{-10} &    1.646\E{- 9} &         \NoData \\
\I{145}{Pm} &    2.768\E{-11} &    5.087\E{-12} &    1.251\E{-10} &    5.526\E{-11} &    5.467\E{-11} &    6.277\E{-11} &    2.555\E{-11} &    1.078\E{-10} &    4.544\E{-11} &         \NoData \\
\I{146}{Sm} &    3.940\E{-10} &    6.339\E{-12} &    3.383\E{-10} &    8.494\E{-10} &    1.236\E{- 9} &    1.286\E{- 9} &    3.914\E{-10} &    3.175\E{-10} &    1.122\E{- 9} &         \NoData \\
\I{151}{Sm} &    1.377\E{-10} &    1.063\E{-10} &    6.109\E{-11} &    1.557\E{-10} &    3.660\E{-10} &    3.988\E{-10} &    1.109\E{-10} &    5.250\E{-11} &    3.038\E{-10} &         \NoData \\
\I{163}{Ho} &    1.280\E{-11} &    1.549\E{-11} &    2.677\E{-10} &    3.368\E{-11} &    4.198\E{-11} &    4.658\E{-11} &    1.173\E{-11} &    2.325\E{-10} &    3.250\E{-11} &         \NoData \\
\I{166}{Ho} &    1.191\E{-10} &    2.297\E{-10} &    9.670\E{-11} &    2.926\E{-10} &    5.326\E{-10} &    5.550\E{-10} &    9.767\E{-11} &    8.617\E{-11} &    4.292\E{-10} &         \NoData \\
\I{182}{Hf} &    1.359\E{-10} &    1.486\E{- 9} &    2.497\E{-10} &    5.522\E{-10} &    4.314\E{-10} &    4.454\E{-10} &    1.217\E{-10} &    2.322\E{-10} &    3.361\E{-10} &         \NoData \\
\I{194}{Os} &    1.546\E{-10} &    4.107\E{-10} &    2.817\E{-10} &    3.332\E{-10} &    6.270\E{-10} &    6.048\E{-10} &    1.179\E{-10} &    2.732\E{-10} &    5.553\E{-10} &         \NoData \\
\I{194}{Hg} &    7.598\E{-11} &    7.171\E{-11} &    2.284\E{- 9} &    1.165\E{-10} &    1.756\E{-10} &    1.946\E{-10} &    8.995\E{-11} &    2.483\E{- 9} &    1.711\E{-10} &         \NoData \\
\I{204}{Tl} &    1.018\E{-10} &    3.335\E{-10} &    5.447\E{-11} &    9.407\E{-10} &    6.076\E{-10} &    5.764\E{-10} &    1.074\E{-10} &    5.980\E{-11} &    6.417\E{-10} &         \NoData \\
\I{202}{Pb} &    5.515\E{-10} &    6.993\E{-10} &    1.108\E{- 8} &    1.293\E{- 9} &    2.041\E{- 9} &    2.224\E{- 9} &    5.124\E{-10} &    1.007\E{- 8} &    1.913\E{- 9} &         \NoData \\
\I{205}{Pb} &    4.651\E{-10} &    6.602\E{-10} &    8.095\E{-10} &    1.471\E{- 9} &    1.703\E{- 9} &    1.698\E{- 9} &    4.837\E{-10} &    7.900\E{-10} &    2.120\E{- 9} &         \NoData \\
\I{207}{Bi} &    4.216\E{-11} &    5.185\E{-11} &    2.401\E{-10} &    1.400\E{-10} &    2.063\E{-10} &    2.193\E{-10} &    3.678\E{-11} &    2.092\E{-10} &    1.614\E{-10} &         \NoData \\
\I{210}{Bi} &    5.650\E{-11} &    2.067\E{-12} &    3.125\E{-11} &    2.095\E{-11} &    1.347\E{-10} &    1.647\E{-10} &    5.063\E{-11} &    2.882\E{-11} &    1.135\E{-10} &         \NoData \\
\hline
\PPI&\PPE&\PPE&\PPE&\PPE&\PPE&\PPE&\PPE&\PPE&\PPE&\PPE \\
\end{tabular}
}% end scale box
\vspace{-1.5\baselineskip}
\end{table}

}% END OF TABLE ENVIRONMENT

\end{document}